\documentclass[11pt,a4paper]{article}

\usepackage[utf8]{inputenc}    % Encoding
\usepackage[T1]{fontenc}       % Font encoding
\usepackage{lmodern}           % Font
\usepackage{geometry}          % Page layout
\usepackage{graphicx}          % For including graphics
\usepackage{amsmath, amssymb}  % Math packages
\usepackage{physics}           % Physics notation
\usepackage{hyperref}          % Hyperlinks
\usepackage{siunitx}           % SI units
\usepackage{float}             % Improved float control
\usepackage{caption}           % Captions
\usepackage{subcaption}        % Subfigures
\usepackage{booktabs}          % Professional tables
\usepackage[numbers,sort&compress]{natbib}            % Bibliography
\usepackage{doi}               % DOI links in bibliography
\usepackage{xcolor}            % Colors
\usepackage{authblk}           % For author affiliations
\usepackage{microtype}         % Better typography
\usepackage{enumitem}          % Better lists
\usepackage{url}               % URL formatting
\usepackage{cancel}               % URL formatting
\usepackage[font=footnotesize]{caption}
\usepackage{comment}
\usepackage{tikz}
\usepackage{pgfplots}
\pgfplotsset{compat=1.18}
\usepackage{pdflscape}

\usepackage{tikz}
\usepackage{pgfplots}
\pgfplotsset{compat=1.18}

\pgfmathdeclarefunction{capabilitypeak}{2}{%
  \pgfmathparse{
    (
      (#1*#2+1)*(2*#2-1)
      +sqrt(max(0,(#1*#2+1)^2-4*#2*(1-#2)))
    )
    /(2*#2*(#1*#2+2))
  }%
}

\title{The Dynamic Theory of Economic Complexity}

\author[1,2,3]{C\'esar A. Hidalgo\thanks{Corresponding author: cesar.hidalgo@tse-fr.eu}}
\author[1,2,4]{Viktor Stojkoski}
\affil[1]{\small Center for Collective Learning, IAST, Toulouse School of Economics}
\affil[2]{\small Center for Collective Learning, CIAS, Corvinus University of Budapest}
\affil[3]{\small Alliance Manchester Business School, University of Manchester}
\affil[4]{\small Ss. Cyril and Methodius University in Skopje}
\date{}
\begin{document}

\maketitle

\begin{abstract}
Why do some economic activities grow faster in some economies than others? We develop a microfounded dynamic model of economic growth by sector in which output is determined by a weak-link production function and investment contributes to the accumulation of shared capabilities. The model generates two sets of predictions. First, as activities become more complex, the economies in which they grow the fastest shift from lower- to higher-complexity economies. Second, shared capabilities generate path-dependent diversification, providing a parsimonious explanation for the principle of relatedness. Moreover, the model predicts that relatedness matters more for complex activities and less in more complex economies. We test these predictions using international trade data by product and employment by industry data for cities in the United States and Brazil, finding empirical patterns consistent with both sets of predictions. Together, these results connect production complementarities, conditional convergence, and path-dependent diversification through a common mechanism of capability accumulation, providing a dynamic theoretical foundation for economic complexity methods.
\end{abstract}  

\section{Introduction}

Why do some activities grow faster in some economies than in others?  \\

Growth theory provides a starting point for understanding these differences. Classic growth theory explains the ability of poorer economies to catch up to richer ones as a consequence of capital accumulation and diminishing returns ~\cite{solow_contribution_1956,barro1992convergence} while endogenous growth theory~\cite{romer_endogenous_1990,aghion_model_1992} explains divergence through knowledge spillovers and increasing returns. But while both traditions explain aggregate growth, these models generally abstract from the granular structure of production, where missing inputs, coordination failures, and bottlenecks can become key constraints.\\  

This is where O-ring or ``weak-link'' models come in~\cite{kremer_o-ring_1993,jones2011intermediate,jones2013ring}. These models bring complementarities and bottlenecks explicitly into growth theory, by emphasizing the idea that development depends on the joint presence of complementary and/or interdependent inputs. These complementarities can explain, for instance, why small differences in local factors can translate into large income gaps \cite{jones2011intermediate} or why less error-prone countries specialize in more complex products \cite{kremer_o-ring_1993}. Yet, the most commonly used versions of these models do not typically include heterogeneous input requirements across activities, limiting their ability to model fine-grained productive structures.\\

Economic complexity methods build on these traditions by focusing largely on specialization patterns and their evolution ~\cite{hidalgo_building_2009,hausmann_network_2011,hidalgo_economic_2021}. Like ``weak-link'' models, they assume complementary inputs, but they also assume each activity requires combinations of highly-specific inputs, or capabilities, that can be shared among activities. These shared inputs allow these models to explain fine-grained changes in the productive structure of an economy. But while these methods have been applied extensively in empirical work~\cite{hidalgo_building_2009,hidalgo_product_2007,jara-figueroa_role_2018,chavez_economic_2017,guevara_research_2016,chinazzi_mapping_2019}, their dynamical formulation is still incomplete.\\

In this paper, we develop a dynamic model of economic complexity that incorporates endogenous capability accumulation. In our model, output is reinvested in activities, helping build the capabilities required by these activities and by those sharing capabilities with them. Shared capabilities provide a natural mechanism for spillovers and for the study of activity specific growth and path dependencies. We then solve several versions of this model that we test empirically by exploring two sets of implications.\\

First, the model predicts that the fastest growing economies in an activity switch from low- to high-complexity as the activity becomes more complex. This provides a mechanism through which convergence becomes increasingly conditional as activities become more demanding of capabilities. Empirically, we test this pattern by looking at international and sub-national data on the growth of activities, finding that in both cases more complex activities tend to grow faster in economies with higher levels of economic complexity.\\

Second, the model provides a mechanistic explanation of path-dependent diversification, also known as the principle of relatedness \cite{hidalgo_product_2007,hidalgo_principle_2018}. This is the observation that an activity grows faster in economies with a larger number of related activities. Here, the model goes beyond the standard implication of the empirical literature, by predicting the signs of interaction terms connecting relatedness and complexity. In particular, the model predicts that relatedness matters more for more complex activities and less in more complex economies. That is, relatedness matters less when economies are well endowed with capabilities, and matters more for activities that are more demanding of capabilities. Both findings are again consistent with international trade data by product and employment data by industry for cities in the United States and Brazil.\\

These findings help show that a model of shared complementary capabilities can simultaneously explain product-level differences in conditional convergence and path-dependent structural transformation. The model also provides a theoretical foundation for key empirical methods in economic complexity that are used widely in international development policy and strategy.\\  

\section{Capability Complementarity and Conditional Convergence}

Our contribution is related to several strands of literature in both economics and complex systems. In economics, it is related to models focused on \emph{weak-links} or complementarities and to the broader literature on endogenous growth and conditional convergence. In complex systems, it is related to the literature on economic complexity, and in particular to work on economic diversification and relatedness.\\ 

Weak-link models emphasize the importance of complementarities and/or linkages between inputs. They are inspired in the historical work from Leontief (1936)~\cite{leontief_quantitative_1936} and Hirschman (1958)~\cite{hirschman_strategy_1958} focused on input-output relationships, but extend to other forms of complementarity. In economics, their formalization is attributed to the \emph{O-Ring} model introduced by Kremer (1993)~\cite{kremer_o-ring_1993}, and expanded by Charles Jones (2011)~\cite{jones2011intermediate} and Garret Jones (2013)~\cite{jones2013ring}. The same mechanism was also present in a model published in 1957 by William Shockley in a paper explaining research productivity~\cite{shockley1957statistics}. These models assume that output is a product of inputs, and thus, falls dramatically if one of these inputs is missing or in short supply. The name of the model comes from the 1986 explosion of the space shuttle Challenger, which involved a defective gasket or O-ring (the ``weak-link'' in a complex product).\\

Weak-link models are used to explain large cross-country differences in income~\cite{jones2011intermediate,jones2013ring} because input complementarities can lead to large multipliers while keeping development difficult. The latter comes from the idea that weak-links can act as bottlenecks, limiting the growth of productivity in economies when a few of these inputs are rare~\cite{jones2011intermediate}.\\ 

More recently, weak-link models have gained attention in work focused on AI and automation. Models using weak-link production functions have been used to argue that weak links can constrain the aggregate productivity gains from AI~\cite{jones2026ai} and to explore how complementarities among tasks shape the returns to automation~\cite{gans2026ring}. These models are also being used to model organizations as combinations of human and AI agents with multidimensional capabilities~\cite{imas2026organizing}, emphasizing the importance of complementarities and coordination.\\

Our work builds on an expanded version of this production function which assumes that activities differ in the intensity with which they require each input or factor~\cite{hausmann_network_2011,hidalgo2025theory}. This extra parameter (the intensity with which an input is required by an activity), allows us to model the fine-grained structure~\cite{hidalgo2025theory} and dynamics of economy-activity matrices (such as exports by country and product or employment by city and industry) of arbitrary sizes.\\

In this paper, we contribute to weak-link models by providing a dynamic version of them that we can use to explain patterns of growth by economy and activity. But the mechanisms that we will explore are also connected to other branches of economics, including economic growth~\cite{romer_endogenous_1990,aghion_model_1992}, conditional convergence~\cite{barro1992convergence,kremer2022converging}, and unified growth theory~\cite{galor2024unified,galor2005stagnation,galor2011unified}.\\

Let us start with convergence.\\ 

Convergence is observed when less developed economies grow faster than more developed ones. It is the default behavior of neoclassical growth models~\cite{solow_contribution_1956}. But convergence can be conditional on economies having first achieved a certain level of development.\\ 

Empirical work using cross-country and cross-regional regressions, like those of Baumol (1986)~\cite{baumol1986productivity} or Barro and Sala-i-Martin (1992)~\cite{barro1992convergence}, have long explored the conditionality of convergence. Baumol (1986)~\cite{baumol1986productivity} presented evidence of unconditional convergence matching the expectations of neoclassical growth models, which was then overturned by the work of Barro and Sala-i-Martin (1992)~\cite{barro1992convergence} who introduced the notion of convergence clubs.\\

In convergence clubs, convergence is conditional on the presence of other factors. For example, it is observed only after holding constant ``initial school enrollment rates and the ratio of government consumption to GDP'' ~\cite{barro1992convergence}.\footnote{There is also evidence of divergence, like the one presented in Pritchett (1997). He shows that ``the ratio of richest to poorest countries' income increased from roughly 9 to 1 to 45 to 1 [...] and the average income gap between the richest and all other countries grew nearly tenfold from \$1,286 to \$12,000.''} That is, that less developed economies converge only if they meet some minimum conditions.\\

Our work is related to the conditional convergence literature because this literature has long emphasized economic structure or industry mix as a conditional convergence factor~\cite{barro1992convergence,chatterji1992convergence,bernard1996productivity}. For example, Barro and Sala-i-Martin (1992)~\cite{barro1992convergence} used agriculture/industry mix as a convergence factor and Bernard and Jones~\cite{bernard1996productivity} explored convergence within industries in the United States.\\

During the last couple of decades, economic complexity indicators have become a go-to method to quantify these economic structures, showing that convergence is conditional on an economy's initial level of complexity after controlling for their income~\cite{hidalgo_building_2009,hausmann_atlas_2014,chavez_economic_2017,domini_patterns_2022,koch_economic_2021,stojkoski_impact_2016,stojkoski_relationship_2017,ourens_can_2012,poncet_economic_2013,stojkoski_multidimensional_2022,tacchella_new_2012,bustos_production_2022,teixeira2022economic,hoeriyah2022economic,mao2021economic,basile_economic_2022,cardoso2023export,romero2021economic,perez2019measuring}. In this paper, we continue this tradition, but focused on convergence patterns at the fine-grained level of an economy-activity (e.g. country-product, city-industry, etc.)\\

\begin{comment}
On the other hand, our work provides a formal way to derive a tunable conditional convergence model, where growth rates rise and fall with an economy's level of development\footnote{Providing a theoretical foundation for charts like the one presented in  Chatterji\cite{chatterji1992convergence}}.This contributes to efforts to adapt O-ring models to investigate convergence/divergence dynamics.\\

Examples of such efforts include Costinot (2009)~\cite{costinot2009origins}, who links task complementarities to institutions and human capital as a way to explain patterns of comparative advantage, and Krishna and Levchenko (2013)~\cite{krishna2013comparative} who use a similar framework to explain why poorer countries specialize in less complex and more volatile goods. Similarly, Bi et. al (2021)~\cite{demir2024ring} integrated O-ring production into models of firm size distribution, highlighting the role of human capital in scaling organizations. Finally, Demir et al. (2021)~\cite{demir2024ring} extended O-ring complementarities across firm networks, showing how quality upgrading and assortative matching can amplify divergence under shocks. \\
\end{comment}

In the simplest versions of our model, the fastest growing economy in an activity depends on the activity's demand for capabilities (activity-specific stocks of inputs or factors). When an activity is not very demanding, the fastest growers are the economies with the lowest stock of the capability. That is unconditional convergence. But as soon as an activity's demand for capabilities crosses a critical threshold, growth shifts towards economies with a non-zero stock of the capability and conditional convergence kicks in \footnote{In more complex versions of the model, a similar qualitative behavior takes place}.\\ 

These unconditional-conditional convergence dynamics are consistent with Galor's unified growth theory or UGT~\cite{galor2011unified,galor2005stagnation,galor2024unified}. In Galor's UGT the transition away from the Malthusian trap and the onset of modern economic growth comes from a transition where resources are shifted gradually towards more human capital intensive activities starting in the late eighteenth century~\cite{galor2024unified}. Our model provides a simple explanation for that gradual transition, by showing that this conditional convergence mechanism can kick-in naturally once activities become complex enough.\\

Our work also dialogues with the literature on economic growth, particularly the endogenous growth tradition, which builds on the work of Romer (1990)~\cite{romer_endogenous_1990}, Aghion and Howitt 1992~\cite{aghion_model_1992}, and Jones 1995~\cite{jones1995r}.\\

In Romer, growth comes from the non-rivalry of knowledge as an input, which translates into spillovers that make researchers increasingly more productive. In Aghion and Howitt, growth is a consequence of innovation producing waves of Schumpeterian creative destruction, as new varieties outcompete older ones. In Jones, growth comes from growth in the population of high human capital workers (e.g. innovators, researchers, etc.). \\

Together, these branches explain why frontier economies can sustain growth without focusing too much on differences among activities\footnote{Extension of this literature include work focused on heterogeneities in the types of innovation (like in~\cite{akcigit2018growth}) and the research intensity of firms}. Our framework complements these traditions by providing an activity-specific weak-link dynamical mechanism that allows unconditional (or less conditional) convergence for simpler activities, while providing conditional convergence or bounded divergence for more complex activities. That is, the model naturally provides a ``Solow''-like world for simpler activities and a more ``Romer''-like world for more complex activities.\\

Finally, our work speaks to the literature on economic complexity \cite{hidalgo_product_2007,hidalgo_building_2009,hausmann_atlas_2014,balland_new_2022,hidalgo_economic_2021}, and in particular to the literature on  relatedness~\cite{hidalgo_product_2007,hidalgo_principle_2018,boschma_emergence_2013,guevara_research_2016,hidalgo_economic_2021,neffke_skill_2013,neffke_how_2011,jara-figueroa_role_2018, poncet_product_2015}, a well-documented empirical fact showing that activities tend to grow faster in economies specialized in related activities. Weak-link models provide a natural way to connect activities based on their shared input requirements, allowing us to derive a theoretically grounded version of this principle that we test empirically.\\

The literature on economic complexity has been building on vectors of capabilities and weak-link inspired functions from its earliest contributions,~\cite{hidalgo_building_2009}, using these to explain the shape of output matrices and their dynamics \cite{hidalgo_dynamic_2009,hausmann_network_2011,bustos_dynamics_2012}. Yet, despite some important theoretical contributions\cite{fink_serendipity_2017,alshamsi_optimal_2018,schetter2019structural,hidalgo2025theory}, there have been few dynamical approaches to the theory. One exception is a recent contribution by McNerney et al. (2025)~\cite{mcnerney2025bridging}, who take a product–product proximity network as given and explore the dynamics it implies, connecting short-run diversification to long-run structural change and estimates of economic complexity. Another is Atkin et al. (2021)~\cite{atkin2021globalization}, who model capability growth as a function of a country’s pattern of specialization. In their model, countries are ranked by a scalar capability and goods by a scalar complexity, capability grows faster when employment is concentrated in more complex goods, and the focus is on aggregate capability and the dynamic welfare effects of international trade. Our approach differs from both. Instead of taking the network of related activities as given, or summarizing capabilities with a single number, we assume that each activity requires a specific combination of capabilities, many of which are shared with other activities, and that these capabilities accumulate through the reinvestment of output. This allows us to study growth at the level of economy–activity pairs, showing how convergence becomes conditional as activities become more demanding of capabilities, and to derive the principle of relatedness together with predictions for how it interacts with the complexity of economies and activities.\\

The remainder of the paper is structured as follows. The next section introduces the framework under which our capability accumulation model operates. Then, we move to the capability accumulation model, its dynamics, and solution. We then test the empirical predictions of the model using international trade and sub-national industry data. The last section puts these results back in the context of the literature and concludes.\\ 

\section{Basic framework}

\subsection{Production Function}

We start from the production function used in Hidalgo-Stojkoski (2025)~\cite{hidalgo2025theory}, which is a generalization of the Kremer-Shockley model (named after Kremer's 1993 O-Ring model~\cite{kremer_o-ring_1993} and Shockley's 1957 research production model~\cite{shockley1957statistics}). This is a non ad-hoc weak-link production function derived from the assumption that activities $p$ require capabilities $b$ with a probability $q_{pb}$ and that economies $c$ have these capabilities with probability $r_{cb}$.\footnote{We retain the location subscript $c$ to annotate that capabilities vary across economies, which is important for empirical cross-country analyses, even though it is not essential for the theoretical dynamics presented here.} While the underlying O-Ring and Shockley functions operate at the level of the firm, where output is an expectation over many production steps, we apply the function at the level of the economy by assuming each activity is populated by a large number of homogeneous firms. In this large-firm limit, sectoral output is an average across firms and the law of large numbers applies.\\

In this setup, output in an activity $Y_{cp}$ comes from economies having the inputs required by an activity, which can be written as them not lacking the capabilities required by the activity.\footnote{These probabilities can also be conceptualized as relative factor intensities by assuming specific forms of human or physical capital that economies are endowed with ($K_{cb}$) or that activities require $K_{pb}$. In that case $r_{cb}=K_{cb}/\textrm{max}_c(K_{cb})$ and $q_{pb}=K_{pb}/\textrm{max}_p(K_{pb})$. Going forward, we proceed with the more compact notation $r_{cb}$ and $q_{pb}$.}\\

We can construct this function by considering extensive and intensive components. The extensive component is based on the ability of a firm in an economy to find the inputs they need. We assume the availability of each input $b$ in economy $c$ is proportional to $r_{cb}$. The intensive component is based on whether the inputs will ``work'' given how demanding a task is. That is the weak-link component which is proportional to the probability that the inputs meet the level of demand required by each capability. That is $1-q_{pb}(1-r_{cb})$), which assumes that it is harder to satisfy inputs that are demanded more intensely by an activity (e.g. higher $q_{pb}$s are only satisfied in deeper labor pools with higher $r_{cb}$s). That is, we assume output is generated using the function:\\

\begin{equation}
Y_{cp}=\prod_{b}r_{cb}^{\alpha q_{pb}}(1-q_{pb}(1-r_{cb})),
\label{basic_output}
\end{equation}
\\

Where $\alpha$ is a parameter that we use to interpolate between different cases, the main ones being $\alpha=0$ (no extensive or search component) and $\alpha=1$.\\

The $\alpha=0$ case assumes economies have a ``cold-start'' ability and are able to produce some output even even when they lack the capabilities (a ``cold-start'' output that is larger for less capability intense activities and equal to $\prod_{b}(1-q_{pb})$). When $\alpha>0$ this ``cold-start'' is not allowed, meaning that economies without one of the required capabilities have zero output. The $q_{pb}$ exponent in the term is used to ensure that the capability is required by the activity. When the capability is not required $q_{pb}=0$ the search term vanishes $r_{cb}^{\alpha q_{pb}}=1$. When it is maximally required $q_{pb}=1$ the term becomes a search friction modulated by the parameter alpha $r_{cb}^{\alpha}$.\\

The specific form of equation~\eqref{basic_output} allows us to obtain explicit results. Nevertheless, as we will show later on, many of our results also hold for a more general function $Y_{cp} = F(\left\{r_{cb}\right\}, \left\{q_{pb}\right\})$, where $F$ is a nonnegative, twice-differentiable production function that is non-decreasing in the availability of each capability. \\

Before we proceed, it is worth clarifying what we mean by a capability. We understand a capability as a non-fungible input to production. Most naturally a form of specialized human capital, such as particular skills or knowledge, that an economy accumulates and that directly enables output. The model is nonetheless agnostic about the precise nature of these inputs: a capability could equally represent a specialized institution, a norm, or a form of physical capital that is specific to certain activities (e.g. an oil refinery cannot be easily adapted to produce semiconductors). What matters for the theory is that capabilities are granular, activity-specific, and held by economies in varying degrees, rather than being freely substitutable across uses. We deliberately abstract from the more traditional aggregate factors of production, such as raw labor and undifferentiated physical capital, and focus on capabilities as the primary source of variation across economies and activities. This is a modeling choice that honors the idea that factors are aggregate representation of otherwise differentiated inputs, similar in nature to the idea of states of matter in physics (e.g. solid, liquid, gas) which, while true in fact and theory, does not negate the presence of more finely grained forms, like alcohol, oil, and water in the case of liquids. By using capabilities instead of factors we can create a model that focus on the fine grained combinatorial complexity that is lost in more aggregate descriptions of inputs.

\subsection{Model Dynamics}

We begin exploring the dynamics of this model by using chain-rule differentiation. That is:\\

\begin{equation}
\frac{dY_{cp}}{dt}=\sum_{b}\frac{\partial Y_{cp}}{\partial r_{cb}}\frac{dr_{cb}}{dt}+\sum_{b}\frac{\partial Y_{cp}}{\partial q_{pb}}\frac{dq_{pb}}{dt}.
\end{equation}\\

Going forward, we will introduce a model for the rate of change of the capability endowments ($dr_{cb}/dt$), but first, we will focus on the partial derivatives of the output function and their interpretation.\\

To differentiate $Y_{cp}$ more easily it is convenient to express it in the form:\\

\begin{equation}
Y_{cp}=r_{cb}^{\alpha q_{pb}}(1-q_{pb}(1-r_{cb}))\prod_{b'\ne b}r_{cb'}^{\alpha q_{pb'}}(1-q_{pb'}(1-r_{cb'})).
\end{equation}\\

and define:\\

\begin{equation}
E_{cpb}\equiv\prod_{b'\ne b}r_{cb'}^{\alpha q_{pb'}}(1-q_{pb'}(1-r_{cb'})).
\end{equation}\\

The term $E_{cpb}$ represents the contribution of capabilities other than $b$ to the output of economy $c$ in activity $p$. Thus, $E_{cpb}$ is a measure of complementarities. When capability requirements are large ($q\rightarrow1$), $E_{cpb}\sim r^{(\alpha+1)N^b_p}$, where $N^b_p$ is the number of other capabilities (not $b$) used by activity $p$ and $r$ is the geometric mean of $r_{cb}$ over $b$. This means this is a highly convex term that is zero when the capabilities are missing and that grows exponentially with the availability of capabilities.\\ 

Using this notation, the partial derivatives for equation~\eqref{basic_output} take the form:\\

\begin{align}
\frac{\partial Y_{cp}}{\partial r_{cb}}&=q_{pb}r_{cb}^{\alpha q_{pb}-1}(\alpha(1-q_{pb}+r_{cb}q_{pb})+r_{cb})E_{cpb},\\
\frac{\partial Y_{cp}}{\partial q_{pb}}&=r_{cb}^{\alpha q_{pb}}\big(\alpha \ln(r_{cb})(1-q_{pb}(1-r_{cb})) +r_{cb}-1\big)E_{cpb}.
\label{eq:dYdr}
\end{align}\\

And the rates of change of output with respect to $r_{cb}$ and $q_{pb}$ are:\\

\begin{align}
\frac{\partial Y_{cp}}{\partial r_{cb}}\frac{dr_{cb}}{dt}&=q_{pb}r_{cb}^{\alpha q_{pb}-1}(\alpha(1-q_{pb}+r_{cb}q_{pb})+r_{cb})E_{cpb}\frac{dr_{cb}}{dt},\\
\frac{\partial Y_{cp}}{\partial q_{pb}}\frac{dq_{pb}}{dt}&=r_{cb}^{\alpha q_{pb}}\big(\alpha \ln(r_{cb})(1-q_{pb}(1-r_{cb})) +r_{cb}-1\big)E_{cpb}\frac{dq_{pb}}{dt}.
\end{align}\\

These expressions simplify for the cases we explore in more detail: $\alpha=0$ and $\alpha=1$.\\

When $\alpha=0$:\\

\begin{align}
\frac{dY_{cp}}{dt}&=\sum_bq_{pb}E_{cpb}\frac{dr_{cb}}{dt}+\sum_{b}(r_{cb}-1)E_{cpb}\frac{dq_{pb}}{dt}.
\end{align}\\

And when $\alpha=1$:\\

\begin{align}
\frac{dY_{cp}}{dt}
&=\sum_b q_{pb}r_{cb}^{q_{pb}-1}
\left[1-q_{pb}+r_{cb}(1+q_{pb})\right]
E_{cpb}\frac{dr_{cb}}{dt}\nonumber\\
&\quad+\sum_b r_{cb}^{q_{pb}}
\left[\ln(r_{cb})(1-q_{pb}+r_{cb}q_{pb})
+r_{cb}-1\right]
E_{cpb}\frac{dq_{pb}}{dt}.
\end{align}\\

We are now at a point where we can highlight a few general ideas.\\

First, in all cases the rate of change is amplified by the complementarity of an input $b$ with other inputs ($E_{cpb}$).\\ 

Second, in the ``cold-start'' model ($\alpha=0$) the impact of the rate of capability accumulation ($dr_{cb}/dt$) in the change in output ($dY_{cp}/dt$) grows when activities require more of the capability (larger $q_{pb}$). This makes intuitive sense, since output on these products is more constrained when there is a lack of the capability. It also means that, at the same rate of capability accumulation (e.g. when $dr_{cb}/dt = dr_{cb'}/dt$), output grows faster in activities that are more intense in the requirement of the capability.\\

In the non-``cold-start'' model ($\alpha=1$) the impact of the rate of capability accumulation ($dr_{cb}/dt$) in the change in output ($dY_{cp}/dt$) includes the term ($q_{pb}r_{cb}^{q_{pb}-1}(1-q_{pb}+r_{cb}(1+q_{pb}))$). This expression requires a bit more attention. In this case, when $q_{pb}=0$ the coefficient is zero, meaning the accumulation of a non-required capability doesn't lead to growth in output. But when $q_{pb}=1$ this term becomes $2r_{cb}$, which means that the growth increases with the stock of the capability $r_{cb}$ when activities are more demanding of it. This means the non-``cold-start'' model contains a built-in increasing returns mechanisms. Again, this response increases with complementarity $E_{cpb}$\\

And third, in both cases, the second term shows that changes in the capabilities required by an activity ($dq_{pb}/dt$) are negatively correlated with growth in output, since $r_{cb}-1$ and $\ln(r_{cb})$ can only be zero or negative. So, when activities require more of a capability ($dq_{pb}/dt>0$) output decreases, and when they require less ($dq_{pb}/dt<0$) output increases. This represents a second channel of growth, coming from activities being less bound by the availability of local capabilities, which we can interpret as technological diffusion. In this interpretation, economies grow by discovering an easier way to produce the same activity, which is represented by the activity requiring less of the capability (lower $q_{pb}$).\\

In the next section, we introduce a model for $dr_{cb}/dt$ and use it to derive testable predictions about which economies experience the fastest growth in a given activity. But before doing that we would like to provide a few additional notes of interpretation focused on the two probability matrices $r_{cb}$ and $q_{pb}$.\\

We can think of $r_{cb}$ as an estimate of how difficult it is for a firm or an entrepreneur to find capability in an economy. This is compatible with the more verbose capital intensity definition introduced earlier (where $r_{cb}=K_{cb}/\textrm{max}_c(K_{cb})$ and $K_{cb}$ is a form of specific human or physical capital). \\

Imagine a firm looking for an expert in innate immunity. A firm in city $c=Boston$ will have an easier time finding a qualified candidate than a firm in city $c'=Fargo$. So if $b=$ ``innate immunity expertise'' then we would say that $r_{c,b}>r_{c',b}$. In this example we used a skill or occupation as the interpretation of a capability, but the idea is more general. For instance, someone in Nashville, Tennessee should have an easier time finding a great recording studio than someone in Mobile, Alabama ($r_{Nashville,RecordingStudio}>r_{Mobile,RecordingStudio}$).\\

Now let us move to the intensity with which an activity requires a capability $q_{pb}$. This is an estimate of how binding that capability is (how strongly it acts as a bottleneck or weak-link). This matrix is a key differentiating factor from Kremer' O-ring model tradition which assumes inputs are always binding ($q=1\:\forall\:p,b$)\\

Consider the capabilities needed to produce a short film in 1950 or in 2025. In 1950, the capabilities needed to film and edit a short movie were difficult to obtain. For example, a movie camera or an editing room, which are specific forms of physical capital, were rare and expensive. Today, everyone has a good camera and video editing software in their pocket. So, if filming and editing are capabilities needed to produce a short film, then these have become easier to obtain (E.g. $q_{Film,Editing}(1950)>q_{Film,Editing}(2025)$). In the model, this means these capabilities became less binding ($dq_{pb}/dt<0$).\footnote{These examples should also make clear that were we are looking at changes in the potential for output, since we have not introduced demand side considerations. In fact, the latter depends on a price that decreases when the capability needed to produce a product becomes less demanding and the production of that product grows more ubiquitous. In Hidalgo and Stojkoski (2025)\cite{hidalgo2025theory} we show that the price of a product behaves roughly as $\sim1/(1-q).$}\\ 

Going back to the model, we start by focusing on the marginal contribution of a capability to change in output. That is, we will fix $q_{pb}$ and all capability endowments except for one ($r_{cb}=\textrm{const,} \:\textrm{for}\:b'\ne b$). In that case, we only need to worry about the dynamics of capability change.\\ 

In the next section we define a model for $dr_{cb}/dt$ and present an endogenous investment rule that we will use to explore its kinematics. For the cold-start case ($\alpha=0$) this model predicts convergence when activities do not intensely require capabilities and conditional convergence when activities are intense in their capability requirements. When cold-start is not allowed ($\alpha>0$) the model predicts tunable conditional convergence, with well-endowed economies growing faster than poorly endowed ones when activities are intense in their capability requirements.\\

\section{The Dynamics of Capability Growth}
\label{sec:dynamics}

We consider a dynamical model where the accumulation of capabilities comes from investing in the industries or products that utilize them. In this model, the rate of growth of capability $b$ in economy $c$ is given by the investment of proceeds the economy $c$ obtains from activity $p$ and allocates into activities $p'$ ($i_{cpp'}$). These investments increase the availability of capability $b$ based on the intensity with which it is used in the activity receiving the investment ($p'$), representing a learning-by-doing term \cite{arrow_economic_1962}. We also include a saturation term $(1-r_{cb})$ that limits the growth of $r_{cb}$ beyond one. In the appendix, we consider a depreciation term ($\delta \sum_p q_{pb}r_{cb}$) to capture the notion that knowledge has a forgetting rate, as studied in \cite{mason2007interrupted,david2011determinants,darr1995acquisition,argote_learning_1990,benkard2000learning,thompson_how_2007}. But here, we concentrate on the stripped-down version of the model. That is, we let: \\

\begin{equation}
    \frac{dr_{cb}}{dt}=(1- r_{cb})^\beta\sum_{p,p'}i_{cpp'}Q_{p'b},
    \label{dynamic_eqn}
\end{equation}\\

where $Q_{pb}$ is the relative intensity of use of capability $b$ in activity $p$ compared to other capabilities used the same activity ($Q_{pb}=q_{pb}/q_p$ and $q_p=\sum_b q_{pb}$) and $\beta$ governs the strength of saturation in capability accumulation. For $\beta=1$, growth declines linearly as $r_{cb}$ approaches one. When $0<\beta<1$, saturation is weaker, meaning that accumulation remains stronger even at higher levels of $r_{cb}$, so capabilities can continue to deepen more easily as they get closer to their upper limit. By contrast, higher values of $\beta$ make accumulation more concave and slow growth more strongly near that upper limit.\\

Similarly to the production function, we can replace $(1-r_{cb})^\beta$ with a general function $G(r_{cb})$, which is positive for $r_{cb}<1$ and satisfies $G(1)=0$. The power function allows us to obtain explicit results. \\

To operationalize the model we use an endogenous investment rule where an entrepreneur with a home sector $p$ at each point in time $t$ optimizes by distributing its investment across multiple sectors $p'$ subject to information and knowledge constraints. We assume that entrepreneurs allocate their investments (net of friction losses) as $i_{cpp'}$ subject to a budget constraint: \\
\begin{equation}
    \sum_{p'} \big(i_{cpp'}(t) + \eta_{cpp'}(t)\big)=\gamma Y_{cp}(t),
\end{equation}\\
where $\eta_{cpp'}(t)$ is a cost that comes from knowledge and information frictions and grows with the \emph{distance} between two sectors $p$ and $p'$.\\

Knowledge and information frictions reflect the notion that entrepreneurs have a limited understanding of the sectors where they do not operate (their non home sectors). For example, an entrepreneur working on consumer software will not be as effective at evaluating investments in biotech. This information friction incorporates the notion that entrepreneurs are agents with bounded rationality \cite{simon_bounded_1991} who observe other sectors imperfectly. Also, entrepreneurs experience knowledge frictions, incorporating the idea that knowledge is non-fungible~\cite{hidalgo_knowledge_2024}. For example, an entrepreneur with knowledge of natural gas extraction cannot simply redeploy their expertise to pharmaceutical drug development.\\

We model these frictions using a kernel that is related to the concept of proximity in the economic complexity literature~\cite{hidalgo_product_2007}. That is, we assume the existence of a function:\\

\begin{align}
    \phi_{p,p'} =1 \quad if \:p=p', \\ 
    \phi_{p,p'} <1 \quad if \:p \ne p',
\end{align}\\
    
that monotonically decreases with increasing separation between sectors $p$ and $p'$.\\

We use this proximity kernel to model frictions by assuming that an entrepreneur investing away from their home sector experiences a loss due to misallocation that is proportional to $1-\phi_{p,p'}$ (so there is no loss when investing in their home sector). This kernel is not necessarily symmetric, since asymmetries incorporate the idea that some sectors are easier to invest \emph{from} than \emph{to}. For example, we could assume that an investment from semiconductors to photovoltaics is more effective than an investment from photovoltaics to semiconductors by setting $\phi_{p,p'}>\phi(p',p)$. This kernel also allows us to potentially model the government sector $g$ by assuming that it has a proximity $\phi_{g,p'}$ to other sectors.\\

To derive the optimal investment rule, we consider the version of the problem in which an entrepreneur in home sector $p$ chooses a single destination sector $p^*(t)$ for investment at each point in time $t$. In this case, the proximity kernel determines the share of investment that is translated into effective capability accumulation, while the remaining share is dissipated through information and knowledge frictions. Formally, if the entrepreneur directs investment toward sector $p'$, the effective investment is\\

\begin{equation}
    i_{cpp'}(t)=\phi_{p,p'}\gamma Y_{cp}(t),
\end{equation}\\

while the residual friction cost is\\

\begin{equation}
    \eta_{cpp'}(t)=(1-\phi_{p,p'})\gamma Y_{cp}(t).
\end{equation}\\

Since a fraction $\gamma$ of output is devoted to investment, contemporaneous profits are\\

\begin{equation}
    \pi_{cp}(t)=(1-\gamma)Y_{cp}(t).
\end{equation}\\

The entrepreneur chooses the investment destination over time so as to maximize the present discounted value of profits:\\

\begin{equation}
    \max_{p^*(t)} \int_0^{\infty} e^{-\rho t} (1 - \gamma) Y_{cp}(t) \, dt,
    \label{profit_max}
\end{equation}\\

where $\rho > 0$ is the discount rate. Under this decision rule, the law of motion for capability $b$ becomes\\

\begin{equation}
   \frac{dr_{cb}}{dt}=(1-r_{cb}(t))^\beta\phi_{p,p'}\gamma Y_{cp}Q_{p'b}.
    \label{rdot}
\end{equation}\\

And the problem reduces to identifying the sector $p'$ that is receiving the investment.\\

Let $V(\{r_{cb}(t)\})$ denote the value function. The associated Hamilton--Jacobi--Bellman equation is\\

\begin{equation}
    \rho V(\{r_{cb}\})=
    \max_{p'}\left[
    (1-\gamma)Y_{cp}
    +
    \sum_b \frac{\partial V}{\partial r_{cb}}\frac{dr_{cb}}{dt}
    \right].
    \label{HJB}
\end{equation}\\

Substituting \eqref{rdot} into \eqref{HJB} yields\\

\begin{equation}
    \rho V(\{r_{cb}\})=
    \max_{p'}\left[
    (1-\gamma)Y_{cp}
    +
    \phi_{p,p'}\gamma Y_{cp}
    \sum_b
    \frac{\partial V}{\partial r_{cb}}
    (1-r_{cb}(t))^{\beta}Q_{p'b}
    \right].
    \label{HJB_full}
\end{equation}\\

The contemporaneous profit term $(1-\gamma)Y_{cp}$ does not depend on $p'$. The term $\gamma Y_{cp}$ multiplying the accumulation term is positive and common to all destinations $p'$, since current output depends on the current capability stocks and the choice of $p'$ affects only their future evolution. Therefore, for $Y_{cp}>0$, the optimal investment rule reduces to\\

\begin{equation}
    p^*(t)=
    \arg\max_{p'}
    \left[
    \phi_{p,p'}
    \sum_b
    \frac{\partial V}{\partial r_{cb}}
    (1-r_{cb}(t))^{\beta}Q_{p'b}
    \right].
    \label{pstar_profit}
\end{equation}\\

This rule states that the entrepreneur invests in the sector that maximizes friction-discounted capability accumulation, weighted by the marginal value of each capability. The term $\frac{\partial V}{\partial r_{cb}}$ captures how much an additional unit of capability $b$ contributes to the discounted stream of future profits.\\

Defining

\begin{equation}
\tilde m_{cp'}(t) \equiv
\sum_b
\frac{\partial V}{\partial r_{cb}}
(1-r_{cb}(t))^{\beta}Q_{p'b},
\end{equation}\\

the rule can be written more compactly as\\

\begin{equation}
    p^*(t)=
    \arg\max_{p'}
    \left[
    \phi_{p,p'}\tilde m_{cp'}(t)
    \right].
\end{equation}\\

This expression also makes clear when the entrepreneur deviates from the home sector. Since \(\phi(p,p)=1\), the value of investing in the home sector is simply \(\tilde m_{cp}(t)\). An outside sector \(p'\neq p\) is chosen only if its friction-discounted value-weighted return exceeds that benchmark, that is, only if\\

\begin{equation}
    \phi_{p,p'}\tilde m_{cp'}(t) > \tilde m_{cp}(t).
\end{equation}\\

Thus, entrepreneurs remain in their home sector unless some outside sector offers a sufficiently large expected gain in future value to compensate for the knowledge and information frictions associated with investing away from home. \\

Under this rule, realized investment from sector $p$ is directed to the chosen destination $p^*$, so that\\

\begin{equation}
i_{cpp'}=\gamma Y_{cp}\phi_{p,p'}\delta_{p'p^*(t)}.
\end{equation}\\

Substituting this rule into the accumulation equation gives \\

\begin{equation}
    \frac{dr_{cb}}{dt}
    =
    \gamma(1-r_{cb})^{\beta}
    \sum_p
    Y_{cp}Q_{p^*b}\phi_{p,p^*(t)},
    \label{dynamic_eqn_with}
\end{equation}\\

and, expanding the output function $Y_{cp}$,\\

\begin{equation}
    \frac{dr_{cb}}{dt}
    =
    \gamma(1-r_{cb})^{\beta}
    \sum_p
    \phi_{p,p^*(t)}Q_{p^*(t)b}
    \prod_{b'} r_{cb'}^{\alpha q_{pb'}}\left(1-q_{pb'}(1-r_{cb'})\right).
    \label{dynamic_eqn_Y}
\end{equation}\\

In the next section, we explore the dynamical implications of this model starting from the case of a single capability and strong knowledge and information frictions ($\phi_{p,p}\gg\phi_{p,p^*}$). We then generalize this to multiple capabilities and weaker knowledge and information frictions.\\

\subsection{The Single Capability Model}
\label{sec:single-capability}

We now solve the single capability version of this model with strong knowledge and information frictions. In this version of the model $p^*(t)=p$ because there is only one capability we have $\frac{\partial V}{\partial r_{cb}} > 0$ and $Q_{pb}= 1$ (meaning that $\arg\max_{p'}\left[\phi_{p,p'}\tilde m_{cp'}(t)\right] = \arg\max_{p'}\phi_{p,p'}$). That is, in this case entrepreneurs stick to their home sector. This case will provide us with valuable intuition that we will later extend to a model with an arbitrary number of capabilities and with more relaxed frictions. In this case, equation~\eqref{dynamic_eqn_with} simplifies to:\\

\begin{equation}
    \frac{dr_{c}}{dt}=\gamma(1-r_{c})^\beta \sum_{p}r_c^{\alpha q_p} (1-q_{p}(1-r_{c})).
    \label{one_cap_raw}
\end{equation}\\

which we can reorganize equation~\eqref{one_cap_raw} into:\\

\begin{equation}
    \frac{dr_{c}}{dt}=\gamma(1-r_{c})^\beta  \big(\sum_{p} r_c^{\alpha q_p}(1-q_{p})+\sum_{p}r_c^{\alpha q_p+1}q_{p}\big).
    \label{one_cap_dyn}
\end{equation}\\

Next we bring this to a simpler form by noting that when the $c$ and $p$ terms are linear and separable (e.g. $\alpha=0$) the sums can be transformed into averages of each of these terms by factoring out the number of activities $N(p)$ (which we denote with the bra-ket notation $\langle \:\rangle$). That is, using the identity:\\

\begin{equation}
\langle q \rangle \equiv\frac{\sum_pq_p}{N(p)}
\label{q_average}
\end{equation}\\

For $\alpha=0$ we obtain:\\

\begin{equation}
    \frac{dr_{c}}{dt}=\gamma N(p)(1-r_{c})^\beta \big((1-\langle q\rangle)+r_c\langle  q \rangle \big).
    \label{growth_average}
\end{equation}\\

Which upon absorbing $N(p)$ into the time unit and setting $\beta=1$ becomes a Riccati equation (named after the Italian mathematician Jacopo Riccati (1676-1754)) of the form:\\

\begin{equation}
    \frac{dr_c}{dt}=Ar_c^{2}+Br_c+C,
    \label{Riccati_poly}
\end{equation}\\
with:\\
\begin{align}
    A &= -\gamma \langle q \rangle, \\
    B &=  \gamma (2\langle q\rangle - 1),\\
    C &= \gamma (1 - \langle q\rangle).
\end{align}\\

For the more general case ($\alpha>0$ and $\beta>0$) the average is over the output across all activities $\langle Y_{c}\rangle$:\\

\begin{equation}
    \frac{dr_{c}}{dt}=\gamma(1-r_{c})^\beta  \langle Y_{c}\rangle.
    \label{one_cap_avgY}
\end{equation}\\

where again we have absorbed $N(p)$ into the time unit and where $\langle Y_{c}\rangle=\sum_p Y_{cp}/N(p)$.\\

In this section, we focus first on the cold-start case, which provides good basic intuition and then move to non cold-start situations  ($\alpha>0$ and $\alpha=1$). For simplicity, we also fix $\beta=1$.

\begin{figure}
\begin{tikzpicture}
  \begin{axis}[
    axis lines = left,
    xlabel = $r_c$,
    ylabel = {$dr_c/dt$},
    domain=0:1,
    samples=200,
    ymin=0,
    ymax=0.1,
    xmax=1,
    width=13cm,
    height=9cm,
    tick label style={/pgf/number format/fixed},
  ]
  % Define constants
  \pgfmathsetmacro{\A}{0.5}
  \pgfmathsetmacro{\B}{0.2}
  \pgfmathsetmacro{\C}{0.3}

  \pgfmathsetmacro{\D}{0.1}
  \pgfmathsetmacro{\E}{0.1}
  \pgfmathsetmacro{\F}{0.9}

  \pgfmathsetmacro{\G}{0.1}
  \pgfmathsetmacro{\H}{0.9}
  \pgfmathsetmacro{\I}{0.1}

  \pgfmathsetmacro{\J}{0.1}
  \pgfmathsetmacro{\K}{0.5}
  \pgfmathsetmacro{\L}{0.5}
  % Plot the function
  %\addplot[
  %  thick,
  %  blue,
  %] { \A*(1 - x)*(\B + \C*x) } node[pos=0.6, above, blue, xshift=25pt, yshift=20pt] {$\sum_pq_pQ_p>\sum_p (1-q_p)Q_p$};

  \addplot[
    thick,
    blue,
  ] { \D*(1 - x)*(\E + \F*x) } node[pos=0.3, above, black, xshift=-60pt, yshift=-10pt] {$\langle q \rangle \gg \frac{1}{2}$};

 \addplot[
    thick,
    blue,
  ] { \J*(1 - x)*(\K + \L*x) } node[pos=0.3, above, black, xshift=-60pt, yshift=9pt] {$\langle q \rangle = \frac{1}{2}$};

    \addplot[
    thick,
    blue,
  ] { \G*(1 - x)*(\H + \I*x) } node[pos=0.6, above, black, xshift=-157pt, yshift=100pt]  {$\langle q \rangle \ll \frac{1}{2}$};
  \end{axis}
\end{tikzpicture}
\caption{Behavior of the capability accumulation or learning rate $dr/dt$ for a fixed investment rate of $\gamma=0.1$ for the cold start case where $\alpha=0$ and $\beta=1$.}
\label{Riccati_fig1}
\end{figure}

To characterize the dynamics we need to look at the shape of the right-hand side of the equation in the $[0,1]$ interval. In particular, we are interested in finding the peak of the Riccati parabola ($r_c=-B/2A$) and understanding when it falls within the $[0,1]$ interval. Because $A=-\gamma\langle q\rangle<0$, the parabola opens downward and has a unique maximum at $r_c^*=-\frac{B}{2A}$. If this peak falls to the left of the feasible region ($r_c^*\le 0$), the growth rate of the capability ($dr_c/dt$) is maximal at $r_c=0$ and decreases monotonically with $r_c$. By contrast, conditional convergence requires an interior maximum $r_c^*\in(0,1)$, so that economies with very low initial capabilities must first accumulate enough $r_c$ to reach the peak before converging. \\

Given $A<0$, the condition $r_c^*>0$ is equivalent to $B>0$. Writing\\

\begin{equation}
B = \gamma (2\langle q\rangle - 1) 
\end{equation}\\

thus the condition is simply that $B>0$ when\\

\begin{align}
\langle q \rangle &> 1/2,
\end{align}

Thus, the interior peak of the growth rate (and hence conditional convergence) arises when the intensity with which the average activity requires capabilities is above the middle of the interval  $\langle q\rangle>1/2$. \\

Figure~\ref{Riccati_fig1} illustrates this transition by plotting the rate of growth of the capability ($dr_c/dt$) for three representative cases.\\

When the capability is not highly required ($\langle q\rangle \ll 1/2$), the maximum growth rate occurs at $r_c=0$, indicating that less endowed economies catch up more rapidly---a ``Solow-type'' convergence pattern. As $\langle q\rangle$ increases toward the threshold, the maximum growth rate moves inward from the boundary, and once $\langle q\rangle$ exceeds the threshold, it occurs at higher values of $r_c$, signifying that convergence becomes conditional on the level of capability already accumulated. \\

We can observe this transition again in Figure~\ref{bifurcation_fig_cold_start}, which shows the point at which the growth rate of the capability ($dr_c/dt$) reaches its maximum within the $[0,1]$ interval ($\mathrm{argmax}(dr_c/dt)$) as a function of $\langle q \rangle$ of the Riccati equation. \\

When the average intensity with which capabilities are required is low ($\langle q\rangle < 1/2$), the maximum occurs at the boundary ($r_c = 0$). As $\langle q\rangle$ crosses the $\langle q\rangle=1/2$ threshold the maximum shifts smoothly toward the interior of the interval. This relationship can be written as:\\

\begin{equation}
    r_c^*=\mathrm{argmax}\!\left(\frac{dr_c}{dt}\right) =
    \begin{cases}
        0, & \text{if } \langle q\rangle < \dfrac{1}{2}, \\[8pt]
        1 - \dfrac{1}{2\langle q\rangle} , & \text{if } \langle q\rangle > \dfrac{1}{2}.
    \end{cases}
\label{argmax_cold_start}
\end{equation}\\

This diagram resembles a \emph{second-order phase transition}, where the order parameter---in this case $\mathrm{argmax}(dr_c/dt)$---changes continuously but its derivative is discontinuous at the critical point $\langle q\rangle = 1/2$. The gradual nature of this transition implies that as activities become more capability-intensive, the advantage in capability accumulation shifts smoothly from economies with no capabilities to those with a higher stock of capabilities. This mirrors the mechanisms of gradual takeoff and divergence described in Galor's unified growth theory~\cite{galor2005stagnation,galor2024unified,galor2011unified}.\\

\begin{figure}[ht]
\centering
\begin{tikzpicture}
  % --- Set parameters (edit these as needed) ---
  \pgfmathsetmacro{\g}{1.0}      % gamma
  \pgfmathsetmacro{\d}{0}     % delta
  \pgfmathsetmacro{\qc}{\g/(2*\g-\d)} % critical threshold q_c
  \pgfmathsetmacro{\ymax}{0.6}

  \begin{axis}[
    axis lines = left,
    xlabel = {$\langle q\rangle$},
    ylabel = {$\arg\max\!\big(dr_{c}/dt\big)$},
    xmin=0, xmax=1,
    ymin=0, ymax=\ymax,
    xtick=\empty,
    ytick=\empty,
    domain=0:1,
    samples=400,
    width=12cm,
    height=9cm,
    tick label style={font=\small},
    label style={font=\small},
  ]

    % Left branch: maximum at r_c = 0 when <q> < q_c
    \addplot[
      thick,
      blue,
      domain=0:\qc
    ] {0};

    % Right branch: interior argmax r_c* = 1 - 1/(2<q>) - delta/(2 gamma) for <q> >= q_c
    \addplot[
      thick,
      blue,
      domain=\qc:1
    ] {1 - 1/(2*x) - (\d)/(2*\g)};

    % Dashed vertical line at the critical threshold q_c
    \addplot[dashed, gray] coordinates {(\qc,0) (\qc,\ymax)};
    \node[anchor=west] at (axis cs:\qc,0.52*\ymax)
      {$\langle q\rangle = 1/2 $};

    % Optional annotation of the right-branch formula
    \node[anchor=west] at (axis cs:0.62,0.12*\ymax)
      {$1 - \dfrac{1}{2\langle q\rangle}$};

  \end{axis}
\end{tikzpicture}
\caption{Location of the maximum growth rate of the capability, $\arg\max(dr_c/dt)$, for the cold start case ($\alpha=0$ and $\beta=1$) as a function of the average intensity $\langle q\rangle$. The critical threshold is $\langle q\rangle=1/2 $. For $\langle q\rangle<1/2$ the maximum occurs at $r_c=0$ (unconditional convergence); for $\langle q\rangle\ge1/2$ it shifts to the interior at $1 - \dfrac{1}{2\langle q\rangle}$ (conditional convergence).}
\label{bifurcation_fig_cold_start}
\end{figure}

Next, we explore the case when $\alpha>0$ and $\beta>1$. Again, we are looking for an expression describing the maximum growth rate of a capability  as a function of the current stock of capabilities. So we differentiate equation \ref{one_cap_raw} with respect to $r_c$. After some algebra, that is:

\begin{equation}
\frac{d}{dr_c}\frac {dr_c}{dt}=\gamma (1-r_c)^\beta \sum_pr_c^{\alpha q_p}\bigg((1-q_p(1-r_c))(\frac{\alpha q_p}{r_c}+\frac{q_p}{1-q_p(1-r_c)}-\frac{\beta}{1-r_c})\bigg)
\label{foc_general}
\end{equation}

Since this expression lacks a closed form solution, we study the behavior of this function after assuming $q_p=q\:\forall\:p$. That will allow us to explore cases when all activities are demanding or not demanding of the capability. \\

On the interval $(0,1)$ every factor in front of the last parenthesis is strictly positive. This means that the derivative will be zero when the last parenthesis is zero. That is, when \\
\begin{equation}
\frac{\alpha q}{r_c}+\frac{q}{1-q(1-r_c)}=\frac{\beta}{1-r_c}.
\label{foc_homogeneous}
\end{equation}\\
Multiplying both sides of equation~\eqref{foc_homogeneous} by the positive
quantity $r_c(1-r_c)\left(1-q(1-r_c)\right)$ and collecting powers of $r_c$
yields the quadratic equation \\
\begin{equation}
q(\alpha q+\beta+1)\,r_c^2
-\left[q(2\alpha q+\beta+1)-(\alpha q+\beta)\right]r_c
-\alpha q(1-q)=0.
\label{quadratic_homogeneous}
\end{equation}\\
The quadratic has a single positive root, which gives the location of the
maximum growth rate\\
\begin{equation}
r_c^*=\operatorname*{arg\,max}_{r_c}\frac{dr_c}{dt}
=\frac{(2\alpha q+\beta+1)q-(\alpha q+\beta)
+\sqrt{\left[(\alpha q+\beta)-(\beta+1)q\right]^2+4\alpha q^2}}
{2q(\alpha q+\beta+1)}.
\label{argmax_alpha_beta}
\end{equation}

These curves provide a prediction that we can validate by exploring data on the growth of activities such as products or industries%
\footnote{For heterogeneous $q_p$, the maximizer satisfies
\[
\begin{aligned}
0={}&\sum_p (r_c^*)^{\alpha q_p-1}
\Bigl[
-q_p(\alpha q_p+\beta+1)(r_c^*)^2
+\bigl((2\alpha q_p+\beta+1)q_p
-(\alpha q_p+\beta)\bigr)r_c^*
+\alpha q_p(1-q_p)
\Bigr].
\end{aligned}
\]
If several roots lie in $(0,1)$, select the one yielding
the largest accumulation rate.}. They predict a positive correlation between how demanding an activity is of the capability, and the economies experiencing the fastest growth in it. In the empirical section of this paper we will look at this relationship across a variety of datasets (proxying capability requirements and stocks using $PCI$ and $ECI$, and looking at growth in output instead of growth in capability stocks, since the latter are not directly observable). This will help us provide estimates for $\alpha$ and $\beta$. But before that, we will finalize the theoretical section of the paper by extending the single-capability result to the case of multiple capabilities.\\

\begin{figure}[htbp]
\centering
\begin{tikzpicture}
  \begin{axis}[
    axis lines=left,
    xlabel={$q$},
    ylabel={$\operatorname*{arg\,max}_{r_c}(dr_c/dt)$},
    xmin=0,
    xmax=1,
    ymin=0,
    ymax=0.8,
    xtick={0,0.2,0.4,0.6,0.8,1},
    ytick={0,0.2,0.4,0.6,0.8},
    width=13cm,
    height=9cm,
    tick label style={/pgf/number format/fixed},
    legend style={
      at={(0.03,0.97)},
      anchor=north west,
      draw=none,
      fill=white,
      font=\small
    },
    declare function={
      peak(\q,\a)=
        (
          (\a*\q+1)*(2*\q-1)
          +sqrt(max(0,(\a*\q+1)^2-4*\q*(1-\q)))
        )
        /(2*\q*(\a*\q+2));
    },
  ]

    % alpha = 0: boundary maximum for q <= 1/2
    \addplot[
      thick,
      blue,
      domain=0:0.5,
      samples=2,
    ] {0};
    \addlegendentry{$\alpha=0$}

    \addplot[
      thick,
      blue,
      domain=0.5:1,
      samples=200,
      forget plot,
    ] {1-1/(2*x)};

    % alpha = 0.1
    \addplot[
      thick,
      red,
      domain=0.0001:1,
      samples=400,
    ] {peak(x,0.1)};
    \addlegendentry{$\alpha=0.1$}

    % alpha = 0.5
    \addplot[
      thick,
      orange,
      domain=0.0001:1,
      samples=400,
    ] {peak(x,0.5)};
    \addlegendentry{$\alpha=0.5$}

    % alpha = 1
    \addplot[
      thick,
      violet,
      domain=0.0001:1,
      samples=400,
    ] {peak(x,1)};
    \addlegendentry{$\alpha=1$}

    % alpha = 2
    \addplot[
      thick,
      teal!80!black,
      domain=0.0001:1,
      samples=400,
    ] {peak(x,2)};
    \addlegendentry{$\alpha=2$}

    % Cold-start transition
    \addplot[
      dashed,
      gray,
      forget plot,
    ] coordinates {(0.5,0) (0.5,0.8)};

  \end{axis}
\end{tikzpicture}
\caption{Capability stock at which the accumulation of a capability is fastest
as a function of the capability intensity requirement $q_p=q$, for
$\beta=1$ and $\alpha=0,0.1,0.5,1,2$.
The dashed line marks the cold-start transition at $q=1/2$..}
\label{bifurcation_alpha}
\end{figure}

\begin{figure}[htbp]
\centering
\begin{tikzpicture}
  \begin{axis}[
    axis lines=left,
    xlabel={$q$},
    ylabel={$\operatorname*{arg\,max}_{r_c}(dr_c/dt)$},
    xmin=0,
    xmax=1,
    ymin=0,
    ymax=1,
    xtick={0,0.2,0.4,0.6,0.8,1},
    ytick={0,0.2,0.4,0.6,0.8,1},
    width=13cm,
    height=9cm,
    tick label style={/pgf/number format/fixed},
    legend style={
      at={(0.03,0.97)},
      anchor=north west,
      draw=none,
      fill=white,
      font=\small
    },
    declare function={
      peakbeta(\q,\a,\b)=
        (
          (2*\a*\q+\b+1)*\q-(\a*\q+\b)
          +sqrt(
            ((\a*\q+\b)-(\b+1)*\q)^2
            +4*\a*\q^2
          )
        )
        /(2*\q*(\a*\q+\b+1));
    },
  ]

    \addplot[
      thick,
      blue!10!red,
      domain=0.001:1,
      samples=400,
    ] {peakbeta(x,0.1,3)};
    \addlegendentry{$\alpha=0.1,\ \beta=3$}

    \addplot[
      thick,
      blue!30!red,
      domain=0.001:1,
      samples=400,
    ] {peakbeta(x,0.1,2)};
    \addlegendentry{$\alpha=0.1,\ \beta=2$}

    \addplot[
      thick,
      blue!50!red,
      domain=0.001:1,
      samples=400,
    ] {peakbeta(x,0.1,1)};
    \addlegendentry{$\alpha=0.1,\ \beta=1$}

    \addplot[
      thick,
      blue!70!red,
      domain=0.001:1,
      samples=400,
    ] {peakbeta(x,0.1,0.5)};
    \addlegendentry{$\alpha=0.1,\ \beta=0.5$}

    \addplot[
      thick,
      blue!90!red,
      domain=0.001:1,
      samples=400,
    ] {peakbeta(x,0.1,0.2)};
    \addlegendentry{$\alpha=0.1,\ \beta=0.2$}

    \addplot[
      thick,
      blue,
      domain=0.001:1,
      samples=400,
    ] {peakbeta(x,0.1,0.1)};
    \addlegendentry{$\alpha=0.1,\ \beta=0.1$}

  \end{axis}
\end{tikzpicture}
\caption{Capability stock at which accumulation is fastest
as a function of the common requirement $q_p=q$, for
$\alpha=0.1$ and different saturation exponents $\beta$.
Lower values of $\beta$ shift the maximum toward higher
capability stocks. All curves approach zero as $q\to0^+$.}
\label{bifurcation_alpha_beta}
\end{figure}

\begin{figure}[htbp]
\centering
\begin{tikzpicture}
  \begin{axis}[
    axis lines=left,
    xlabel={$q$},
    ylabel={$\operatorname*{arg\,max}_{r_c}(dr_c/dt)$},
    xmin=0,
    xmax=1,
    ymin=0,
    ymax=1,
    xtick={0,0.2,0.4,0.6,0.8,1},
    ytick={0,0.2,0.4,0.6,0.8,1},
    width=13cm,
    height=9cm,
    tick label style={/pgf/number format/fixed},
    legend style={
      at={(0.03,0.97)},
      anchor=north west,
      draw=none,
      fill=white,
      font=\small
    },
    declare function={
      peakbeta(\q,\a,\b)=
        (
          (2*\a*\q+\b+1)*\q-(\a*\q+\b)
          +sqrt(
            ((\a*\q+\b)-(\b+1)*\q)^2
            +4*\a*\q^2
          )
        )
        /(2*\q*(\a*\q+\b+1));
    },
  ]

    \addplot[
      thick,
      blue!10!red,
      domain=0.001:1,
      samples=400,
    ] {peakbeta(x,1,3)};
    \addlegendentry{$\alpha=1,\ \beta=3$}

    \addplot[
      thick,
      blue!30!red,
      domain=0.001:1,
      samples=400,
    ] {peakbeta(x,1,2)};
    \addlegendentry{$\alpha=1,\ \beta=2$}

    \addplot[
      thick,
      blue!50!red,
      domain=0.001:1,
      samples=400,
    ] {peakbeta(x,1,1)};
    \addlegendentry{$\alpha=1,\ \beta=1$}

    \addplot[
      thick,
      blue!70!red,
      domain=0.001:1,
      samples=400,
    ] {peakbeta(x,1,0.5)};
    \addlegendentry{$\alpha=1,\ \beta=0.5$}

    \addplot[
      thick,
      blue!90!red,
      domain=0.001:1,
      samples=400,
    ] {peakbeta(x,1,0.2)};
    \addlegendentry{$\alpha=1,\ \beta=0.2$}

    \addplot[
      thick,
      blue,
      domain=0.001:1,
      samples=400,
    ] {peakbeta(x,1,0.1)};
    \addlegendentry{$\alpha=1,\ \beta=0.1$}

  \end{axis}
\end{tikzpicture}
\caption{Capability stock at which accumulation is fastest as a function of the common requirement $q_p=q$, for $\alpha=1$ and different saturation exponents $\beta$. Lower values of $\beta$ shift the maximum toward higher capability stocks. All curves approach zero as $q\to0^+$.}
\label{bifurcation_alpha_beta2}
\end{figure}

\subsubsection{Generalized expressions}

We can generalize these results by replacing $(1-r_c)^\beta$ with $G(r_c)$ and using a more general production function $Y_{cp} = F(\left\{r_{cb}\right\}, \left\{q_{pb}\right\})$. With a single capability, the activity output is $Y_{cp}=F(r_c,q_p)$. Using the same time units as in equation~\eqref{one_cap_avgY}, we have \\
\begin{equation}
\langle Y_c\rangle=\frac{1}{N(p)}\sum_p F(r_c,q_p),\qquad
\frac{dr_c}{dt}=\gamma G(r_c)\langle Y_c\rangle.
\label{eq:general-single}
\end{equation} \\

We assume $G(r_c)$ and each $F(r_c,q_p)$ are continuous on $[0,1]$, positive and twice differentiable on $(0,1)$, and that $G(1)=0$. Differentiating with respect to $r_c$, an interior maximum of capability accumulation must satisfy \\
\begin{equation}
\frac{dG(r_c)}{dr_c}\langle Y_c\rangle
+G(r_c)\frac{d\langle Y_c\rangle}{dr_c}=0,
\end{equation} \\
Substituting the definition of average output from equation~\eqref{eq:general-single}, and noting that its derivative is the average of the derivatives of $F(r_c,q_p)$ across activities, we can divide through by $G(r_c)\langle Y_c\rangle$ and cancel the common normalization factor $1/N(p)$ to obtain:\\\\
\begin{equation}
\frac{1}{G(r_c)}\frac{dG(r_c)}{dr_c}
=
-\frac{\sum_p \frac{\partial F(r_c,q_p)}{\partial r_c}}
       {\sum_p F(r_c,q_p)}.
\label{eq:general-capability-peak}
\end{equation} \\
Thus, the maximum occurs where the proportional increase in average output equals the proportional decrease in the saturation term.\\

More simply, the condition is:
\begin{equation}
\frac{d\ln{\langle Y_c\rangle}}{dr_c}=-\frac{d\ln{G(r_c)}}{dr_c}
\end{equation} \\
That is, when the growth in output implied by the accumulation of the capability balances with the reduction associated with saturation.\\

Since $\gamma G(r_c)\langle Y_c\rangle>0$ on $(0,1)$, the derivative of capability accumulation has the same sign as \\
\begin{equation}
\frac{1}{G(r_c)}\frac{dG(r_c)}{dr_c}
+\frac{1}{\langle Y_c\rangle}
\frac{d\langle Y_c\rangle}{dr_c}.
\end{equation}\\
Thus, capability accumulation increases where this expression is positive and decreases where it is negative. When the expression decreases strictly and crosses zero, the crossing gives the unique global maximum. Otherwise, we locate the global maximum by comparing the accumulation rates at the interior solutions and the boundaries.\\
\subsection{The Multiple Capabilities Model}
\label{sec:multiple-capabilities}

We now return to the case involving an arbitrary number of capabilities. For this exploration, it will be convenient to separate again the term that depends on capability $b$ from all other terms. That is, we focus on:\\

\begin{equation}
    \frac{dr_{cb}}{dt}=\gamma(1-r_{cb})^\beta\sum_{p}\phi_{p,p^*(t)}Q_{p^*(t)b} r_{cb}^{\alpha q_{pb}}(1-q_{pb}(1-r_{cb}))\prod_{b'\neq b} r_{cb'}^{\alpha q_{pb'}} (1-q_{pb'}(1-r_{cb'}))
    \label{dynamic_eqn_separate}
\end{equation}\\

and recall that:\\

\begin{equation}
E_{cpb}=\prod_{b'\neq b} r_{cb'}^{\alpha q_{pb'}}(1-q_{pb'}(1-r_{cb'})).
\label{E_def_multi}
\end{equation}\\

which represents the output that can be attributed to the presence of other capabilities (complementary inputs). With this, we can rewrite \eqref{dynamic_eqn_separate} as:\\

\begin{equation}
    \frac{dr_{cb}}{dt}=\gamma(1-r_{cb})^\beta\bigg(\sum_{p}r_{cb}^{\alpha q_{pb}}\phi_{p,p^*(t)}(1-q_{pb})Q_{p^*(t)b}E_{cpb} + \sum_p r_{cb}^{\alpha q_{pb}+1}\phi_{p,p^*(t)}q_{pb}Q_{p^*(t)b}E_{cpb}\bigg)
    \label{eq:Riccati-multiple-cap}
\end{equation}\\

Defining the activity weights $w_p(t) = Q_{p^*(t)b}E_{cpb}\phi_{p,p^*(t)}$, their sum $W(t) = \sum_p w_p(t)$, and the weighted average capability intensity\\

\begin{equation}
    \langle q \rangle_w(t) = \frac{\sum_p q_{pb}\, w_p(t)}{W(t)}
    \label{q_weighted}
\end{equation}\\

we can factor out $W(t)$. For the cold-start case ($\alpha=0$) this gives\footnote{We note that we could have also extracted an $N(p)$ factor and instead use the average of the effective capability intensity: $q_{pb}Q_{p^*b}E_{cpb}\phi_{p,p^*(t)}$ across all activities}:\\

\begin{equation}
    \frac{dr_{cb}}{dt} = \gamma W(t)(1-r_{cb})^\beta\big((1-\langle q\rangle_w(t)) + r_{cb}\langle q\rangle_w(t)\big)
    \label{growth_average_multi}
\end{equation}\\

For $\alpha>0$ the factor $r_{cb}^{\alpha q_{pb}}$ varies across activities and remains inside the sums. When activities require capability $b$ with a common intensity $q_{pb}=q_b$, the simplification used to derive equation~\eqref{argmax_alpha_beta}, we have $\langle q\rangle_w=q_b$ and:\\

\begin{equation}
    \frac{dr_{cb}}{dt} = \gamma W(t)(1-r_{cb})^\beta r_{cb}^{\alpha q_b}\big((1-q_b) + q_b r_{cb}\big)
    \label{growth_common_q_multi}
\end{equation}\\

Equation~\eqref{growth_average_multi} is structurally identical to equation~\eqref{growth_average} from the single-capability case, except that we are now dealing with weighted averages $\langle q \rangle_w(t)$. Moreover, equation~\eqref{growth_common_q_multi} is structurally identical to equation~\eqref{one_cap_raw} with $q_p=q_b$, so the location of the maximum growth rate is given by equation~\eqref{argmax_alpha_beta} with $q=q_b$. In both cases $W(t)$ scales the speed of accumulation and leaves the location of the maximum unchanged.\\

This weighted average $\langle q \rangle_w(t)$ has a natural economic interpretation which is easy to understand for the case in which $p^*=p$\footnote{In other cases, the interpretation is about the sector $p^*$ receiving the investment and it is weighted by an additional $\phi_{p,p^*}$ factor}. The intensity with which capability $b$ is demanded is weighted by how intensely that capability is used in those activities ($Q_{p^*(t)b}$) and by how complementary that capability is to the other capabilities in the economy ($E_{cpb}$). Because $E_{cpb}$ differs across economies, $\langle q \rangle_w$ is specific to each economy.\\

The key difference from the single-capability case is that $W(t)$ and $\langle q\rangle_w(t)$ evolve over time, for two reasons. First, from equation~\eqref{pstar_profit}, the optimal investment sector $p^*(t)$ depends on the current capability vector $\{r_{cb}(t)\}$. Second, the complementarity term $E_{cpb}(t)$ changes as the other capabilities of the economy accumulate. As a result, the equivalence to the single-capability solutions holds exactly only at each instant, and the full trajectory of the system depends on how $p^*(t)$ and the capability vector co-evolve.\\

To make further progress, we characterize three cases of increasing complexity.\\

The simplest case arises when knowledge and information frictions are strong, so that $\phi_{p,p} \gg \phi_{p,p'}$ for all $p' \neq p$. In this regime, the entrepreneur never finds it worthwhile to invest outside their home sector, and $p^*(t) = p$ for all $t$. The weights become $w_p(t)=Q_{pb}E_{cpb}(t)$, which are positive only for activities that use capability $b$ and still evolve through $E_{cpb}(t)$. The argmax results of Section~\ref{sec:single-capability} hold at each instant given the current complementary endowments, which is the cross-sectional comparison we test in the empirical section. Under a common requirement $q_{pb}=q_b$, we have $\langle q \rangle_w = q_b$ for every economy and period, and the time change $d\tau = W(t)\,dt$ reduces the system exactly to the single-capability model of Section~\ref{sec:single-capability}. All analytical results, including the bifurcation at $q_b = 1/2$ for $\alpha=0$, the argmax formula, and the phase transition diagrams, then carry over without modification.\\

A richer dynamics emerge when frictions are weaker and one capability acts as a binding constraint on production. To characterize this case, we consider the situation when value function $V(\{r_{cb}\}) \propto \sum_b r_{cb}$. This assumption implies $\frac{\partial V}{\partial r_{cb}} \propto 1$ for all $b$ and captures a setting in which the entrepreneur values all capabilities equally at the margin. The marginal return to investing in sector $p'$ is then\\

\begin{equation}
    \tilde{m}_{cp'}(t) = \sum_b (1-r_{cb}(t))^\beta Q_{p'b}.
\end{equation}\\

Now suppose that one capability $b^*$ is much scarcer than all others, so that its gap $(1-r_{cb^*})^\beta$ dominates the gaps of the remaining capabilities. The marginal return then collapses to\\

\begin{equation}
    \tilde{m}_{cp'}(t) \approx (1-r_{cb^*})^\beta Q_{p'b^*},
\end{equation}\\

and the optimal investment rule simplifies to\\

\begin{equation}
    p^*(t) = \arg\max_{p'}\left[\phi_{p,p'} Q_{p'b^*}\right].
    \label{pstar_bottleneck}
\end{equation}\\

That is, the investment goes to a sector that is close to the home sector of the entrepreneur yet intense in the use of the scarce capability. This rule is time-invariant within the bottleneck phase, so $\phi_{p,p^*}$ and $Q_{p^*b^*}$ are fixed and the weights vary only through $E_{cpb^*}$. Each factor in $E_{cpb^*}$ approaches one as the corresponding capability approaches one, so the weights vary slowly while the other capabilities remain abundant relative to $b^*$. The single-capability solutions then provide a good approximation, and the dynamics reduce to a single ordinary differential equation governing the accumulation of $b^*$. This phase persists until the gap of $b^*$ shrinks to the size of the gap of the next scarcest capability $b^{**}$, at which point $b^{**}$ dominates the marginal return and $p^*(t)$ switches discontinuously to\\

\begin{equation}
    p^{**} = \arg\max_{p'}\left[\phi_{p,p'} Q_{p'b^{**}}\right].
    \label{pstar_switch}
\end{equation}\\

The economy thus moves through a succession of capability-building phases ordered by scarcity, each governed by a fixed investment destination and well-approximated by the single-capability dynamics, with discrete switches between phases as successive bottlenecks are resolved.\\

In the general case, neither of these simplifications might apply. When no single capability dominates and frictions are sufficiently weak, $p^*(t)$ will shift continuously as the full capability vector $\{r_{cb}(t)\}$ evolves, and both $W(t)$ and $\langle q\rangle_w(t)$ become genuinely time-varying. The instantaneous bifurcation structure derived above continues to hold at each moment, characterizing which economies grow fastest given their current endowments and the current investment destination. However, the global trajectory of the system depends on the full co-evolution of capabilities and investment decisions and does not reduce to a single autonomous equation. A complete characterization of this general case, including the conditions under which $p^*(t)$ converges to a steady investment pattern or cycles among destinations, could represent an important direction for future work.\\

Finally, these results do not require the particular production and saturation functions used here. With $Y_{cp}=F(\{r_{cb}\},\{q_{pb}\})$ and saturation $G(r_{cb})$, the conditions for fastest capability accumulation follow as in the single-capability case, discussed in the previous section. \\

\section{Empirical Validation}
\subsection{Sectoral Growth}
\label{sec:sectoral-growth}

In this section, we compare the model with data by examining the economies experiencing the fastest growth in each sector. Our model predicts that the capability endowment at which capability accumulation is fastest increases with the capability requirements of an activity. Since output $Y_{cp}$ increases monotonically with capability endowments, and previous work establishes a monotonic relationship between the Economic Complexity Index ($ECI$) and capability stocks ($r_{cb}$)~\cite{hidalgo2025theory}, we use $ECI$ and the Product Complexity Index $PCI$ as empirical proxies for capability endowments and requirements.

Thus, we examine whether an analogous positive relationship exists between the complexity of an activity ($PCI$) and the complexity of the economies ($ECI$) experiencing the fastest growth in it. This is not an exact test, but one that recovers the directionality of the effect identified by our theory.\\

We test this prediction using data on international exports by product (UN Comtrade, 2002--2022), payroll by industry for U.S. metropolitan statistical areas (MSAs, BEA, 2003--2022), and formal employment by sector for Brazilian microregions (RAIS, 2013--2023). These datasets are fine-grained, comprehensive, and well-harmonized across economies and activities, making them a common choice in empirical work on economic complexity~\cite{hidalgo_economic_2021,stojkoski_multidimensional_2023,balland_new_2022,hidalgo_policy_2023}. The trade data cover all countries in the world, while the MSA and RAIS datasets provide ``local'' lenses on industry specialization within a single country.\\

\begin{figure}[ht]
    \centering
    \includegraphics[width=\linewidth]{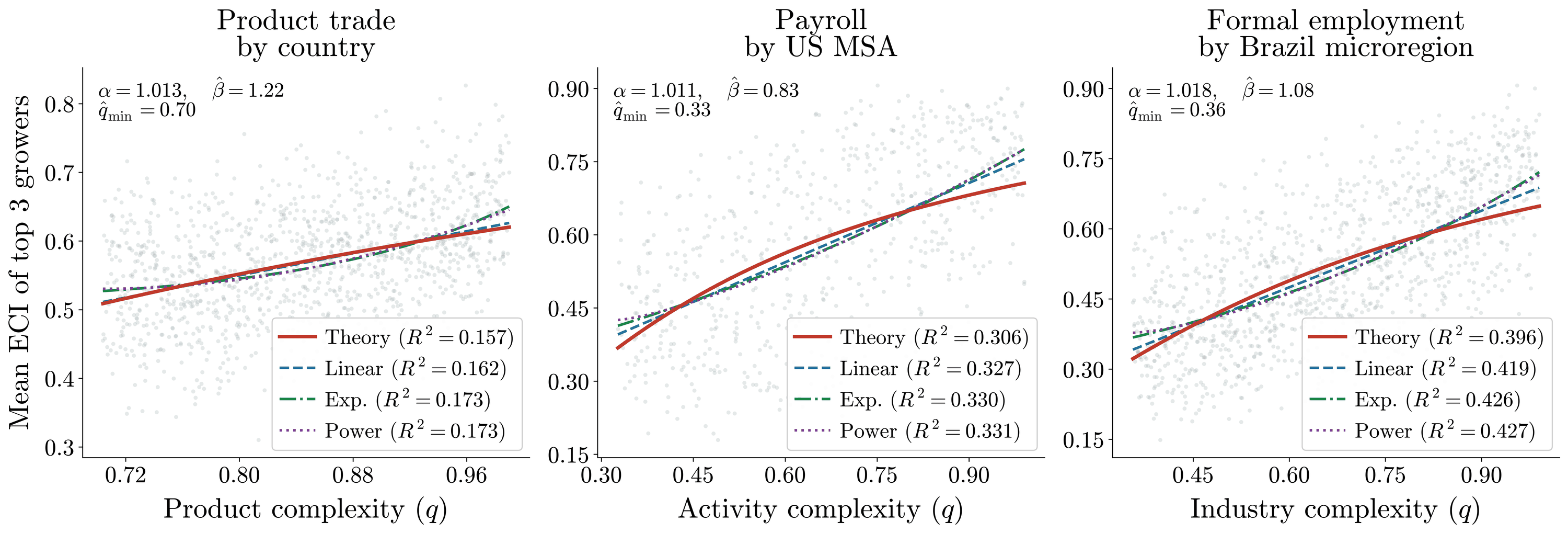}
    \caption{
        Mean ECI of the three fastest-growing economies (MSAs or microregions) in each activity as a function of the activity's complexity index (PCI), for international trade, US metropolitan areas (MSAs), and Brazilian microregions. Growth is measured over a single long period for each dataset: 2002--2022 for international trade, 2003--2022 for US MSAs, and 2013--2023 for Brazilian employment, reflecting data availability. PCI is estimated on a random $80\%$ subsample of economies and the relationship is evaluated on the remaining $20\%$, averaged over 20 independent splits. Each panel overlays the fit of equation~\eqref{argmax_alpha_beta} (solid line) against linear, exponential, and power benchmarks (dashed lines).
    }
    \label{fig:oos_fastest_growers}
\end{figure}

Figure~\ref{fig:oos_fastest_growers} plots the mean $ECI$ of the three fastest-growing economies in each activity (our proxy for $r_c$, as validated by the theoretical work in~\cite{hidalgo2025theory}), as a function of the activity's complexity index ($PCI$, our proxy for $\langle q\rangle$). To address the endogeneity that arises from using the same data to estimate both $PCI$ and $ECI$, we adopt a split-sample strategy: in each of 20 independent random splits, $PCI$ is estimated on a random $80\%$ subsample of economies and $ECI$ on the remaining $20\%$, so the two indices are never estimated from the same economies, and results are averaged over the splits. Both indices are the second eigenvector of the complexity matrices ($M_{cc'}$ and $M_{pp'}$, see~\cite{hausmann_atlas_2014,hidalgo2025theory})and can be derived iteratively through the method of reflections~\cite{hidalgo_building_2009}. They are rank-normalized to $[0.01,0.99]$. Since our model abstracts from differences in economy size, we measure output of economy $c$ in activity $p$ as $\tilde{Y}_{cp} = R_{cp}/(1+R_{cp})$, which lies in $[0,1]$ as $Y_{cp}$ does in the model, and growth as its change over the period, $\Delta\tilde{Y}_{cp}$. The fastest-growing economies in an activity are those with the largest $\Delta\tilde{Y}_{cp}$.\\

We fit the model to this relationship with equation~\eqref{argmax_alpha_beta}, which can be extended to the multiple capability model more easily when frictions are large enough that entrepreneurs stick to their home sector. Note that this equation locates the fastest growth of the capability stock $r_c$, whereas $\Delta\tilde{Y}_{cp}$ measures the change in normalized specialization . This means that the curve we fit describes a different dependent variable from the one we observe. Nevertheless, equation~\eqref{argmax_alpha_beta} has at least two advantages over fitting the location of the fastest growth of output directly. First, it is analytically tractable, since it gives the location of the fastest growth in closed form in $\alpha$, $\beta$, and $q$. Second, it provides interpretable parameters instead of an ad-hoc fit (See Appendix~\ref{app:output-growth} for the derivation of the rates of change of $\Delta Y_{cp}$). In Appendix~\ref{app:alternative_sectoral_growth} we also investigate this relationship with the share of total output and activity output per capita as approximation $\Delta\tilde{Y}_{cp}$.\\

The curve in equation~\eqref{argmax_alpha_beta} has three parameters: $\alpha$, $\beta$, and the lower end of the range of capability intensities, $q_{\min}$. Because the curve vanishes as $q\to 0$ for any $(\alpha,\beta)$ while the data do not (the mean $ECI$ of top growers is $0.3$--$0.5$ even for the simplest activities), we map the $PCI$ rank into capability intensity as $q=q_{\min}+\mathrm{rank}\times(0.99-q_{\min})$, which shifts the fitted domain away from the structural zero. The growth curve alone cannot identify $\alpha$ (Appendix~\ref{app:growth_identification}), so we estimate it from an independent source, a cross-sectional fit of the production function to a single snapshot of each dataset, which yields $\alpha=1.01$--$1.02$ in every dataset. We hold $\alpha$ fixed and estimate $\beta$ and $q_{\min}$ on the growth curve.\\

Figure~\ref{fig:oos_fastest_growers} compares equation~\eqref{argmax_alpha_beta} with linear, exponential, and power benchmarks. Across all three datasets the relationship is positive meaning that economies with lower $ECI$s grow faster in low complexity activities and economies with higher $ECI$s grow faster in more complex activities. The theory captures this positive relationship, with $R^2$ values of $0.157$ for trade, $0.306$ for US MSAs, and $0.396$ for Brazilian microregions. The best benchmarks yield somewhat higher values of $0.173$, $0.331$, and $0.427$, respectively. But the key here is that, unlike the alternative functional forms, our fitted curve is motivated by the theoretical mechanism of capability accumulation. Its parameters therefore have an interpretation within the model, even though these estimates remain conditional on the chosen proxy variables and mappings.\\

The parameter $\alpha$ is the exponent applied to $r_{cb}$ when a capability is maximally required ($q_{pb}=1$) (technically $r_{cb}^{\alpha+1}$ when $q_{pb}=1$), so it penalizes economies that lack a capability in the activities that require it most. Because $\alpha$ is essentially the same across datasets, differences in the curvature of the fitted curves are carried by $\beta$.\\

The estimates of $\beta$ range from $0.83$ for US MSAs to $1.22$ for trade, with $1.08$ for Brazilian microregions. Since $\alpha$ is around one, we can use Figure~\ref{bifurcation_alpha_beta2}, which shows the capability stock at which growth is fastest as a function of $q$ for $\alpha=1$ and various $\beta$, to interpret our results. The value well below one for US MSAs indicates that this optimum starts saturating earlier, so there is not a large difference between activities with intermediate capability requirements and those with higher requirements. As $\hat\beta$ increases for the other datasets, we move toward situations where the highest-$q$ activities also display much larger values. This indicates that in trade and Brazilian microregions there is a much larger growth premium of capability intensive activities. \\

Overall, the positive relationship between the complexity of an activity and the $ECI$ of the economies that grow fastest in it points to conditional convergence. Whether less-endowed economies catch up depends on the capability requirements of the activity, and the fastest growth shifts toward better-endowed economies as those requirements rise. \\

\subsection{Relatedness}
\label{sec:relatedness}

To conclude, we show that this model provides a theoretical explanation for the principle of relatedness, the idea that economies are more likely to enter activities that are similar to the ones they are already specialized in. This principle is one of the most well documented facts in economic geography~\cite{hidalgo_product_2007,neffke_how_2011,neffke_skill_2013,hidalgo_principle_2018,boschma_emergence_2013,kogler_mapping_2015,guevara_research_2016,hidalgo_economic_2021,balland_new_2022,jara-figueroa_role_2018}, yet work focused on this principle remains largely empirical.\\ 

In econometric terms, the principle of relatedness is usually expressed as a regression where the output of an economy in an activity (usually measured in terms of specialization) increases as a function of the number of related activities the economy is already specialized in. Formally, let the specialization $R_{cp}$ of economy $c$ in activity $p$ be defined using Balassa's revealed comparative advantage as\cite{balassa_trade_1965}:\\

\begin{equation}
R_{cp}=Y_{cp}Y/Y_cY_p
\end{equation}

And let $\omega_{cp}$ stand for relatedness, a number between 0 and 1 estimating the number of related activities an economy is specialized in. Then the principle of relatedness can be stated in regression form as\cite{hidalgo_economic_2021} as the fact that $B_2>0$ and significant in regressions of the form:\\

\begin{align}
Y_{cp}(t+\Delta t)&=B_1Y_{cp}(t)+B_2w_{cp}(t)+\ldots \\
log(Y_{cp}(t+\Delta t))&=B_1log(Y_{cp}(t))+B_2w_{cp}(t)+\ldots \\
R_{cp}(t+\Delta t)&=B_1R_{cp}(t)+B_2w_{cp}(t)+\ldots \\
log(R_{cp}(t+\Delta t))&=B_1log(R_{cp}(t))+B_2w_{cp}(t)+\ldots \\
\end{align}

Here we derive this principle by showing that the model we have been exploring naturally predicts this principle. That is, that the rate of growth of a capability $r_{cb}$ (and consequentially its output and specialization) increases in economies where other complementary capabilities are present.\\

To show this, we will start by unpacking a few well-known definitions of complementarity using the language of this model and then move on to the demonstration.\\

In general, we can say two capabilities $b$ and $b'$ are complementarity when activities tend to require both of them. Using the capability requirement matrix $q_{pb}$ this is simply the weighted dot product between the capability requirement vectors $q_{pb}$ and $q_{pb'}$:\\

\begin{equation}
    C_{bb'}=\sum_p W_p q_{pb}q_{pb'}
\end{equation}

where $W_p>0$ are weights.\\

We can also use a standard second derivative definition of complementary to obtain a similar expression. That is for $\alpha=0$ this is cleanly:\\

\begin{equation}
    \frac{\partial^2 Y_{cp}}{\partial r_{cb}\partial r_{cb'}}=E_{cpbb'} q_{pb}q_{pb'}
\end{equation}\\

where $E_{cpbb'}$ includes all terms in the production function that are unrelated to capabilities $b$ and $b'$:

\begin{equation}
E_{cpbb'}=\prod_{b''\neq b,b'}r_{cb''}^{\alpha q_{pb''}}(1-q_{pb''}(1-r_{cb''}))
\end{equation}

And for $\alpha>0$ this results in:

\begin{equation}
    \frac{\partial^2 Y_{cp}}{\partial r_{cb}\partial r_{cb'}}=E_{cpbb'} q_{pb}q_{pb'}r_{cb}^{\alpha q_{pb}-1}r_{cb'}^{\alpha q_{pb'}-1}\big(r_{cb}+\alpha (1-q_{pb}(1-r_{cb}))\big)\big( r_{cb'}+\alpha(1-q_{pb'}(1-r_{cb'}))\big)
\end{equation}\\

Which means that the response of output $Y_{cp}$ to the presence of capability $b$ increases with capability $b'$ when these are required by the same activities and are available.\\

Another way to derive the principle of relatedness is to study how the growth of a capability depends on the presence of other capabilities. That is, we explore how the rate of change of capability $b$ in economy $c$ changes with the presence of capability $b'$:\\

\begin{equation}
    \frac{\partial }{\partial r_{cb'}}\frac{d r_{cb}}{dt}
\end{equation}\\

which applied to the multiple capability model with $\alpha=0$ and strong frictions (investors focus on home sector) is:\\

\begin{equation}
    \frac{\partial }{\partial r_{cb'}}\frac{d r_{cb}}{dt}=\frac{\partial }{\partial r_{cb'}}\big(\gamma(1-r_{cb})^\beta\sum_{p}Q_{pb}(1-q_{pb}(1-r_{cb}))\prod_{b''\neq b}(1-q_{pb''}(1-r_{cb''}))\big)
\end{equation}

Here, we have unpacked the $E_{cpb}$ term coupling the rate of change of different capabilities. Taking the derivative yields:\\

\begin{equation}
    \frac{\partial }{\partial r_{cb'}}\frac{d r_{cb}}{dt}=\gamma(1-r_{cb})^\beta\sum_{p}Q_{pb}(1-q_{pb}(1-r_{cb}))q_{pb'}\prod_{b''\neq b,b'}(1-q_{pb''}(1-r_{cb''}))
    \label{eq:partialrcb}
\end{equation}\\

Which we can rearrange by including the definition of $E_{cpbb'}$ and unpacking $Q_{pb}=q_{pb}/q_{p}$. After some algebra we obtain:\\

\begin{equation}
\frac{\partial }{\partial r_{cb'}}\frac{d r_{cb}}{dt}=\gamma(1-r_{cb})^\beta\sum_{p}E_{cpbb'}\frac{q_{pb}q_{pb'}}{q_p}(1-q_{pb}(1-r_{cb}))
\end{equation}\\

Finally, we can fold the contribution of capability $b$ to output $Y_{cpb}=1-q_{pb}(1-r_{cb})$ back into $E_{cpbb'}$ to obtain $E_{cpb'}$. With that we arrive at the compact expression:\\

\begin{equation}
\frac{\partial }{\partial r_{cb'}}\frac{d r_{cb}}{dt}=\gamma(1-r_{cb})^\beta\sum_{p}E_{cpb'}\frac{q_{pb}q_{pb'}}{q_p},
\end{equation}\\

which contains the complementarity term $q_{pb}q_{pb'}$ explicitly in the sum. This expression says that the growth of capability $b$ in economy $c$ ($dr_{cb}/dt$) increases with the presence of capability $b'$ in the same economy when the capabilities $b$ and $b'$ tend to overlap across many products $\sum_{p}E_{cpb'}\frac{q_{pb}q_{pb'}}{q_p}$.\\

This shows that our production function provides a parsimonious explanation for the principle of relatedness, the idea that output in an economy grows faster for the activities that share inputs with other complementary capabilities.\footnote{We also notice that the derived functional form $\frac{q_{pb}q_{pb'}}{q_{p}}$ is very similar to the empirical form that has been used to estimate proximity for the best part of the last two decades~\cite{hidalgo_product_2007}, which focuses on conditional probabilities.}\\

To provide a derivation that is closer to what is being used in the empirical literature, which deals with output by economy and activity instead of directly unobservable capabilities, we consider the log-output $y_{cp} = \ln Y_{cp}$:\\

\begin{equation}
y_{cp}=\sum_b\big(\alpha q_{pb}\textrm{ln}(r_{cb})+\textrm{ln}\big({1-q_{pb}(1-r_{cb})}\big)\big),
\end{equation}\\
and its rate of change:\\
\begin{align}
\frac{dy_{cp}}{dt}
&= \sum_{b}q_{pb}\left[\frac{\alpha}{r_{cb}}
   + \frac{1}{1-q_{pb}(1-r_{cb})}\right]\frac{dr_{cb}}{dt},
   \label{eq:growth_full}
\end{align}\\

going forward it will be useful to define:\\

\begin{equation}
L_{cpb}=q_{pb}\left[\frac{\alpha}{r_{cb}}
   + \frac{1}{1-q_{pb}(1-r_{cb})}\right]
   \label{eq:L_def}
\end{equation}

Where $L_{cpb}$ captures the marginal contribution of the change in capability $b$ to output growth in activity $p$ in economy $c$.\\

The capability dynamics in the strong-friction regime (when investors have a strong home sector bias $\phi(p,p')=1\:for\:p=p'\:and\:0\:otherwise$) is:\\

\begin{align}
\frac{dr_{cb}}{dt}
= \gamma(1-r_{cb})^{\beta}\sum_{p'} Y_{cp'}\,Q_{p' b},
\end{align}\\

where $Q_{p' b} = q_{p' b}/q_{p'}$ is the relative intensity with which capability $b$ is required by activity $p'$. That is:\\

\begin{equation}
\frac{dy_{cp}}{dt}= \gamma \sum_b L_{cpb}(1-r_{cb})^{\beta}\sum_{p'} Y_{cp'}\,Q_{p' b},
\end{equation}\\

Which upon expanding $L_{cpb}$ and $Q_{p'b}$ results in:\\

\begin{equation}
\frac{dy_{cp}}{dt}
=
\gamma\sum_{p'}Y_{cp'}
\sum_b
\frac{q_{pb}q_{p'b}}{\sum_{b'}q_{p'b'}}
(1-r_{cb})^\beta
\left[
\frac{\alpha}{r_{cb}}
+\frac{1}{1-q_{pb}(1-r_{cb})}
\right].
\label{eq:growth_emp}
\end{equation}
Which we can interpret as output in economy $c$ in activity $p$ grows as a function of output in activity $p'$, when activity $p'$ shares capabilities with activity $p$. To bring this to a standard principle of relatedness form, we can approximate $dy_{cp}/dt \sim (y_{cp}(t+\Delta t)-y_{cp}(t))/\Delta t$ to obtain:

\begin{equation}
      y_{cp}(t+\Delta t) = y_{cp}(t)+\omega_{cp}\Delta t+...  
\end{equation}\\

where $\omega_{cp}$ is:

\begin{equation}
\omega_{cp}=\gamma\sum_{p'}Y_{cp'}
\sum_b
\frac{q_{pb}q_{p'b}}{\sum_{b'}q_{p'b'}}
(1-r_{cb})^\beta
\left[
\frac{\alpha}{r_{cb}}
+\frac{1}{1-q_{pb}(1-r_{cb})}
\right].
\end{equation}

Equation~\eqref{eq:growth_emp} generates several predictions about the role of relatedness in the growth dynamics of a location-activity pair. \\

First, the responsiveness of growth to relatedness should decline with capability stocks. That is, if we write relatedness as:\\

\begin{equation}
    \omega_{cp}=\sum_{p'}Y_{cp'}K_{cpp'}
\end{equation}\\

We can show that:
\begin{equation}
\frac{\partial K_{cpp'}}{\partial r_{cb}}\leq 0
\end{equation}\\

Second, the model predicts that relatedness is transmitted more through products that use the capability more intensively. That is:

\begin{equation}
\frac{\partial K_{cpp'}}{ \partial q_{pb}}\geq0
\end{equation}

Third, holding an economy's overall level of capabilities, activities whose relatedness is higher should grow faster within that economy.\\

We can test these predictions by estimating a regression model of the form:\\

\begin{equation}
    y_{cp}(t+\Delta t)\sim B_1y_{cp}(t)+B_2\tilde{\omega}_{cp}+B_3\tilde{\omega}_{cp}ECI_{c}+B_4\tilde{\omega}_{cp}PCI_{p}+\mu_{c}+\nu_{p}+\ldots
\end{equation}\\

where capability stocks $r_{cb}$ are proxied by an economy's economic complexity index $ECI_c$, capability requirements by the product complexity index $PCI_p$ (both of which are Z-score centered by definition) and where $\mu_{c}$ and $\nu_{p}$ are respectively country and product fixed-effects. For relatedness, we use the approximation:\\

\begin{equation}
    \tilde{\omega}_{cp}= \frac{\sum_{p'\neq p} \varphi_{pp'}M_{cp'}}{\sum_{p'\neq p} \varphi_{pp'}}
\end{equation}\\

where $\varphi_{pp'}$ is a measure of similarity among activities (e.g. proximity as defined in \cite{hidalgo_product_2007}) and $M_{cp'}=1$ if economy $c$ is specialized in activity $p'$.\\ 

Using ECI and PCI as proxies for capability stocks and requirements,
respectively, the first prediction motivates the hypothesis $B_3<0$
and the second motivates $B_4>0$. The third prediction motivates a
positive marginal association between relatedness and subsequent
log output, given by $B_2+B_3ECI_c+B_4PCI_p$. In particular, $B_2>0$
indicates a positive association at the reference values
$ECI_c=PCI_p=0$. Country and product fixed effects absorb additive
country-wide and product-specific differences.\\

Finally, to bring the model to the data, we adjust our measurement by incorporating scale factors for the size of an economy and that of an activity. That is, we consider $X_{cp}$ where:\\

\begin{equation}
    X_{cp}=a_ca_pY_{cp}
\end{equation}\\

And $a_c$ and $a_p$ accommodate the fact that, even at the same level of capabilities, some economies and activities can be bigger than others (e.g. Japan is much bigger than Singapore, and total output in apples is larger than in pears). This motivates the use of a specialization index like Balassa's revealed comparative advantage or $R_{cp}$ where:\\

\begin{equation}
R_{cp}=\frac{X_{cp}X}{X_cX_p}
\end{equation}\\

We note that:

\begin{equation}
    \ln{R_{cp}}=\ln{Y_{cp}}+\ln{a_c}+\ln{a_p}+\ln{X}-\ln{X_c}-\ln{X_p}.
\end{equation}

So $\ln{R_{cp}}$ differs from $y_{cp}=\ln{Y_{cp}}$ by additive ``economy'' and ``activity'' terms.\\

In addition, we use the normalized $R_{cp}$ estimator:\\

\begin{equation}
z_{cp}=\ln(1+R_{cp}/(1+R_{cp})),
\end{equation}\\

which is defined at $0$ and compresses variation in $R_{cp}$. We note that the derivatives of $z_{cp}$ and $R_{cp}$ are connected by a positive multiplier:

\begin{equation}
\frac{dz_{cp}}{dt}=\frac{R_{cp}}{(1+R_{cp})(1+2R_{cp})}\frac{d\ln(R_{cp})}{dt}\quad \textrm{for}\:R_{cp}>0.
\end{equation}\\

Tables~\ref{tab:reg-results-rca-ratio} and~\ref{tab:reg-results-log-rca} show the predictions of the model using the same three datasets of the previous section: international trade data by country and product, payroll by city (MSA) and industry in the United States, and formal employment by city (microregions) and industry in Brazil. \\

Table~\ref{tab:reg-results-rca-ratio} shows that when we use a bounded specialization measure as the dependent variable ($\ln(1+R_{cp}/(1+R_{cp}))$) the signs work as expected by the theory for all cases in the multi period panel.\\

\begin{table}[htbp]\centering\small
\caption{Regression results for RCA-ratio $y_{cp}=\ln(1+R_{cp}/(1+R_{cp}))$, five-year multi-period panel}
\resizebox{\textwidth}{!}{%
\begin{tabular}{lr@{}lr@{}lr@{}lr@{}lr@{}lr@{}l}
\toprule
 & \multicolumn{4}{c}{Product Trade} & \multicolumn{4}{c}{Payroll USA MSA} & \multicolumn{4}{c}{Employment Brazil Microregions} \\
\cmidrule(lr){2-5}\cmidrule(lr){6-9}\cmidrule(lr){10-13}
 & \multicolumn{2}{c}{(1)} & \multicolumn{2}{c}{(2)} & \multicolumn{2}{c}{(3)} & \multicolumn{2}{c}{(4)} & \multicolumn{2}{c}{(5)} & \multicolumn{2}{c}{(6)} \\
\midrule
$y_{cp,t}$ ($B_1$) & 0 & .841$^{***}$ & 0 & .804$^{***}$ & 0 & .248$^{***}$ & 0 & .239$^{***}$ & 0 & .724$^{***}$ & 0 & .688$^{***}$ \\
 & \textit{(0} & \textit{.001)} & \textit{(0} & \textit{.001)} & \textit{(0} & \textit{.001)} & \textit{(0} & \textit{.001)} & \textit{(0} & \textit{.001)} & \textit{(0} & \textit{.001)} \\
Relatedness density ($B_2$) &  &  & 0 & .066$^{***}$ &  &  & 0 & .031$^{***}$ &  &  & 0 & .084$^{***}$ \\
 &  &  & \textit{(0} & \textit{.001)} &  &  & \textit{(0} & \textit{.001)} &  &  & \textit{(0} & \textit{.001)} \\
Density $\times$ ECI$_c$ ($B_3$) &  &  & $-$0 & .015$^{***}$ &  &  & $-$0 & .010$^{***}$ &  &  & $-$0 & .001$^{**}$ \\
 &  &  & \textit{(0} & \textit{.001)} &  &  & \textit{(0} & \textit{.001)} &  &  & \textit{(0} & \textit{.001)} \\
Density $\times$ PCI$_p$ ($B_4$) &  &  & 0 & .023$^{***}$ &  &  & 0 & .027$^{***}$ &  &  & 0 & .024$^{***}$ \\
 &  &  & \textit{(0} & \textit{.001)} &  &  & \textit{(0} & \textit{.001)} &  &  & \textit{(0} & \textit{.001)} \\
\midrule
Observations & \multicolumn{4}{c}{515,160} & \multicolumn{4}{c}{957,846} & \multicolumn{4}{c}{1,116,050} \\
Within $R^2$ & 0 & .707 & 0 & .709 & 0 & .062 & 0 & .063 & 0 & .525 & 0 & .531 \\
\bottomrule
\end{tabular}}
\par\vspace{2pt}\parbox{\textwidth}{\footnotesize\textit{Notes:} Robust standard errors in parentheses. $^{*}p<0.1$; $^{**}p<0.05$; $^{***}p<0.01$. Each regression includes location, activity, and period fixed effects. Within each dataset the first column omits $B_2$--$B_4$ (baseline) and the second is the full model. All coefficients are standardized (z-scored after demeaning); $R^2$ is the within $R^2$ on the FE-demeaned data. Periods --- Trade: 2002-2007, 2007-2012, 2012-2017, 2017-2022 (4 periods); MSA: 2003-2008, 2008-2013, 2013-2018 (3 periods); RAIS: 2013-2018, 2018-2023 (2 periods).\label{tab:reg-results-rca-ratio}}
\end{table}

Table~\ref{tab:reg-results-log-rca} repeats the exercise with ($\ln(1+R_{cp})$) as the dependent variable. The signs here also work mostly as expected in the multi period panel, except for the $\omega_{cp}ECI_c$ interaction in the US MSA dataset. 

\begin{table}[htbp]\centering\small
\caption{Regression results for log of RCA. $y_{cp}=\ln(1+R_{cp})$, five-year multi-period panel}
\resizebox{\textwidth}{!}{%
\begin{tabular}{lr@{}lr@{}lr@{}lr@{}lr@{}lr@{}l}
\toprule
 & \multicolumn{4}{c}{Product Trade} & \multicolumn{4}{c}{Payroll USA MSA} & \multicolumn{4}{c}{Employment Brazil Microregions} \\
\cmidrule(lr){2-5}\cmidrule(lr){6-9}\cmidrule(lr){10-13}
 & \multicolumn{2}{c}{(1)} & \multicolumn{2}{c}{(2)} & \multicolumn{2}{c}{(3)} & \multicolumn{2}{c}{(4)} & \multicolumn{2}{c}{(5)} & \multicolumn{2}{c}{(6)} \\
\midrule
$y_{cp,t}$ ($B_1$) & 0 & .872$^{***}$ & 0 & .857$^{***}$ & 0 & .217$^{***}$ & 0 & .204$^{***}$ & 0 & .774$^{***}$ & 0 & .754$^{***}$ \\
 & \textit{(0} & \textit{.002)} & \textit{(0} & \textit{.002)} & \textit{(0} & \textit{.002)} & \textit{(0} & \textit{.002)} & \textit{(0} & \textit{.002)} & \textit{(0} & \textit{.002)} \\
Relatedness density ($B_2$) &  &  & 0 & .029$^{***}$ &  &  & 0 & .049$^{***}$ &  &  & 0 & .054$^{***}$ \\
 &  &  & \textit{(0} & \textit{.002)} &  &  & \textit{(0} & \textit{.001)} &  &  & \textit{(0} & \textit{.001)} \\
Density $\times$ ECI$_c$ ($B_3$) &  &  & $-$0 & .007$^{***}$ &  &  & $-$0 & .018$^{***}$ &  &  & $-$0 & .004$^{***}$ \\
 &  &  & \textit{(0} & \textit{.001)} &  &  & \textit{(0} & \textit{.001)} &  &  & \textit{(0} & \textit{.001)} \\
Density $\times$ PCI$_p$ ($B_4$) &  &  & 0 & .018$^{***}$ &  &  & 0 & .040$^{***}$ &  &  & 0 & .020$^{***}$ \\
 &  &  & \textit{(0} & \textit{.001)} &  &  & \textit{(0} & \textit{.001)} &  &  & \textit{(0} & \textit{.001)} \\
\midrule
Observations & \multicolumn{4}{c}{515,160} & \multicolumn{4}{c}{957,846} & \multicolumn{4}{c}{1,116,050} \\
Within $R^2$ & 0 & .761 & 0 & .762 & 0 & .047 & 0 & .051 & 0 & .599 & 0 & .602 \\
\bottomrule
\end{tabular}}
\par\vspace{2pt}\parbox{\textwidth}{\footnotesize\textit{Notes:} Robust standard errors in parentheses. $^{*}p<0.1$; $^{**}p<0.05$; $^{***}p<0.01$. Each regression includes location, activity, and period fixed effects. Within each dataset the first column omits $B_2$--$B_4$ (baseline) and the second is the full model. All coefficients are standardized (z-scored after demeaning); $R^2$ is the within $R^2$ on the FE-demeaned data. Periods --- Trade: 2002-2007, 2007-2012, 2012-2017, 2017-2022 (4 periods); MSA Payroll: 2003-2008, 2008-2013, 2013-2018 (3 periods); Brazil Microregion Formal Employment: 2013-2018, 2018-2023 (2 periods).\label{tab:reg-results-log-rca}}
\end{table}

Overall, the results show a reasonable level of agreement with the theory. The density prediction ($B_2$) and the density times $PCI$ interaction works across all datasets and the two specifications. The density $ECI$ interaction also shows the desired sign. This suggests that our model captures not only the most commonly reported effect of the principle of relatedness (the effect of density), but also makes predictions for the sign of the interaction between relatedness and measures of complexity for economies and activities. 

We can also derive relatedness using the general production function $Y_{cp}=F(\{r_{cb}\},\{q_{pb}\})$. For this, we retain the home-sector investment rule $p^*(t)=p$ considered in Section~\ref{sec:multiple-capabilities}, under which equation~\eqref{dynamic_eqn_with} directs investment back into the activity generating the output. We further assume that output is differentiable and non-decreasing in every capability, holding capability requirements fixed. For $Y_{cp}>0$, the term $L_{cpb}$ introduced in equation~\eqref{eq:L_def} can be written as \\
\begin{equation}
L_{cpb}
=\frac{1}{F(\{r_{cb}\},\{q_{pb}\})}
\frac{\partial F(\{r_{cb}\},\{q_{pb}\})}{\partial r_{cb}}.
\label{eq:general-L}
\end{equation} \\
Replacing $(1-r_{cb})^\beta$ with $G(r_{cb})$, capability accumulation becomes \\
\begin{equation}
\frac{dr_{cb}}{dt}
=\gamma G(r_{cb})\sum_{p'}Y_{cp'}Q_{p'b}.
\end{equation} \\
Substituting this expression into the rate of change of log output gives \\
\begin{equation}
\frac{d\ln Y_{cp}}{dt}
=\gamma\sum_b L_{cpb}G(r_{cb})
\sum_{p'}Y_{cp'}Q_{p'b}.
\end{equation} \\
Rearranging the sums, we obtain \\
\begin{equation}
\frac{d\ln Y_{cp}}{dt}
=\sum_{p'}Y_{cp'}K_{cpp'},
\qquad
K_{cpp'}=\gamma\sum_b L_{cpb}G(r_{cb})Q_{p'b}\geq0.
\label{eq:general-relatedness-kernel}
\end{equation} \\
This is the same coefficient $K_{cpp'}$, now written for a general production function. It is positive when investment in activity $p'$ builds a capability that increases output in activity $p$, provided that capability can still accumulate. Thus, investment in activities sharing capabilities with $p$ contributes to its growth. This result depends on the first derivatives of production with respect to capabilities. It does not require positive mixed second derivatives, which describe complementarities in production.\\

We can also examine how the accumulation of capability $b$ depends on the presence of another capability $b'\ne b$. Differentiating the accumulation equation gives \\
\begin{equation}
\frac{\partial}{\partial r_{cb'}}
\frac{dr_{cb}}{dt}
=\gamma G(r_{cb})\sum_p Q_{pb}
\frac{\partial F(\{r_{cb}\},\{q_{pb}\})}{\partial r_{cb'}}
\geq0.
\label{eq:general-cross-capability}
\end{equation} \\
The presence of capability $b'$ increases output in activities that use it. Investment in these activities also builds capability $b$ when they use $b$ as well. Hence, this connection between capabilities arises even when they are not complementary in production.\\

The interaction predictions derived from equation~\eqref{eq:growth_emp} require additional conditions. Namely, the contribution of capability $b$ to $K_{cpp'}$ decreases with its stock when
\begin{equation}
\frac{\partial}{\partial r_{cb}}
\left[G(r_{cb})L_{cpb}\right]\leq0,
\label{eq:general-relatedness-attenuation}
\end{equation}
holding the other capabilities fixed. This holds if both $G(r_{cb})$ and $L_{cpb}$ are nonincreasing in $r_{cb}$. In this case, a larger capability stock reduces accumulation through saturation and reduces the marginal contribution of that capability to output growth. For the overall effect of relatedness to decline with capability stocks, the weighted sum defining $K_{cpp'}$ must also decline as those stocks increase. Similarly, a stronger effect of relatedness in activities with higher capability requirements requires this weighted sum to increase with those requirements.\\

The condition $G(1)=0$ ensures that capability accumulation stops at the upper limit, but does not determine how accumulation changes as that limit is approached. Nor does it determine how output responds to capability stocks or requirements. Shared capabilities and investment can thus generate relatedness under general production and accumulation functions, while the interaction signs depend on additional conditions on these functions. \\

\section{Discussion}
\label{sec:discussion}

Increasing returns, conditional convergence, and escaping the Malthusian trap are all recurrent themes in the economic growth and international development literature. Here, we presented a model that informs these questions by using a product specific weak-link production function and showing this can be used to induce a dynamical system that can help explain the onset of conditional convergence.\\

This simple insight provides an interesting twist to the growth literature, suggesting that conditional divergence can emerge spontaneously as the production of activities crosses a threshold of capability requirements. In simple terms, in a world where activities are simple enough to not require much learning or specialized knowledge, we should expect convergence, as those entering late into the game will accumulate the capability at a rate that is largely independent of their proficiency or use. But suddenly, when an activity becomes intense enough in the use of a capability, divergence will emerge, since at that point those with the capability (e.g. those with the experience) will start accumulating the capability faster than those without it.\\

This idea is consistent with the onset of growth at the time of the industrial revolution. The industrial revolution involved a transition from manual to mechanized activities that is well illustrated by the history of water-powered cotton spinning technology.\\

Water-powered cotton spinning technology was developing in the midlands of the UK during the late eighteenth century.\footnote{For a detailed description of the history, see Chapter 6 of The Infinite Alphabet\cite{hidalgo2025infinite} and Ref.~\cite{bagnall_samuel_1890}.} It was a fiendishly difficult technology to master, with patents predating the development of the first successful commercial mills by decades. It was also a technology that came of age around the time of the American revolution. In the UK, the first successful mills were created by Richard Arkwright and Jedediah Strutt, who became incredibly successful as entrepreneurs. In the United States, attempts to replicate their technology based on hearsay failed to produce successful commercial mills. That is, until Samuel Slater, a young man with several years of experience in one of Strutt's mills, arrived in the United States.\\

Slater was able to achieve in months what no American entrepreneur had been able to do until his arrival. Within a year, he developed the first viable cotton spinning mill in the United States, kick-starting the economic divergence of Rhode Island and New England. In the following years, several cotton spinning mills were created not far from Slater's first mill. According to the 1810 Tabular Statements of American Manufactures, Rhode Island and Massachusetts become key centers for the production of spun cotton. Their combined output in both yarns and dollars was more than half that of the entire country.\footnote{See page 6 in \url{https://www2.census.gov/library/publications/decennial/1810/1810v2/1810v2-06.pdf}.} By some accounts, ``Rhode Island was the richest state in the union'' by 1840~\cite{lagerlof2005geography}. These historical facts are consistent with the idea that divergence starts in the presence of complex production process, since it is in the presence of these processes that previous experience, like that brought by Samuel Slater, sets on the onset of divergence. It is also in the presence of these processes that laggards (in this case hand spinners or those without experience building mills) struggle to catch up.\\

Our model provides a simple and parsimonious way to recover this behavior. But it also provides a mean to explain why a divergence like that experienced by Rhode Island is not guaranteed an eternal life. While this paper focused only on the accumulation of capabilities, the model provides mechanisms for the return of unconditional convergence. For example, in activities where changes in technology reduce the intensity with which capabilities are required (e.g. some production processes become easier) then we should experience a return to convergence. That is, as cotton spinning technology becomes more common knowledge (e.g., factories can buy the machines without the need to have someone like Slater design and build them), then convergence in that activity should become stronger.\\

Our model also provides a parsimonious explanation for path dependencies in industrial development, which is one of the best document facts in economic geography. The dynamics of the model naturally leads to a cross dependency between capabilities in which the rate at which one capability is accumulated depends on the presence of others, and where this effect grows when these other capabilities are complementary (tend to be inputs to the same activities). This provides a theoretical foundation for a fact that has been documented empirically, but that has hitherto lacked a formal deductive explanation:a theoretical foundation for the principle of relatedness that derives its effects from underlying shared capabilities, production complementarities, and capability accumulation, rather than taking the network of related activities as given.\\

The empirical analysis lends credibility to the model's core mechanism, and helps situate the world within the theoretical landscape the model describes. What the model adds to the existing empirical literature on conditional convergence is a structural interpretation, suggesting that convergence is conditional because of the intensity with which activities require capabilities. In simple terms, it tells us that a world where activities are more intense in their capability requirements is a world where convergence depends more on capability stocks.\\

Our work is an additional step towards a dynamic understanding of economic complexity that opens many additional questions.\\

First, each capability $r_{cb}$ rises toward its ceiling whenever an economy produces in the activities that use it. Conditional on producing across all activities, every economy eventually approaches a common frontier, at which output is maximized and growth ceases, so that the conditional convergence we emphasize is ultimately a property of the transition path rather than of the steady state. One could argue that no economy is ever near $r=1$ in practice, since technologies and knowledge continually evolve and the frontier itself keeps advancing, so that all economies remain on the transition path and grow at rates that depend on their position relative to the maximum. Even so, it remains relevant to embed our framework in a more complete model of long-run growth. A natural route, rather than capturing catch-up alone, is to nest the capability-based production function in a larger structure with accumulable factors, for instance by treating the capability term as an efficiency component multiplying a standard production function in physical capital and labor, so that long-run growth is driven by capital accumulation and technological progress as in the Solow model. \\

Second, our current implementation does not consider changes in the number of capabilities available to economies or required by activities. Adding capabilities is bound to provide an interesting dynamics, since products can become simpler or more complex depending on how these capabilities interact with others.\footnote{E.g., a new general capability that makes many other capabilities redundant would make a product simpler.} \\

Third, our model operates at a partial equilibrium level with respect to capabilities, treating them as available to all sectors simultaneously without accounting for potential inter-sectoral competition for access to a limited pool of skilled workers—something that could be explored in a fully general equilibrium framework. \\

Finally, we did not explore equations governing changes in the capabilities required by an activity ($dq_{pb}/dt$), which could be in themselves an exciting avenue of research. \\

All of these extensions would help move the theory closer to a general dynamical framework for structural transformation: one capturing not only how economies grow through their existing activities, but how they reshape the very space of possibilities in which development unfolds.\\

%\bibliographystyle{unsrt}
%\bibliography{eci-bib,BibTexJun4}

\section*{Appendix}

\subsection*{Appendix I: The Single Capability Model with Depreciation}
\label{app:single-dep}

Here we present the single-capability version of the model with a depreciation term $\delta$. In  this case, equation~\eqref{one_cap_raw} generalizes to\\

\begin{equation}
\frac{dr_c}{dt} = \gamma (1-r_c)^{\beta} \sum_p r_c^{\alpha q_p}\bigl(1-q_p(1-r_c)\bigr) \;-\; \delta\, r_c \sum_p q_p ,
\label{app:eq:single-dep}
\end{equation}\\

which we can reorganize, as in equation~\eqref{one_cap_dyn}, into\\

\begin{equation}
\frac{dr_c}{dt} = \gamma (1-r_c)^{\beta}\left(\sum_p r_c^{\alpha q_p}(1-q_p) + \sum_p r_c^{\alpha q_p+1} q_p\right) - \delta\, r_c \sum_p q_p .
\label{app:eq:single-dep-reorg}
\end{equation}\\

For $\alpha=0$ the terms are linear and separable, so the sums become averages after factoring out $N(p)$. Using the definition of $\langle q\rangle$ in equation~\eqref{q_average} and absorbing $N(p)$ into the time unit, we obtain\\

\begin{equation}
\frac{dr_c}{dt} = \gamma (1-r_c)^{\beta}\bigl[(1-\langle q\rangle) + \langle q\rangle r_c\bigr] - \delta \langle q\rangle r_c .
\label{app:eq:single-dep-a0}
\end{equation}\\

The depreciation term $\delta\langle q\rangle r_c$ removes capability at a rate proportional to the current stock and to the average intensity with which activities require it. As in Section~\ref{sec:single-capability}, we first study the cold-start case and then the case $\alpha>0$.\\

\subsubsection*{Case $\alpha=0$, $\beta=1$ (cold-start accumulation)}

Setting $\beta=1$ in equation~\eqref{app:eq:single-dep-a0} gives a Riccati equation\\

\begin{equation}
\frac{dr_c}{dt} = A r_c^2 + B r_c + C,
\label{app:eq:riccati-dep}
\end{equation}\\

with\\

\begin{equation}
A = -\gamma\langle q\rangle, \qquad
B = \gamma(2\langle q\rangle - 1) - \delta\langle q\rangle, \qquad
C = \gamma(1-\langle q\rangle).
\label{app:eq:riccati-coef}
\end{equation}\\

Depreciation enters only the linear coefficient $B$. Because $A<0$ the parabola opens downward and has a unique maximum at $r_c^{*}=-B/2A$. This maximum is interior to $[0,1]$ exactly when $B>0$, since $-B/2A<1$ holds for every admissible parameter value. Writing $B=(2\gamma-\delta)\langle q\rangle-\gamma$, the condition $B>0$ is\\

\begin{equation}
\langle q\rangle > \frac{\gamma}{2\gamma-\delta}, \qquad \text{for } \delta<2\gamma .
\label{app:eq:threshold-a0}
\end{equation}\\

For $\delta\ge 2\gamma$ the coefficient $B$ is negative for every $\langle q\rangle\in[0,1]$. The threshold in equation~\eqref{app:eq:threshold-a0} lies inside $(0,1)$ when $\delta<\gamma$, and it rises from $1/2$ at $\delta=0$ to $1$ at $\delta=\gamma$. An interior peak of the accumulation rate, and hence conditional convergence, is possible only when reinvestment outpaces depreciation ($\delta<\gamma$). For $\delta\ge\gamma$ the maximum stays at $r_c=0$ for all feasible $\langle q\rangle$. The location of the maximum is\\

\begin{equation}
r_c^{*} = \arg\max_{r_c}\frac{dr_c}{dt} =
\begin{cases}
0, & \text{if } \langle q\rangle \le \dfrac{\gamma}{2\gamma-\delta},\\[2ex]
1 - \dfrac{1}{2\langle q\rangle} - \dfrac{\delta}{2\gamma}, & \text{if } \langle q\rangle > \dfrac{\gamma}{2\gamma-\delta}.
\end{cases}
\label{app:eq:argmax-a0}
\end{equation}\\

Relative to equation~\eqref{argmax_cold_start}, depreciation shifts the interior branch down by the constant $\delta/(2\gamma)$, so the peak moves toward less endowed economies by an amount proportional to $\delta/\gamma$. The order parameter $r_c^{*}$ remains continuous at the threshold and its derivative jumps, so the transition keeps the character of a second-order phase transition with a critical point at $\langle q\rangle_c=\gamma/(2\gamma-\delta)$.\\

Depreciation also changes the long-run level of the capability. At $r_c=0$ the accumulation rate equals $C>0$, and at $r_c=1$ it equals $A+B+C=-\delta\langle q\rangle<0$. Since the right-hand side of equation~\eqref{app:eq:riccati-dep} is concave, there is a unique steady state in $(0,1)$, and it is globally stable on $[0,1]$,\\

\begin{equation}
\bar r_c = \frac{B + \sqrt{B^2 + 4\gamma^2\langle q\rangle(1-\langle q\rangle)}}{2\gamma\langle q\rangle} \;<\; 1 \quad \text{for } \delta>0 .
\label{app:eq:ss-a0}
\end{equation}\\

The steady state falls as $\delta$ rises and equals one at $\delta=0$.\\

\subsubsection*{Case $\alpha>0$ (no cold start)}

For $\alpha>0$ the factor $r_c^{\alpha q_p}$ varies across activities and remains inside the sums. Differentiating equation~\eqref{app:eq:single-dep} with respect to $r_c$, as in equation~\eqref{foc_general}, gives\\

\begin{equation}
\frac{d}{dr_c}\frac{dr_c}{dt} = \gamma(1-r_c)^{\beta}\sum_p r_c^{\alpha q_p}\left[\bigl(1-q_p(1-r_c)\bigr)\left(\frac{\alpha q_p}{r_c} + \frac{q_p}{1-q_p(1-r_c)} - \frac{\beta}{1-r_c}\right)\right] - \delta\sum_p q_p .
\label{app:eq:foc}
\end{equation}\\

The first term is equation~\eqref{foc_general}, and the second term is the derivative of the depreciation term. As in the main text, this expression lacks a closed-form solution, so we study it after assuming $q_p=q$ for all $p$. After absorbing $N(p)$ into the time unit, the maximum $r_c^{*}$ then satisfies\\

\begin{equation}
\gamma(1-r_c^{*})^{\beta} (r_c^{*})^{\alpha q}\bigl(1-q(1-r_c^{*})\bigr)\left(\frac{\alpha q}{r_c^{*}} + \frac{q}{1-q(1-r_c^{*})} - \frac{\beta}{1-r_c^{*}}\right) = \delta q .
\label{app:eq:foc-dep}
\end{equation}\\

For $\delta=0$ the factors in front of the parenthesis are strictly positive on $(0,1)$, so the condition reduces to equation~\eqref{foc_homogeneous}, whose solution is equation~\eqref{argmax_alpha_beta}. For $\delta>0$ the right-hand side is positive, so the parenthesis must be positive at the maximum. The parenthesis is positive to the left of the no-depreciation maximum and negative to its right, meaning that depreciation moves the peak of the accumulation rate toward less endowed economies. The size of the shift grows with $\delta/\gamma$. For $\delta=0$ the left-hand side of equation~\eqref{app:eq:foc-dep} reduces to the quadratic in equation~\eqref{quadratic_homogeneous} times the positive factor $(1-r_c)^{\beta-1}r_c^{\alpha q-1}$, which drops out and yields the closed form in equation~\eqref{argmax_alpha_beta} for any $\alpha\ge 0$ and $\beta>0$. For $\delta>0$ this factor stays in the condition, so equation~\eqref{app:eq:foc-dep} has no closed-form solution for general $\alpha$ and $\beta$, and $r_c^{*}$ is obtained numerically. The cold-start case $\alpha=0$, $\beta=1$ is the exception, since the factor equals $r_c^{-1}$ and the condition becomes linear.\\

Depreciation also changes the long-run level of the capability. With $q_p=q$, steady states of equation~\eqref{app:eq:single-dep} with $r_c>0$ solve\\

\begin{equation}
\gamma(1-r_c)^{\beta} r_c^{\alpha q-1}\bigl[(1-q)+q r_c\bigr] = \delta q .
\label{app:eq:ss-general}
\end{equation}\\

For $\alpha q<1$ the left-hand side diverges as $r_c\to 0$ and vanishes at $r_c=1$, so at least one steady state lies in $(0,1)$, the state $r_c=0$ is unstable, and every economy with a positive stock accumulates toward a level below one. For $\alpha q>1$ the left-hand side vanishes at both ends of the interval, so $r_c=0$ is locally stable. Equation \eqref{app:eq:ss-general} then has either no interior solution, in which case the capability decays to zero from every initial stock, or two solutions, a lower unstable threshold and an upper stable steady state. Economies starting below the threshold lose the capability and economies above it converge to the upper steady state. With the estimated $\alpha\approx 1.01$, this regime applies only to activities with $q$ close to one, which are the activities in which conditional convergence is strongest in the main text. Depreciation thus sharpens the conditional convergence of the main text into a capability trap in the most demanding activities.\\

\subsection*{Appendix II: Analytical Solution for Multiple Capabilities with Depreciation}
\label{app:multi-dep}

We now extend the capability model with depreciation to an arbitrary number of capabilities. Using the depreciation term introduced in Section~\ref{sec:dynamics} and separating the contribution of capability $b$ as in equation~\eqref{dynamic_eqn_separate}, the dynamics are\\

\begin{equation}
\frac{dr_{cb}}{dt} = \gamma(1-r_{cb})^{\beta}\sum_p \phi_{p,p^*(t)}\,Q_{p^*(t)b}\, r_{cb}^{\alpha q_{pb}}\bigl(1-q_{pb}(1-r_{cb})\bigr)E_{cpb}(t) \;-\; \delta\, r_{cb}\sum_p q_{pb},
\label{app:eq:multi-dep}
\end{equation}\\

with $E_{cpb}$ defined in equation~\eqref{E_def_multi}. Capability $b$ depreciates at a rate set by its own stock and by the intensity with which activities use it, independently of the current investment destination. This keeps forgetting a property of the capability, consistent with the literature cited in Section~\ref{sec:dynamics} \cite{mason2007interrupted,david2011determinants,darr1995acquisition,argote_learning_1990,benkard2000learning,thompson_how_2007}.\\

Using the activity weights $w_p(t)=Q_{p^*(t)b}\,E_{cpb}(t)\,\phi_{p,p^*(t)}$, their sum $W(t)=\sum_p w_p(t)$, the weighted intensity $\langle q\rangle_w(t)$ of equation~\eqref{q_weighted}, and the unweighted average $\langle q\rangle=\sum_p q_{pb}/N(p)$, the cold-start case ($\alpha=0$, $\beta=1$) becomes the Riccati equation\\

\begin{equation}
\frac{dr_{cb}}{dt} = \gamma W(t)(1-r_{cb})\bigl[(1-\langle q\rangle_w(t)) + \langle q\rangle_w(t)\, r_{cb}\bigr] - \delta N(p)\langle q\rangle\, r_{cb},
\label{app:eq:multi-dep-a0}
\end{equation}\\

which is the analog of equation~\eqref{growth_average_multi}. Its coefficients are $A=-\gamma W\langle q\rangle_w$, $B=\gamma W(2\langle q\rangle_w-1)-\delta N(p)\langle q\rangle$, and $C=\gamma W(1-\langle q\rangle_w)$. Investment is weighted by $W(t)$ and $\langle q\rangle_w(t)$, while depreciation is weighted by $N(p)$ and $\langle q\rangle$. The interior maximum of the accumulation rate is\\

\begin{equation}
r_{cb}^{*} = 1 - \frac{1}{2\langle q\rangle_w} - \frac{\delta}{2\gamma}\,\frac{N(p)\langle q\rangle}{W(t)\langle q\rangle_w},
\label{app:eq:argmax-multi-a0}
\end{equation}\\

and it lies in $(0,1)$ when $B>0$. Relative to equation~\eqref{app:eq:argmax-a0}, depreciation enters through the effective rate $\delta N(p)\langle q\rangle/(W(t)\langle q\rangle_w)$, which depends on the economy's complementary capabilities through $W(t)$ and $\langle q\rangle_w(t)$.\\

For $\alpha>0$ the factor $r_{cb}^{\alpha q_{pb}}$ varies across activities and remains inside the sums. When activities require capability $b$ with a common intensity $q_{pb}=q_b$, as in equation~\eqref{growth_common_q_multi}, we have $\langle q\rangle_w=\langle q\rangle=q_b$ and\\

\begin{equation}
\frac{dr_{cb}}{dt} = \gamma W(t)(1-r_{cb})^{\beta} r_{cb}^{\alpha q_b}\bigl[(1-q_b) + q_b r_{cb}\bigr] - \delta N(p)\, q_b\, r_{cb}.
\label{app:eq:multi-dep-common}
\end{equation}\\

Measuring time in units of $d\tau=W(t)\,dt$ turns equation~\eqref{app:eq:multi-dep-common} into the single-capability model of Appendix~I with $q=q_b$ and the depreciation rate $\delta$ replaced by $\delta N(p)/W(t)$. In the single-capability model $W=N(p)$ and the two rates coincide. With several capabilities each weight satisfies $w_p\le 1$, so $W(t)\le N(p)$ and the effective rate $\delta N(p)/W(t)$ is at least $\delta$. All results of Appendix~I hold at each instant with this substitution. In the cold-start case the threshold becomes $q_b>\gamma/(2\gamma-\delta N(p)/W(t))$, the interior argmax becomes $1-1/(2q_b)-\delta N(p)/(2\gamma W(t))$, and the steady state is equation~\eqref{app:eq:ss-a0} with $\delta N(p)/W(t)$ in place of $\delta$.\\

The effective depreciation rate adds an economic channel absent from the model without depreciation. The weight $W(t)$ rises with the complementary capabilities summarized in $E_{cpb}$, so economies rich in complements face a lower effective depreciation of capability $b$. They reach the conditional convergence regime at a lower intensity threshold and sustain a higher long-run stock $\bar r_{cb}$. \\

The simplest case arises when knowledge and information frictions are strong ($\phi_{p,p}\gg\phi_{p,p'}$ for $p'\neq p$), so that $p^*(t)=p$ for all $t$. The weights become $w_p(t)=Q_{pb}E_{cpb}(t)$, which still evolve with the complementary capabilities. The argmax results above hold at each instant given the current complementary endowments, which is the cross-sectional comparison used in the empirical section. Under a common requirement $q_{pb}=q_b$, the time change $d\tau=W(t)dt$ reduces the system to the single-capability model of Appendix~I with the time-varying depreciation rate $\delta N(p)/W(t)$, so the threshold $\gamma/(2\gamma-\delta N(p)/W)$ for $\alpha=0$ and the trap structure for $\alpha q_b>1$ also hold at each instant. The full trajectory is a non-autonomous system, since $\delta N(p)/W(t)$ falls as complementary capabilities accumulate.\\

A richer dynamics emerges when frictions are weaker and one capability acts as a binding constraint. With the same approximation $V(\{r_{cb}\})\propto\sum_b r_{cb}$, the marginal return to investing in sector $p'$ is $\tilde m_{cp'}(t)=\sum_b(1-r_{cb}(t))^{\beta}Q_{p'b}$. When one capability $b^*$ has the largest gap, the optimal rule is $p^*(t)=\arg\max_{p'}\phi_{p,p'}Q_{p'b^*}$, as in equation~\eqref{pstar_bottleneck}. Within this phase capability $b^*$ follows the single-capability dynamics of Appendix~I and approaches its steady state $\bar r_{cb^*}<1$. Depreciation changes the phase structure in two ways. First, capabilities used little in the chosen destinations receive little investment while depreciating at the rate $\delta r_{cb}\sum_p q_{pb}$, so their gaps widen during the phase. Second, the phase ends when the gap of $b^*$ shrinks to the size of the largest other gap, which can happen before $b^*$ reaches its steady state, since that other gap is growing. Investment then switches to $p^{**}=\arg\max_{p'}\phi_{p,p'}Q_{p'b^{**}}$, as in equation~\eqref{pstar_switch}. Because neglected capabilities erode, an earlier bottleneck can bind again, and the economy can cycle among investment destinations. Without depreciation every gap shrinks monotonically and phases follow one another in order of scarcity.\\

In the general case, neither simplification applies. When no single capability dominates and frictions are weak, $p^*(t)$ shifts continuously as the capability vector evolves, and $W(t)$ and $\langle q\rangle_w(t)$ vary over time. The instantaneous results above continue to characterize which economies accumulate a capability fastest given their current endowments. Depreciation places every long-run capability level below one, and it ties the level each economy can sustain to its stock of complementary capabilities through $W(t)$.\\\\

\subsection*{Appendix III: Identification of the Growth Curve Parameters}
\label{app:growth_identification}

Here we document why and how $\alpha$ is imported from a cross-sectional fit instead of being estimated on the growth curve.\\

By profiling $R^2$ against $\alpha$ (re-optimizing $\beta$ and $q_{\min}$ at each value), we find that $R^2$ varies by only $0.0001$--$0.003$ across $\alpha\in[0.1,20]$ in every dataset. The reason is for this structural. Namely, $\alpha$ enters the production function only through the exponent $\alpha q_{pb}$. Thus, it becomes invisible wherever $q_{pb}$ is small, and the growth curve, which provides a single point per activity, lacks the variation in $q$ needed to separate $\alpha$ from $\beta$ and $q_{\min}$. By contrast, $R^2$ is sharply peaked in both $\beta$ and $q_{\min}$ once $\alpha$ is fixed, so these two are ``identified'' by the growth curve itself (given the chosen proxies and mappings).\\

\begin{figure}[ht]
    \centering
    \includegraphics[width=\linewidth]{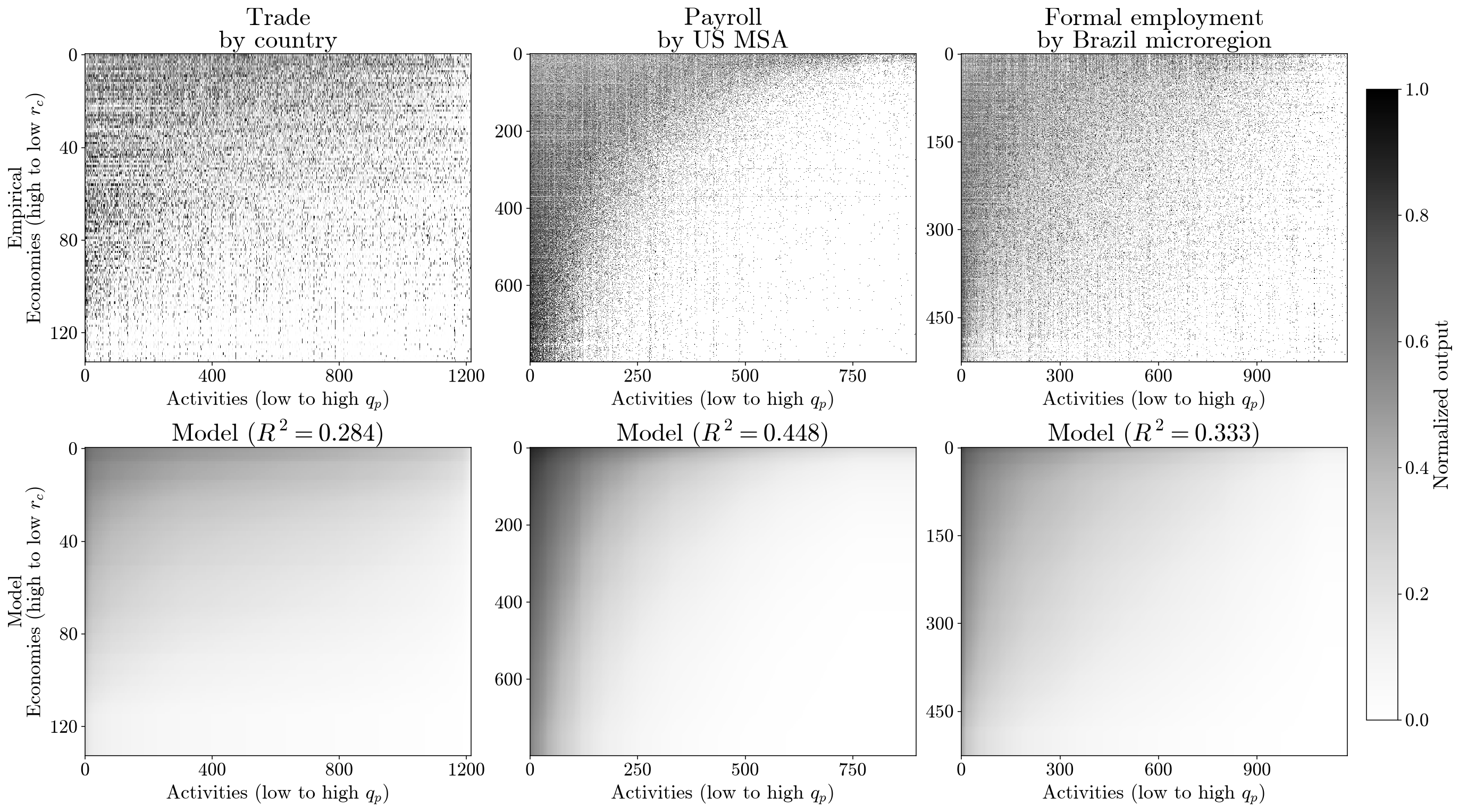}
    \caption{
        Empirical (top) and model-predicted (bottom) $\tilde{Y}_{cp}$ matrices from the cross-sectional fit, for product trade by country, payroll by US metropolitan area, and employment by Brazilian microregion. Rows are sorted from the highest to the lowest fitted $r_c$ and columns from the lowest to the highest fitted $q_p$. The $R^2$ of each fit is reported above the corresponding model panel.
    }
    \label{fig:app_sorted_matrices}
\end{figure}

\begin{figure}[ht]
    \centering
    \includegraphics[width=\linewidth]{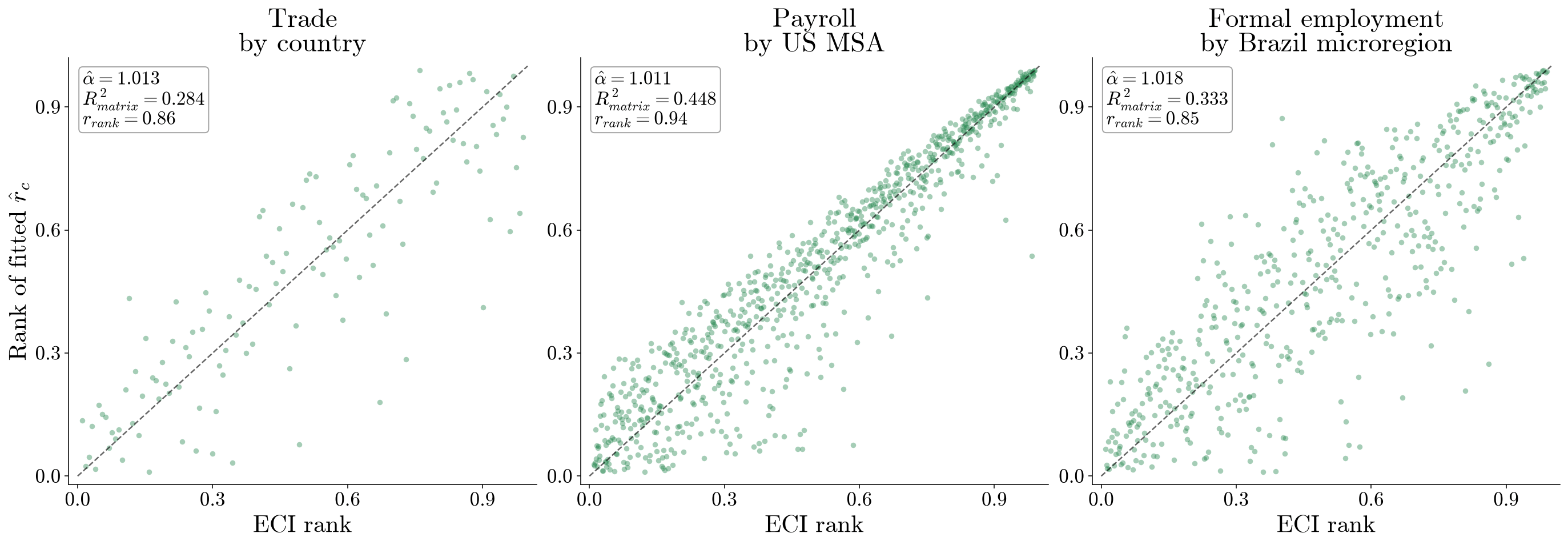}
    \caption{
        Rank of the fitted $r_c$ against the rank of the eigenvector $ECI$, computed independently on the same cross-sectional snapshot, for product trade by country, payroll by US metropolitan area, and employment by Brazilian microregion. The dashed line is the $45^{\circ}$ line and $r$ is the correlation between the two ranks.
    }
    \label{fig:app_rc_eci}
\end{figure}

To tackle this limitation, we estimate $\alpha$ from a cross-sectional fit of the single-capability production function to a single snapshot of each dataset. In this estimation procedure, we first define $\tilde{Y}_{cp}=R_{cp}/(1+R_{cp})$, where $R_{cp}$ is the revealed comparative advantage. This is a smooth, parameter-free mapping of $R_{cp}$ onto $[0,1]$ in which $R_{cp}=1$ (specialization) maps to $\tilde{Y}_{cp}=0.5$. Then, we recover $r_c$, $q_p$, and a single global $\alpha$ by alternating minimization of the distance between the production function and the fitted $\tilde{Y}_{cp}$. This fit compares many economies within each activity and the same economy across many activities, which gives $\alpha$ leverage from cells across the full range of $q_p$. It explains $28\%$ (Trade), $45\%$ (MSA), and $33\%$ (Brazil microregions) of the variance of the empirical $\tilde{Y}_{cp}$ matrix (Figure~\ref{fig:app_sorted_matrices}, and the fitted $r_c$ rank-correlate with the eigenvector $ECI$ computed on the same snapshot at $0.86$ to $0.94$ (Figure~\ref{fig:app_rc_eci}), thus suggesting an overall good fit.\\

\subsection*{Appendix IV: Output Growth in the Single Capability Model}
\label{app:output-growth}

Here we repeat the analysis of Section~\ref{sec:single-capability} for three output-based quantities, the change in activity output $dY_{cp}/dt$, the growth rate of activity output $d\ln Y_{cp}/dt$, and the growth rate of aggregate output $d\ln Y_c/dt$. \\

Due to chain rule, we have \\
\begin{equation}
\frac{dY_{cp}}{dt} = q_p\, r_c^{\alpha q_p - 1}\left[\alpha\left(1 - q_p(1-r_c)\right) + r_c\right]\frac{dr_c}{dt}. \label{eq:Ydot}
\end{equation} \\
Dividing by $Y_{cp}$ gives the growth rate of activity output, \\
\begin{equation}
\frac{dY_{cp}}{dt} \frac{1}{Y_cp} = \frac{d\ln Y_{cp}}{dt} = q_p\left[\frac{\alpha}{r_c} + \frac{1}{1 - q_p(1-r_c)}\right]\frac{dr_c}{dt} \equiv L_{cp}\,\frac{dr_c}{dt}, \label{eq:lnYdot}
\end{equation} \\
with $L_{cp}$ the single capability version of equation~\eqref{eq:L_def}. Aggregate output $Y_c = \sum_p Y_{cp}$ grows at \\
\begin{equation}
\frac{d\ln Y_c}{dt} = \frac{1}{Y_c}\sum_p \frac{dY_{cp}}{dt} = \sum_p \frac{Y_{cp}}{Y_c}\,L_{cp}\,\frac{dr_c}{dt}, \label{eq:lnYcdot}
\end{equation} \\
the output-weighted average of the activity growth rates. As in Section~\ref{sec:single-capability}, we first study the cold-start case and then move to $\alpha > 0$. \\

\subsubsection*{Cold start ($\alpha = 0$, $\beta = 1$)}

For $\alpha = 0$ the accumulation rate is the Riccati equation~\eqref{Riccati_poly}, $dr_c/dt = \gamma(1-r_c)\left(1-\langle q\rangle + \langle q\rangle r_c\right)$, with $N(p)$ absorbed into the time unit. Equation~\eqref{eq:Ydot} reduces to \\
\begin{equation}
\frac{dY_{cp}}{dt} = q_p\,\frac{dr_c}{dt}, \label{eq:Ydot0}
\end{equation} \\
so the change in activity output is the Riccati parabola scaled by the constant $q_p$. Its maximum lies at the same location as the maximum of capability accumulation, given by equation~\eqref{argmax_cold_start}, with the same critical point at $\langle q\rangle = 1/2$. The same holds for the change in aggregate output, $dY_c/dt = N(p)\langle q\rangle\, dr_c/dt$. In the cold-start case the level of output growth inherits the transition from unconditional to conditional convergence exactly. \\

The growth rate of activity output is \\
\begin{equation}
\frac{d\ln Y_{cp}}{dt} = \gamma\, q_p\,(1-r_c)\,\frac{1-\langle q\rangle + \langle q\rangle r_c}{1-q_p + q_p r_c}. \label{eq:lnYdot0}
\end{equation} \\
Setting the derivative of its logarithm to zero and multiplying by $(1-r_c)(1-\langle q\rangle + \langle q\rangle r_c)(1-q_p+q_p r_c)$ yields the quadratic \\
\begin{equation}
\langle q\rangle q_p\, r_c^2 + 2\langle q\rangle(1-q_p)\, r_c - \left[\langle q\rangle(2-q_p) - 1\right] = 0. \label{eq:quad0}
\end{equation} \\
The constant term is negative exactly when $\langle q\rangle(2-q_p) > 1$, and in that case the quadratic has a single positive root. The interior maximum is located at \\
\begin{equation}
r_c^{*} =
\begin{cases}
0, & \text{if } \langle q\rangle \le \dfrac{1}{2-q_p},\\[2ex]
\dfrac{\sqrt{\langle q\rangle\left(\langle q\rangle - q_p\right)} - \langle q\rangle(1-q_p)}{\langle q\rangle\, q_p}, & \text{if } \langle q\rangle > \dfrac{1}{2-q_p}.
\end{cases} \label{eq:argmax0}
\end{equation} \\
The interior branch requires $q_p < 2 - 1/\langle q\rangle \le \langle q\rangle$, so an interior peak exists only for activities that require the capability less intensely than the average activity. As $q_p \to 0$ the interior branch converges to $1 - 1/(2\langle q\rangle)$, the value in equation~\eqref{argmax_cold_start}. As $q_p$ rises toward $2 - 1/\langle q\rangle$ the peak moves continuously to $r_c = 0$. Figure~\ref{fig:A1} shows the growth rate for $\langle q\rangle = 0.75$ and Figure~\ref{fig:A2} shows the location of its maximum as a function of $q_p$. \\

The cold-start growth rate thus reverses the prediction of equation~\eqref{argmax_cold_start} across activities. Capability accumulation is conditional for every activity once $\langle q\rangle > 1/2$, yet the growth rate of output in demanding activities is fastest in the least endowed economies. The factor $1/(1-q_p+q_p r_c)$ in equation~\eqref{eq:lnYdot0} is the elasticity of output with respect to the capability, and it is largest for demanding activities in economies with small stocks. \\

For aggregate output, equation~\eqref{eq:lnYcdot} with $\alpha = 0$ gives \\
\begin{equation}
\frac{d\ln Y_c}{dt} = \frac{\langle q\rangle}{1-\langle q\rangle(1-r_c)}\frac{dr_c}{dt} = \gamma \langle q\rangle (1-r_c), \label{eq:lnYc0}
\end{equation} \\
which declines monotonically in $r_c$ for every $\langle q\rangle$. Aggregate growth converges unconditionally in the cold-start case, in agreement with the discussion at the end of Section~\ref{sec:discussion}. For general $\beta$ the same expression holds with $(1-r_c)^{\beta}$. \\

\begin{figure}[h]
\centering
\begin{tikzpicture}
  \begin{axis}[
    axis lines=left,
    xlabel={$r_c$},
    ylabel={$\dfrac{d\ln Y_{cp}/dt}{\max_{r_c}\, d\ln Y_{cp}/dt}$},
    xmin=0, xmax=1, ymin=0, ymax=1.05,
    xtick={0,0.2,0.4,0.6,0.8,1},
    ytick={0,0.2,0.4,0.6,0.8,1},
    width=0.55\textwidth,
    height=0.45\textwidth,
    tick label style={/pgf/number format/fixed},
    legend style={draw=none, fill=none, font=\footnotesize, at={(0.98,0.98)}, anchor=north east},
    legend cell align=left,
    declare function={
      % growth rate of activity output, equation (eq:lnYdot0), gamma = 0.1, <q> = 0.75
      gY(\r,\qp)=0.1*\qp*(1-\r)*(1-0.75+0.75*\r)/(1-\qp+\qp*\r);
    },
  ]
    % normalised by the value at the maximum, r_c^* from equation (eq:argmax0)
    \addplot[thick, blue,               domain=0:1, samples=200] {gY(x,0.1)/gY(0.3095,0.1)};
    \addlegendentry{$q_p=0.1$}
    \addplot[thick, red,    dashed,     domain=0:1, samples=200] {gY(x,0.3)/gY(0.2487,0.3)};
    \addlegendentry{$q_p=0.3$}
    \addplot[thick, orange, dashdotted, domain=0:1, samples=200] {gY(x,0.5)/gY(0.1547,0.5)};
    \addlegendentry{$q_p=0.5$}
    \addplot[thick, violet, densely dotted, domain=0:1, samples=200] {gY(x,0.7)/gY(0,0.7)};
    \addlegendentry{$q_p=0.7$}
    \addplot[thick, teal!80!black, loosely dashed, domain=0:1, samples=200] {gY(x,0.9)/gY(0,0.9)};
    \addlegendentry{$q_p=0.9$}
  \end{axis}
\end{tikzpicture}
\caption{Growth rate of activity output, $d\ln Y_{cp}/dt$, relative to its maximum, as a function of $r_c$ for the cold-start case ($\alpha = 0$, $\beta = 1$) with $\gamma = 0.1$ and $\langle q\rangle = 0.75$. Activities with $q_p < 2 - 1/\langle q\rangle = 2/3$ display an interior maximum.}
\label{fig:A1}
\end{figure}

\begin{figure}[h]
\centering
\begin{tikzpicture}
  \begin{axis}[
    axis lines=left,
    xlabel={$q_p$},
    ylabel={$\operatorname*{arg\,max}_{r_c}(d\ln Y_{cp}/dt)$},
    xmin=0, xmax=1, ymin=-0.02, ymax=0.6,
    xtick={0,0.2,0.4,0.6,0.8,1},
    ytick={0,0.1,0.2,0.3,0.4,0.5,0.6},
    width=0.55\textwidth,
    height=0.45\textwidth,
    tick label style={/pgf/number format/fixed},
    legend style={draw=none, fill=none, font=\footnotesize, at={(0.98,0.98)}, anchor=north east},
    legend cell align=left,
    declare function={
      % interior branch of equation (eq:argmax0)
      peak(\qp,\Q)=(sqrt(\Q*(\Q-\qp))-\Q*(1-\qp))/(\Q*\qp);
    },
  ]
    % maximum of capability accumulation, 1 - 1/(2<q>), equation (argmax_cold_start)
    \addplot[gray, thin, domain=0:1, samples=2, forget plot] {1-1/(2*0.6)};
    \addplot[gray, thin, domain=0:1, samples=2, forget plot] {1-1/(2*0.75)};
    \addplot[gray, thin, domain=0:1, samples=2, forget plot] {1-1/(2*0.9)};

    % <q> = 0.6: interior peak for q_p < 2 - 1/<q> = 1/3
    \addplot[thick, blue, domain=0.001:0.3333, samples=150] {peak(x,0.6)};
    \addlegendentry{$\langle q\rangle=0.6$}
    % <q> = 0.75: interior peak for q_p < 2/3
    \addplot[thick, red, dashed, domain=0.001:0.6667, samples=150] {peak(x,0.75)};
    \addlegendentry{$\langle q\rangle=0.75$}
    % <q> = 0.9: interior peak for q_p < 8/9
    \addplot[thick, orange, dashdotted, domain=0.001:0.8889, samples=150] {peak(x,0.9)};
    \addlegendentry{$\langle q\rangle=0.9$}

    % peak at r_c = 0 beyond each threshold
    \addplot[thick, blue, domain=0.3333:1, samples=2, forget plot] {0};
    \addplot[thick, red, dashed, domain=0.6667:1, samples=2, forget plot] {0};
    \addplot[thick, orange, dashdotted, domain=0.8889:1, samples=2, forget plot] {0};
  \end{axis}
\end{tikzpicture}
\caption{Location of the maximum of the growth rate of activity output, $\arg\max_{r_c}(d\ln Y_{cp}/dt)$, for the cold-start case as a function of the activity's requirement $q_p$, given by \eqref{eq:argmax0}. Grey horizontal lines mark the location of the maximum of capability accumulation, $1 - 1/(2\langle q\rangle)$ from \eqref{M-argmax_cold_start}, which is common to all activities.}
\label{fig:A2}
\end{figure}

\subsubsection*{The general case ($\alpha > 0$)}

For $\alpha > 0$ the factor $r_c^{\alpha q_p}$ varies across activities and remains inside the sums, so the capability stock of the fastest growers lacks a closed form for heterogeneous requirements. As in equation~\eqref{foc_homogeneous}, we assume $q_p = q$ for all $p$ and write $S \equiv 1-q(1-r_c)$, so that output in each activity is $Y_{cp} = r_c^{\alpha q} S$.\\

We begin with the change in activity output. Substituting the common requirement into equation~\eqref{eq:Ydot} gives\\
\begin{equation}
\frac{dY_{cp}}{dt} = \gamma\, q\,(1-r_c)^{\beta}\, r_c^{2\alpha q - 1}\, S\left(\alpha S + r_c\right). \label{eq:Ydotq}
\end{equation}\\
The change in output is the product of the marginal product of the capability, $\partial Y_{cp}/\partial r_c = q\, r_c^{\alpha q - 1}(\alpha S + r_c)$, which measures the increase in output from a small increase in the capability stock $r_c$, and the accumulation rate, $dr_c/dt = \gamma(1-r_c)^{\beta} r_c^{\alpha q} S$. Near the origin, the marginal product scales as $r_c^{\alpha q - 1}$ and the accumulation rate scales as $r_c^{\alpha q}$, so the change in output scales as\\
\begin{equation}
\frac{dY_{cp}}{dt} \simeq \gamma\,\alpha\, q\,(1-q)^2\, r_c^{2\alpha q - 1} \qquad \text{as } r_c \to 0. \label{eq:Ydot_origin}
\end{equation}\\
The exponent $2\alpha q - 1$ determines which economies grow fastest. In activities with $q < 1/(2\alpha)$ the exponent is negative. The increase in the marginal product as $r_c$ decreases outweighs the decrease in the accumulation rate, so the change in output decreases with $r_c$ near the origin and is largest in the least endowed economies. In activities with $q > 1/(2\alpha)$ the exponent is positive. The change in output converges to zero in the least endowed economies, as it does at the frontier ($r_c = 1$), where the capability stops accumulating, so the fastest growers hold a positive capability stock $r_c$. \\

To identify the capability stock of the fastest growers, we characterize the critical points of the change in output on $(0,1)$. We set the derivative of the logarithm of equation~\eqref{eq:Ydotq} to zero. The change in output is positive on $(0,1)$, so its logarithm has the same critical points, and the logarithm turns the product into a sum of terms that can be differentiated one at a time\\
\begin{equation}
\frac{2\alpha q - 1}{r_c} - \frac{\beta}{1-r_c} + \frac{q}{S} + \frac{\alpha q + 1}{\alpha S + r_c} = 0. \label{eq:focY}
\end{equation}\\
Multiplying by the positive quantity $r_c(1-r_c)\,S\,(\alpha S + r_c)$ yields the cubic $P(r_c) = C_3 r_c^3 + C_2 r_c^2 + C_1 r_c + C_0 = 0$ with\\
\begin{align}
C_3 &= -q(\alpha q + 1)(2\alpha q + \beta + 1), \nonumber\\
C_2 &= (2\alpha q + 1)\left(3\alpha q^2 - 2\alpha q + \beta q - \beta + q\right), \nonumber\\
C_1 &= \alpha(1-q)\left(6\alpha q^2 - 2\alpha q + \beta q - \beta + q + 1\right), \label{eq:cubic}\\
C_0 &= \alpha(1-q)^2(2\alpha q - 1). \nonumber
\end{align}\\
Since the multiplying quantity is positive, $P(r_c)$ has the sign of the derivative of the logarithm of the change in output, so the change in output increases where $P>0$ and decreases where $P<0$. At the origin $P(0)=C_0$, and at the frontier $P(1)=C_3+C_2+C_1+C_0=-\beta(1+\alpha)<0$. The sign of $C_0$ switches at $q=1/(2\alpha)$, and this switch separates the two cases.\\

In activities with $q > 1/(2\alpha)$, $P(0)>0>P(1)$. The change in output increases near the origin and decreases near the frontier, so it attains an interior maximum, which identifies the fastest growers overall.\\

In activities with $q < 1/(2\alpha)$, $P(0)<0$ and $P(1)<0$. The change in output decreases near both ends of the interval, so $P$ has an even number of roots in $(0,1)$, either zero or two. With zero roots, the change in output decreases monotonically and the least endowed economies are the fastest growers. With two roots, the smaller is a local minimum and the larger is an interior local maximum, which identifies the fastest growers among economies with a positive capability stock. The two roots emerge together as a double root of $P$, which occurs when the discriminant of the cubic vanishes. Writing\\
\begin{equation}
u = \frac{3C_3C_1 - C_2^2}{3C_3^2}, \qquad v = \frac{2C_2^3 - 9C_3C_2C_1 + 27C_3^2C_0}{27C_3^3},
\end{equation}\\
this condition is $4u^3 + 27v^2 = 0$, and its solution in $q$ defines the requirement $\underline{q}(\alpha,\beta)$ at which the interior local maximum first appears. The change in output thus displays an interior local maximum for $\underline{q}(\alpha,\beta) < q < 1/(2\alpha)$. In contrast with $1/(2\alpha)$, the onset $\underline{q}$ depends on $\beta$ through all four coefficients.\\

In both cases, whenever the change in output displays an interior maximum, the cubic has three real roots ($4u^3 + 27v^2 < 0$), and the maximum is given by the trigonometric solution of the cubic\\
\begin{equation}
r_c^{*} = -\frac{C_2}{3C_3} + 2\sqrt{-\frac{u}{3}}\,\cos\!\left[\frac{1}{3}\arccos\!\left(\frac{3v}{2u}\sqrt{-\frac{3}{u}}\right)\right]. \label{eq:argmaxY}
\end{equation}\\
For $\alpha = 0$ the cubic factors as $r_c^2\left[(1+\beta)q(1 - r_c) - \beta\right]$, whose root $r_c^{*} = 1 - \beta/\left((1+\beta)q\right)$ equals equation~\eqref{argmax_cold_start} for $\beta = 1$.\\

The change in output thus converges unconditionally in activities with a low requirement and conditionally in activities with a high requirement, in line with capability accumulation. The switch between the two at $q = 1/(2\alpha)$ is discontinuous, since the interior maximum already holds a strictly positive stock when it becomes global. \\

We now turn to the growth rate of activity output. Dividing equation~\eqref{eq:Ydotq} by $Y_{cp}$ gives\\
\begin{equation}
\frac{d\ln Y_{cp}}{dt} = \gamma\, q\,(1-r_c)^{\beta}\, r_c^{\alpha q - 1}\left(\alpha S + r_c\right). \label{eq:lnYdotq}
\end{equation}\\
Under a common requirement $Y_c = N(p)\,Y_{cp}$, so this is also the growth rate of aggregate output. Near the origin, output scales as $r_c^{\alpha q}$, so dividing equation~\eqref{eq:Ydot_origin} by output gives\\
\begin{equation}
\frac{d\ln Y_{cp}}{dt} \simeq \gamma\,\alpha\, q\,(1-q)\, r_c^{\alpha q - 1} \qquad \text{as } r_c \to 0. \label{eq:lnYdot_origin}
\end{equation}\\
The exponent $\alpha q - 1$ now determines which economies grow fastest, and the threshold is $q = 1/\alpha$, again independent of $\beta$. In activities with $q < 1/\alpha$ the exponent is negative, so the growth rate decreases with $r_c$ near the origin and is largest in the least endowed economies. In these economies a small accumulation of the capability translates into a large proportional change on a small output base. For $\alpha \le 1$ the threshold lies above one, so the least endowed economies have the fastest growth rate in every activity. In activities with $q > 1/\alpha$, a range that exists only for $\alpha > 1$, the exponent is positive. The growth rate converges to zero in the least endowed economies, as it does at the frontier, so the fastest growers hold a positive capability stock.\\

To identify the capability stock of the fastest growers, we proceed in the same way and set the derivative of the logarithm of equation~\eqref{eq:lnYdotq} to zero\\
\begin{equation}
\frac{\alpha q - 1}{r_c} - \frac{\beta}{1-r_c} + \frac{\alpha q + 1}{\alpha(1-q) + (\alpha q+1) r_c} = 0. \label{eq:focL}
\end{equation}\\
Compared with equation~\eqref{eq:focY}, the first term is smaller by $\alpha q$ and the term in $q/S$ is absent, since dividing by output removes the response of output in the activity. Multiplying by $r_c(1-r_c)\left[\alpha(1-q) + (\alpha q+1) r_c\right]$ yields the quadratic $Q(r_c) = A r_c^2 + B r_c + C = 0$ with\\
\begin{align}
A &= -(\alpha q + 1)(\alpha q + \beta), \nonumber\\
B &= (\alpha q - 1)(2\alpha q + 1 - \alpha) + \alpha q + 1 - \alpha\beta(1-q), \label{eq:quadL}\\
C &= \alpha(1-q)(\alpha q - 1). \nonumber
\end{align}\\
As for the cubic, $Q$ has the sign of the derivative of the logarithm of the growth rate, with $Q(0)=C$ and $Q(1)=A+B+C=-\beta(1+\alpha)<0$. The sign of $C$ switches at $q=1/\alpha$. In activities with $q>1/\alpha$, $Q(0)>0>Q(1)$, so the quadratic has a single root in $(0,1)$, the growth rate attains an interior maximum there, and this maximum identifies the fastest growers overall. In activities with $q<1/\alpha$, $Q(0)<0$ and $Q(1)<0$, so the quadratic has either zero or two roots in $(0,1)$. With two roots, the smaller is a local minimum and the larger is an interior local maximum, which identifies the fastest growers among economies with a positive capability stock. The two roots emerge together when $B^2 = 4AC$. In both cases the maximum is the larger root\\
\begin{equation}
r_c^{*} = \frac{B + \sqrt{B^2 - 4AC}}{2(\alpha q + 1)(\alpha q + \beta)}. \label{eq:argmaxL}
\end{equation}\\
For $\alpha = q = 1$ it simplifies to $r_c^{*} = 1/(1+\beta)$.\\

\begin{figure}[htbp]
\centering
\begin{minipage}{0.49\linewidth}
\centering
\begin{tikzpicture}
  \begin{axis}[
    axis lines=left,
    xlabel={$q$},
    ylabel={$\operatorname*{arg\,max}_{r_c}(dY_{cp}/dt)$},
    xmin=0,
    xmax=1,
    ymin=-0.04,
    ymax=1,
    xtick={0,0.2,0.4,0.6,0.8,1},
    ytick={0,0.2,0.4,0.6,0.8,1},
    width=\linewidth,
    height=0.85\linewidth,
    tick label style={/pgf/number format/fixed},
    declare function={
      % coefficients of the cubic, equation (eq:cubic)
      cthree(\q,\a,\b)=-\q*(\a*\q+1)*(2*\a*\q+\b+1);
      ctwo(\q,\a,\b)=(2*\a*\q+1)*(3*\a*\q^2-2*\a*\q+\b*\q-\b+\q);
      cone(\q,\a,\b)=\a*(1-\q)*(6*\a*\q^2-2*\a*\q+\b*\q-\b+\q+1);
      czero(\q,\a,\b)=\a*(1-\q)^2*(2*\a*\q-1);
      % trigonometric solution, equation (eq:argmaxY)
      peakYc(\A,\B,\C,\D)=
        -\B/(3*\A)
        +2*sqrt(-((3*\A*\C-\B^2)/(3*\A^2))/3)
        *cos(acos(max(-1,min(1,
          (3*((2*\B^3-9*\A*\B*\C+27*\A^2*\D)/(27*\A^3)))
          /(2*((3*\A*\C-\B^2)/(3*\A^2)))
          *sqrt(-3/((3*\A*\C-\B^2)/(3*\A^2))))))/3);
      peakY(\q,\a,\b)=peakYc(cthree(\q,\a,\b),ctwo(\q,\a,\b),
        cone(\q,\a,\b),czero(\q,\a,\b));
      % larger root of the quadratic, equation (eq:argmaxL)
      peakL(\q,\a,\b)=
        (
          (\a*\q-1)*(2*\a*\q+1-\a)+\a*\q+1-\a*\b*(1-\q)
          +sqrt(max(0,
            ((\a*\q-1)*(2*\a*\q+1-\a)+\a*\q+1-\a*\b*(1-\q))^2
            +4*(\a*\q+1)*(\a*\q+\b)*\a*(1-\q)*(\a*\q-1)))
        )
        /(2*(\a*\q+1)*(\a*\q+\b));
    },
  ]

    % least endowed economies grow fastest overall
    \addplot[
      very thick,
      black,
      domain=0:1,
      samples=2,
      forget plot,
    ] {0};

    % alpha = 0
    \addplot[
      thick,
      blue,
      domain=0.5:1,
      samples=200,
      forget plot,
    ] {1-1/((1+1)*x)};

    % alpha = 0.1
    \addplot[
      thick,
      red,
      dashed,
      domain=0.6432:0.9999,
      samples=150,
      forget plot,
    ] {peakY(x,0.1,1)};

    % alpha = 0.5
    \addplot[
      thick,
      orange,
      dashed,
      domain=0.5340:0.9999,
      samples=150,
      forget plot,
    ] {peakY(x,0.5,1)};

    % alpha = 1
    \addplot[
      thick,
      violet,
      dashed,
      domain=0.4048:0.4999,
      samples=150,
      forget plot,
    ] {peakY(x,1,1)};
    \addplot[
      thick,
      violet,
      domain=0.5001:0.999,
      samples=150,
      forget plot,
    ] {peakY(x,1,1)};
    % jump of the fastest growers overall at q = 0.5000
    \addplot[violet, only marks, mark=*, mark size=1.6pt, forget plot]
      coordinates {(0.5000,0.4768)};
    \addplot[violet, only marks, mark=*, mark options={fill=white}, mark size=1.6pt, forget plot]
      coordinates {(0.5000,0)};

    % alpha = 2
    \addplot[
      thick,
      teal!80!black,
      dashed,
      domain=0.2470:0.2499,
      samples=150,
      forget plot,
    ] {peakY(x,2,1)};
    \addplot[
      thick,
      teal!80!black,
      domain=0.2501:0.999,
      samples=150,
      forget plot,
    ] {peakY(x,2,1)};
    % jump of the fastest growers overall at q = 0.2500
    \addplot[teal!80!black, only marks, mark=*, mark size=1.6pt, forget plot]
      coordinates {(0.2500,0.1547)};
    \addplot[teal!80!black, only marks, mark=*, mark options={fill=white}, mark size=1.6pt, forget plot]
      coordinates {(0.2500,0)};

    % Cold-start transition
    \addplot[
      dotted,
      gray,
      forget plot,
    ] coordinates {(0.5,0) (0.5,1)};
  \end{axis}
\end{tikzpicture}
\end{minipage}\hfill
\begin{minipage}{0.49\linewidth}
\centering
\begin{tikzpicture}
  \begin{axis}[
    axis lines=left,
    xlabel={$q$},
    ylabel={$\operatorname*{arg\,max}_{r_c}(d\ln Y_{cp}/dt)$},
    xmin=0,
    xmax=1,
    ymin=-0.04,
    ymax=1,
    xtick={0,0.2,0.4,0.6,0.8,1},
    ytick={0,0.2,0.4,0.6,0.8,1},
    width=\linewidth,
    height=0.85\linewidth,
    tick label style={/pgf/number format/fixed},
    legend style={
      at={(0.03,0.97)},
      anchor=north west,
      draw=none,
      fill=white,
      font=\small
    },
    declare function={
      % coefficients of the cubic, equation (eq:cubic)
      cthree(\q,\a,\b)=-\q*(\a*\q+1)*(2*\a*\q+\b+1);
      ctwo(\q,\a,\b)=(2*\a*\q+1)*(3*\a*\q^2-2*\a*\q+\b*\q-\b+\q);
      cone(\q,\a,\b)=\a*(1-\q)*(6*\a*\q^2-2*\a*\q+\b*\q-\b+\q+1);
      czero(\q,\a,\b)=\a*(1-\q)^2*(2*\a*\q-1);
      % trigonometric solution, equation (eq:argmaxY)
      peakYc(\A,\B,\C,\D)=
        -\B/(3*\A)
        +2*sqrt(-((3*\A*\C-\B^2)/(3*\A^2))/3)
        *cos(acos(max(-1,min(1,
          (3*((2*\B^3-9*\A*\B*\C+27*\A^2*\D)/(27*\A^3)))
          /(2*((3*\A*\C-\B^2)/(3*\A^2)))
          *sqrt(-3/((3*\A*\C-\B^2)/(3*\A^2))))))/3);
      peakY(\q,\a,\b)=peakYc(cthree(\q,\a,\b),ctwo(\q,\a,\b),
        cone(\q,\a,\b),czero(\q,\a,\b));
      % larger root of the quadratic, equation (eq:argmaxL)
      peakL(\q,\a,\b)=
        (
          (\a*\q-1)*(2*\a*\q+1-\a)+\a*\q+1-\a*\b*(1-\q)
          +sqrt(max(0,
            ((\a*\q-1)*(2*\a*\q+1-\a)+\a*\q+1-\a*\b*(1-\q))^2
            +4*(\a*\q+1)*(\a*\q+\b)*\a*(1-\q)*(\a*\q-1)))
        )
        /(2*(\a*\q+1)*(\a*\q+\b));
    },
  ]

    \addlegendimage{thick,blue}
    \addlegendentry{$\alpha=0$}
    \addlegendimage{thick,red}
    \addlegendentry{$\alpha=0.1$}
    \addlegendimage{thick,orange}
    \addlegendentry{$\alpha=0.5$}
    \addlegendimage{thick,violet}
    \addlegendentry{$\alpha=1$}
    \addlegendimage{thick,teal!80!black}
    \addlegendentry{$\alpha=2$}
    \addlegendimage{very thick,black}
    \addlegendentry{$r_c^{*}=0$}

    % least endowed economies grow fastest overall
    \addplot[
      very thick,
      black,
      domain=0:1,
      samples=2,
      forget plot,
    ] {0};

    % alpha = 0

    % alpha = 0.1
    \addplot[
      thick,
      red,
      dashed,
      domain=0.9739:0.9999,
      samples=150,
      forget plot,
    ] {peakL(x,0.1,1)};

    % alpha = 0.5
    \addplot[
      thick,
      orange,
      dashed,
      domain=0.8558:0.9999,
      samples=150,
      forget plot,
    ] {peakL(x,0.5,1)};

    % alpha = 1
    \addplot[
      thick,
      violet,
      dashed,
      domain=0.7072:0.9999,
      samples=150,
      forget plot,
    ] {peakL(x,1,1)};

    % alpha = 2
    \addplot[
      thick,
      teal!80!black,
      dashed,
      domain=0.4767:0.4999,
      samples=150,
      forget plot,
    ] {peakL(x,2,1)};
    \addplot[
      thick,
      teal!80!black,
      domain=0.5001:0.999,
      samples=150,
      forget plot,
    ] {peakL(x,2,1)};
    % jump of the fastest growers overall at q = 0.5000
    \addplot[teal!80!black, only marks, mark=*, mark size=1.6pt, forget plot]
      coordinates {(0.5000,0.2500)};
    \addplot[teal!80!black, only marks, mark=*, mark options={fill=white}, mark size=1.6pt, forget plot]
      coordinates {(0.5000,0)};

    % Cold-start transition
    \addplot[
      dotted,
      gray,
      forget plot,
    ] coordinates {(0.5,0) (0.5,1)};
  \end{axis}
\end{tikzpicture}
\end{minipage}
\caption{Capability stock of the fastest growers in the change in activity output (left) and in the growth rate of activity output (right) as a function of the common requirement $q_p = q$, for $\beta = 1$ and $\alpha = 0, 0.1, 0.5, 1, 2$. The dotted vertical line marks $q = 1/2$. Solid lines show the fastest growers overall, and the thick black line along $r_c = 0$ marks activities in which the least endowed economies grow fastest overall. Open and filled circles mark the requirement at which the fastest growers overall jump from $r_c = 0$ to an interior capability stock. Dashed lines show the fastest growers among economies with a positive capability stock, in activities where the least endowed economies grow fastest overall. The left panel follows from equation~\eqref{eq:argmaxY} and the right panel from equation~\eqref{eq:argmaxL}. For $\alpha = 0$ the left curve coincides with the cold-start curve in Figure~\ref{bifurcation_alpha}, and the least endowed economies have the fastest growth rate for all $q$.}
\label{fig:A3}
\end{figure}

\begin{figure}[htbp]
\centering
\begin{minipage}{0.49\linewidth}
\centering
\begin{tikzpicture}
  \begin{axis}[
    axis lines=left,
    xlabel={$q$},
    ylabel={$\operatorname*{arg\,max}_{r_c}(dY_{cp}/dt)$},
    xmin=0,
    xmax=1,
    ymin=-0.04,
    ymax=1,
    xtick={0,0.2,0.4,0.6,0.8,1},
    ytick={0,0.2,0.4,0.6,0.8,1},
    width=\linewidth,
    height=0.85\linewidth,
    tick label style={/pgf/number format/fixed},
    declare function={
      % coefficients of the cubic, equation (eq:cubic)
      cthree(\q,\a,\b)=-\q*(\a*\q+1)*(2*\a*\q+\b+1);
      ctwo(\q,\a,\b)=(2*\a*\q+1)*(3*\a*\q^2-2*\a*\q+\b*\q-\b+\q);
      cone(\q,\a,\b)=\a*(1-\q)*(6*\a*\q^2-2*\a*\q+\b*\q-\b+\q+1);
      czero(\q,\a,\b)=\a*(1-\q)^2*(2*\a*\q-1);
      % trigonometric solution, equation (eq:argmaxY)
      peakYc(\A,\B,\C,\D)=
        -\B/(3*\A)
        +2*sqrt(-((3*\A*\C-\B^2)/(3*\A^2))/3)
        *cos(acos(max(-1,min(1,
          (3*((2*\B^3-9*\A*\B*\C+27*\A^2*\D)/(27*\A^3)))
          /(2*((3*\A*\C-\B^2)/(3*\A^2)))
          *sqrt(-3/((3*\A*\C-\B^2)/(3*\A^2))))))/3);
      peakY(\q,\a,\b)=peakYc(cthree(\q,\a,\b),ctwo(\q,\a,\b),
        cone(\q,\a,\b),czero(\q,\a,\b));
      % larger root of the quadratic, equation (eq:argmaxL)
      peakL(\q,\a,\b)=
        (
          (\a*\q-1)*(2*\a*\q+1-\a)+\a*\q+1-\a*\b*(1-\q)
          +sqrt(max(0,
            ((\a*\q-1)*(2*\a*\q+1-\a)+\a*\q+1-\a*\b*(1-\q))^2
            +4*(\a*\q+1)*(\a*\q+\b)*\a*(1-\q)*(\a*\q-1)))
        )
        /(2*(\a*\q+1)*(\a*\q+\b));
    },
  ]

    % least endowed economies grow fastest overall
    \addplot[
      very thick,
      black,
      domain=0:1,
      samples=2,
      forget plot,
    ] {0};

    % beta = 3
    \addplot[
      thick,
      blue!10!red,
      dashed,
      domain=0.8191:0.9999,
      samples=150,
      forget plot,
    ] {peakY(x,0.1,3)};

    % beta = 2
    \addplot[
      thick,
      blue!30!red,
      dashed,
      domain=0.7609:0.9999,
      samples=150,
      forget plot,
    ] {peakY(x,0.1,2)};

    % beta = 1
    \addplot[
      thick,
      blue!50!red,
      dashed,
      domain=0.6432:0.9999,
      samples=150,
      forget plot,
    ] {peakY(x,0.1,1)};

    % beta = 0.5
    \addplot[
      thick,
      blue!70!red,
      dashed,
      domain=0.5175:0.9999,
      samples=150,
      forget plot,
    ] {peakY(x,0.1,0.5)};

    % beta = 0.2
    \addplot[
      thick,
      blue!90!red,
      dashed,
      domain=0.3691:0.9999,
      samples=150,
      forget plot,
    ] {peakY(x,0.1,0.2)};

    % beta = 0.1
    \addplot[
      thick,
      blue,
      dashed,
      domain=0.2825:0.9999,
      samples=150,
      forget plot,
    ] {peakY(x,0.1,0.1)};

  \end{axis}
\end{tikzpicture}
\end{minipage}\hfill
\begin{minipage}{0.49\linewidth}
\centering
\begin{tikzpicture}
  \begin{axis}[
    axis lines=left,
    xlabel={$q$},
    ylabel={$\operatorname*{arg\,max}_{r_c}(d\ln Y_{cp}/dt)$},
    xmin=0,
    xmax=1,
    ymin=-0.04,
    ymax=1,
    xtick={0,0.2,0.4,0.6,0.8,1},
    ytick={0,0.2,0.4,0.6,0.8,1},
    width=\linewidth,
    height=0.85\linewidth,
    tick label style={/pgf/number format/fixed},
    legend style={
      at={(0.03,0.97)},
      anchor=north west,
      draw=none,
      fill=white,
      font=\small
    },
    declare function={
      % coefficients of the cubic, equation (eq:cubic)
      cthree(\q,\a,\b)=-\q*(\a*\q+1)*(2*\a*\q+\b+1);
      ctwo(\q,\a,\b)=(2*\a*\q+1)*(3*\a*\q^2-2*\a*\q+\b*\q-\b+\q);
      cone(\q,\a,\b)=\a*(1-\q)*(6*\a*\q^2-2*\a*\q+\b*\q-\b+\q+1);
      czero(\q,\a,\b)=\a*(1-\q)^2*(2*\a*\q-1);
      % trigonometric solution, equation (eq:argmaxY)
      peakYc(\A,\B,\C,\D)=
        -\B/(3*\A)
        +2*sqrt(-((3*\A*\C-\B^2)/(3*\A^2))/3)
        *cos(acos(max(-1,min(1,
          (3*((2*\B^3-9*\A*\B*\C+27*\A^2*\D)/(27*\A^3)))
          /(2*((3*\A*\C-\B^2)/(3*\A^2)))
          *sqrt(-3/((3*\A*\C-\B^2)/(3*\A^2))))))/3);
      peakY(\q,\a,\b)=peakYc(cthree(\q,\a,\b),ctwo(\q,\a,\b),
        cone(\q,\a,\b),czero(\q,\a,\b));
      % larger root of the quadratic, equation (eq:argmaxL)
      peakL(\q,\a,\b)=
        (
          (\a*\q-1)*(2*\a*\q+1-\a)+\a*\q+1-\a*\b*(1-\q)
          +sqrt(max(0,
            ((\a*\q-1)*(2*\a*\q+1-\a)+\a*\q+1-\a*\b*(1-\q))^2
            +4*(\a*\q+1)*(\a*\q+\b)*\a*(1-\q)*(\a*\q-1)))
        )
        /(2*(\a*\q+1)*(\a*\q+\b));
    },
  ]

    \addlegendimage{thick,blue!10!red}
    \addlegendentry{$\alpha=0.1,\ \beta=3$}
    \addlegendimage{thick,blue!30!red}
    \addlegendentry{$\alpha=0.1,\ \beta=2$}
    \addlegendimage{thick,blue!50!red}
    \addlegendentry{$\alpha=0.1,\ \beta=1$}
    \addlegendimage{thick,blue!70!red}
    \addlegendentry{$\alpha=0.1,\ \beta=0.5$}
    \addlegendimage{thick,blue!90!red}
    \addlegendentry{$\alpha=0.1,\ \beta=0.2$}
    \addlegendimage{thick,blue}
    \addlegendentry{$\alpha=0.1,\ \beta=0.1$}
    \addlegendimage{very thick,black}
    \addlegendentry{$r_c^{*}=0$}

    % least endowed economies grow fastest overall
    \addplot[
      very thick,
      black,
      domain=0:1,
      samples=2,
      forget plot,
    ] {0};

    % beta = 3
    \addplot[
      thick,
      blue!10!red,
      dashed,
      domain=0.9907:0.9999,
      samples=150,
      forget plot,
    ] {peakL(x,0.1,3)};

    % beta = 2
    \addplot[
      thick,
      blue!30!red,
      dashed,
      domain=0.9863:0.9999,
      samples=150,
      forget plot,
    ] {peakL(x,0.1,2)};

    % beta = 1
    \addplot[
      thick,
      blue!50!red,
      dashed,
      domain=0.9739:0.9999,
      samples=150,
      forget plot,
    ] {peakL(x,0.1,1)};

    % beta = 0.5
    \addplot[
      thick,
      blue!70!red,
      dashed,
      domain=0.9522:0.9999,
      samples=150,
      forget plot,
    ] {peakL(x,0.1,0.5)};

    % beta = 0.2
    \addplot[
      thick,
      blue!90!red,
      dashed,
      domain=0.9037:0.9999,
      samples=150,
      forget plot,
    ] {peakL(x,0.1,0.2)};

    % beta = 0.1
    \addplot[
      thick,
      blue,
      dashed,
      domain=0.8515:0.9999,
      samples=150,
      forget plot,
    ] {peakL(x,0.1,0.1)};

  \end{axis}
\end{tikzpicture}
\end{minipage}
\caption{Capability stock of the fastest growers in the change in activity output (left) and in the growth rate of activity output (right) as a function of the common requirement $q_p = q$, for $\alpha = 0.1$ and different saturation exponents $\beta$. Lines and circles as in Figure~\ref{fig:A3}. For $\alpha = 0.1$ the least endowed economies grow fastest overall in every activity and for every $\beta$.}
\label{fig:A4}
\end{figure}

\begin{figure}[htbp]
\centering
\begin{minipage}{0.49\linewidth}
\centering
\begin{tikzpicture}
  \begin{axis}[
    axis lines=left,
    xlabel={$q$},
    ylabel={$\operatorname*{arg\,max}_{r_c}(dY_{cp}/dt)$},
    xmin=0,
    xmax=1,
    ymin=-0.04,
    ymax=1,
    xtick={0,0.2,0.4,0.6,0.8,1},
    ytick={0,0.2,0.4,0.6,0.8,1},
    width=\linewidth,
    height=0.85\linewidth,
    tick label style={/pgf/number format/fixed},
    declare function={
      % coefficients of the cubic, equation (eq:cubic)
      cthree(\q,\a,\b)=-\q*(\a*\q+1)*(2*\a*\q+\b+1);
      ctwo(\q,\a,\b)=(2*\a*\q+1)*(3*\a*\q^2-2*\a*\q+\b*\q-\b+\q);
      cone(\q,\a,\b)=\a*(1-\q)*(6*\a*\q^2-2*\a*\q+\b*\q-\b+\q+1);
      czero(\q,\a,\b)=\a*(1-\q)^2*(2*\a*\q-1);
      % trigonometric solution, equation (eq:argmaxY)
      peakYc(\A,\B,\C,\D)=
        -\B/(3*\A)
        +2*sqrt(-((3*\A*\C-\B^2)/(3*\A^2))/3)
        *cos(acos(max(-1,min(1,
          (3*((2*\B^3-9*\A*\B*\C+27*\A^2*\D)/(27*\A^3)))
          /(2*((3*\A*\C-\B^2)/(3*\A^2)))
          *sqrt(-3/((3*\A*\C-\B^2)/(3*\A^2))))))/3);
      peakY(\q,\a,\b)=peakYc(cthree(\q,\a,\b),ctwo(\q,\a,\b),
        cone(\q,\a,\b),czero(\q,\a,\b));
      % larger root of the quadratic, equation (eq:argmaxL)
      peakL(\q,\a,\b)=
        (
          (\a*\q-1)*(2*\a*\q+1-\a)+\a*\q+1-\a*\b*(1-\q)
          +sqrt(max(0,
            ((\a*\q-1)*(2*\a*\q+1-\a)+\a*\q+1-\a*\b*(1-\q))^2
            +4*(\a*\q+1)*(\a*\q+\b)*\a*(1-\q)*(\a*\q-1)))
        )
        /(2*(\a*\q+1)*(\a*\q+\b));
    },
  ]

    % least endowed economies grow fastest overall
    \addplot[
      very thick,
      black,
      domain=0:0.5,
      samples=2,
      forget plot,
    ] {0};

    % beta = 3
    \addplot[
      thick,
      blue!10!red,
      dashed,
      domain=0.4915:0.4999,
      samples=150,
      forget plot,
    ] {peakY(x,1,3)};
    \addplot[
      thick,
      blue!10!red,
      domain=0.5001:0.999,
      samples=150,
      forget plot,
    ] {peakY(x,1,3)};
    % jump of the fastest growers overall at q = 0.5000
    \addplot[blue!10!red, only marks, mark=*, mark size=1.6pt, forget plot]
      coordinates {(0.5000,0.0883)};
    \addplot[blue!10!red, only marks, mark=*, mark options={fill=white}, mark size=1.6pt, forget plot]
      coordinates {(0.5000,0)};

    % beta = 2
    \addplot[
      thick,
      blue!30!red,
      dashed,
      domain=0.4640:0.4999,
      samples=150,
      forget plot,
    ] {peakY(x,1,2)};
    \addplot[
      thick,
      blue!30!red,
      domain=0.5001:0.999,
      samples=150,
      forget plot,
    ] {peakY(x,1,2)};
    % jump of the fastest growers overall at q = 0.5000
    \addplot[blue!30!red, only marks, mark=*, mark size=1.6pt, forget plot]
      coordinates {(0.5000,0.2287)};
    \addplot[blue!30!red, only marks, mark=*, mark options={fill=white}, mark size=1.6pt, forget plot]
      coordinates {(0.5000,0)};

    % beta = 1
    \addplot[
      thick,
      blue!50!red,
      dashed,
      domain=0.4048:0.4999,
      samples=150,
      forget plot,
    ] {peakY(x,1,1)};
    \addplot[
      thick,
      blue!50!red,
      domain=0.5001:0.999,
      samples=150,
      forget plot,
    ] {peakY(x,1,1)};
    % jump of the fastest growers overall at q = 0.5000
    \addplot[blue!50!red, only marks, mark=*, mark size=1.6pt, forget plot]
      coordinates {(0.5000,0.4768)};
    \addplot[blue!50!red, only marks, mark=*, mark options={fill=white}, mark size=1.6pt, forget plot]
      coordinates {(0.5000,0)};

    % beta = 0.5
    \addplot[
      thick,
      blue!70!red,
      dashed,
      domain=0.3464:0.4999,
      samples=150,
      forget plot,
    ] {peakY(x,1,0.5)};
    \addplot[
      thick,
      blue!70!red,
      domain=0.5001:0.999,
      samples=150,
      forget plot,
    ] {peakY(x,1,0.5)};
    % jump of the fastest growers overall at q = 0.5000
    \addplot[blue!70!red, only marks, mark=*, mark size=1.6pt, forget plot]
      coordinates {(0.5000,0.6831)};
    \addplot[blue!70!red, only marks, mark=*, mark options={fill=white}, mark size=1.6pt, forget plot]
      coordinates {(0.5000,0)};

    % beta = 0.2
    \addplot[
      thick,
      blue!90!red,
      dashed,
      domain=0.2821:0.4999,
      samples=150,
      forget plot,
    ] {peakY(x,1,0.2)};
    \addplot[
      thick,
      blue!90!red,
      domain=0.5001:0.999,
      samples=150,
      forget plot,
    ] {peakY(x,1,0.2)};
    % jump of the fastest growers overall at q = 0.5000
    \addplot[blue!90!red, only marks, mark=*, mark size=1.6pt, forget plot]
      coordinates {(0.5000,0.8551)};
    \addplot[blue!90!red, only marks, mark=*, mark options={fill=white}, mark size=1.6pt, forget plot]
      coordinates {(0.5000,0)};

    % beta = 0.1
    \addplot[
      thick,
      blue,
      dashed,
      domain=0.2449:0.4999,
      samples=150,
      forget plot,
    ] {peakY(x,1,0.1)};
    \addplot[
      thick,
      blue,
      domain=0.5001:0.999,
      samples=150,
      forget plot,
    ] {peakY(x,1,0.1)};
    % jump of the fastest growers overall at q = 0.5000
    \addplot[blue, only marks, mark=*, mark size=1.6pt, forget plot]
      coordinates {(0.5000,0.9240)};
    \addplot[blue, only marks, mark=*, mark options={fill=white}, mark size=1.6pt, forget plot]
      coordinates {(0.5000,0)};

  \end{axis}
\end{tikzpicture}
\end{minipage}\hfill
\begin{minipage}{0.49\linewidth}
\centering
\begin{tikzpicture}
  \begin{axis}[
    axis lines=left,
    xlabel={$q$},
    ylabel={$\operatorname*{arg\,max}_{r_c}(d\ln Y_{cp}/dt)$},
    xmin=0,
    xmax=1,
    ymin=-0.04,
    ymax=1,
    xtick={0,0.2,0.4,0.6,0.8,1},
    ytick={0,0.2,0.4,0.6,0.8,1},
    width=\linewidth,
    height=0.85\linewidth,
    tick label style={/pgf/number format/fixed},
    legend style={
      at={(0.03,0.97)},
      anchor=north west,
      draw=none,
      fill=white,
      font=\small
    },
    declare function={
      % coefficients of the cubic, equation (eq:cubic)
      cthree(\q,\a,\b)=-\q*(\a*\q+1)*(2*\a*\q+\b+1);
      ctwo(\q,\a,\b)=(2*\a*\q+1)*(3*\a*\q^2-2*\a*\q+\b*\q-\b+\q);
      cone(\q,\a,\b)=\a*(1-\q)*(6*\a*\q^2-2*\a*\q+\b*\q-\b+\q+1);
      czero(\q,\a,\b)=\a*(1-\q)^2*(2*\a*\q-1);
      % trigonometric solution, equation (eq:argmaxY)
      peakYc(\A,\B,\C,\D)=
        -\B/(3*\A)
        +2*sqrt(-((3*\A*\C-\B^2)/(3*\A^2))/3)
        *cos(acos(max(-1,min(1,
          (3*((2*\B^3-9*\A*\B*\C+27*\A^2*\D)/(27*\A^3)))
          /(2*((3*\A*\C-\B^2)/(3*\A^2)))
          *sqrt(-3/((3*\A*\C-\B^2)/(3*\A^2))))))/3);
      peakY(\q,\a,\b)=peakYc(cthree(\q,\a,\b),ctwo(\q,\a,\b),
        cone(\q,\a,\b),czero(\q,\a,\b));
      % larger root of the quadratic, equation (eq:argmaxL)
      peakL(\q,\a,\b)=
        (
          (\a*\q-1)*(2*\a*\q+1-\a)+\a*\q+1-\a*\b*(1-\q)
          +sqrt(max(0,
            ((\a*\q-1)*(2*\a*\q+1-\a)+\a*\q+1-\a*\b*(1-\q))^2
            +4*(\a*\q+1)*(\a*\q+\b)*\a*(1-\q)*(\a*\q-1)))
        )
        /(2*(\a*\q+1)*(\a*\q+\b));
    },
  ]

    \addlegendimage{thick,blue!10!red}
    \addlegendentry{$\alpha=1,\ \beta=3$}
    \addlegendimage{thick,blue!30!red}
    \addlegendentry{$\alpha=1,\ \beta=2$}
    \addlegendimage{thick,blue!50!red}
    \addlegendentry{$\alpha=1,\ \beta=1$}
    \addlegendimage{thick,blue!70!red}
    \addlegendentry{$\alpha=1,\ \beta=0.5$}
    \addlegendimage{thick,blue!90!red}
    \addlegendentry{$\alpha=1,\ \beta=0.2$}
    \addlegendimage{thick,blue}
    \addlegendentry{$\alpha=1,\ \beta=0.1$}
    \addlegendimage{very thick,black}
    \addlegendentry{$r_c^{*}=0$}

    % least endowed economies grow fastest overall
    \addplot[
      very thick,
      black,
      domain=0:1,
      samples=2,
      forget plot,
    ] {0};

    % beta = 3
    \addplot[
      thick,
      blue!10!red,
      dashed,
      domain=0.8165:0.9999,
      samples=150,
      forget plot,
    ] {peakL(x,1,3)};

    % beta = 2
    \addplot[
      thick,
      blue!30!red,
      dashed,
      domain=0.7778:0.9999,
      samples=150,
      forget plot,
    ] {peakL(x,1,2)};

    % beta = 1
    \addplot[
      thick,
      blue!50!red,
      dashed,
      domain=0.7072:0.9999,
      samples=150,
      forget plot,
    ] {peakL(x,1,1)};

    % beta = 0.5
    \addplot[
      thick,
      blue!70!red,
      dashed,
      domain=0.6364:0.9999,
      samples=150,
      forget plot,
    ] {peakL(x,1,0.5)};

    % beta = 0.2
    \addplot[
      thick,
      blue!90!red,
      dashed,
      domain=0.5521:0.9999,
      samples=150,
      forget plot,
    ] {peakL(x,1,0.2)};

    % beta = 0.1
    \addplot[
      thick,
      blue,
      dashed,
      domain=0.4994:0.9999,
      samples=150,
      forget plot,
    ] {peakL(x,1,0.1)};

  \end{axis}
\end{tikzpicture}
\end{minipage}
\caption{Same as Figure~\ref{fig:A4} for $\alpha = 1$.}
\label{fig:A5}
\end{figure}

\subsubsection*{General functional forms}

For a more general production function, let $Y_{cp}=F(r_c,q_p)$, where capability requirements $q_p$ can differ across activities but remain constant over time. Average output is \\
\begin{equation}
\langle Y_c\rangle=\frac{1}{N(p)}\sum_p F(r_c,q_p).
\end{equation} \\
Using the same time units as in equation~\eqref{one_cap_avgY}, capability accumulation is given by \\
\begin{equation}
\frac{dr_c}{dt}=\gamma G(r_c)\langle Y_c\rangle,
\end{equation} \\
where $\gamma>0$, $G(r_c)>0$ for $r_c\in(0,1)$, and $G(1)=0$. We assume that $F$ and $G$ are continuous on $[0,1]$, positive and twice differentiable on $(0,1)$.\\

The maximum of capability accumulation need not coincide with the maximum of output growth. By the chain rule, the change in output and its growth rate are \\
\begin{align}
\frac{dY_{cp}}{dt}
&=\gamma G(r_c)\langle Y_c\rangle
\frac{\partial F(r_c,q_p)}{\partial r_c},\\
\frac{d\ln Y_{cp}}{dt}
&=\gamma G(r_c)\langle Y_c\rangle
\frac{1}{F(r_c,q_p)}
\frac{\partial F(r_c,q_p)}{\partial r_c}.
\label{eq:general-output-rates}
\end{align} \\
Where $\partial F(r_c,q_p)/\partial r_c>0$, these expressions are positive. We can therefore locate their interior maxima by differentiating their logarithms with respect to $r_c$. For the change in output, this gives \\
\begin{equation}
\frac{1}{G(r_c)}\frac{dG(r_c)}{dr_c}
+\frac{1}{\langle Y_c\rangle}
\frac{d\langle Y_c\rangle}{dr_c}
+\frac{\partial^2 F(r_c,q_p)/\partial r_c^2}
      {\partial F(r_c,q_p)/\partial r_c}
=0.
\label{eq:general-output-peak}
\end{equation} \\
For the growth rate of output, the corresponding condition is \\
\begin{equation}
\begin{aligned}
\frac{1}{G(r_c)}\frac{dG(r_c)}{dr_c}
&+\frac{1}{\langle Y_c\rangle}
\frac{d\langle Y_c\rangle}{dr_c}
+\frac{\partial^2 F(r_c,q_p)/\partial r_c^2}
      {\partial F(r_c,q_p)/\partial r_c}\\
&-\frac{1}{F(r_c,q_p)}
\frac{\partial F(r_c,q_p)}{\partial r_c}
=0.
\end{aligned}
\label{eq:general-log-output-peak}
\end{equation} \\
In both conditions, the response of average output to the capability stock is \\
\begin{equation}
\frac{d\langle Y_c\rangle}{dr_c}
=\frac{1}{N(p)}\sum_p
\frac{\partial F(r_c,q_p)}{\partial r_c}.
\end{equation} \\
A change in the left-hand side from positive to negative identifies a local maximum. To locate the global maximum, we compare output growth at these solutions and at the boundaries, using limits where necessary. These limits can diverge as $r_c$ approaches zero.\\

\subsection{Appendix V: Alternative Definitions of Sectoral Output}
\label{app:alternative_sectoral_growth}

Here we repeat the sectoral growth analysis using two alternative definitions of $Y_{cp}$. In the first, activity output is normalized by the economy's initial total output. In the second, activity output is expressed per capita, using the population at each endpoint. For each definition, we identify the three economies experiencing the largest increases in $Y_{cp}$ within each activity and examine their mean ECI as a function of activity complexity. \\

\begin{figure}[htbp]
    \centering
    \includegraphics[width=\linewidth]{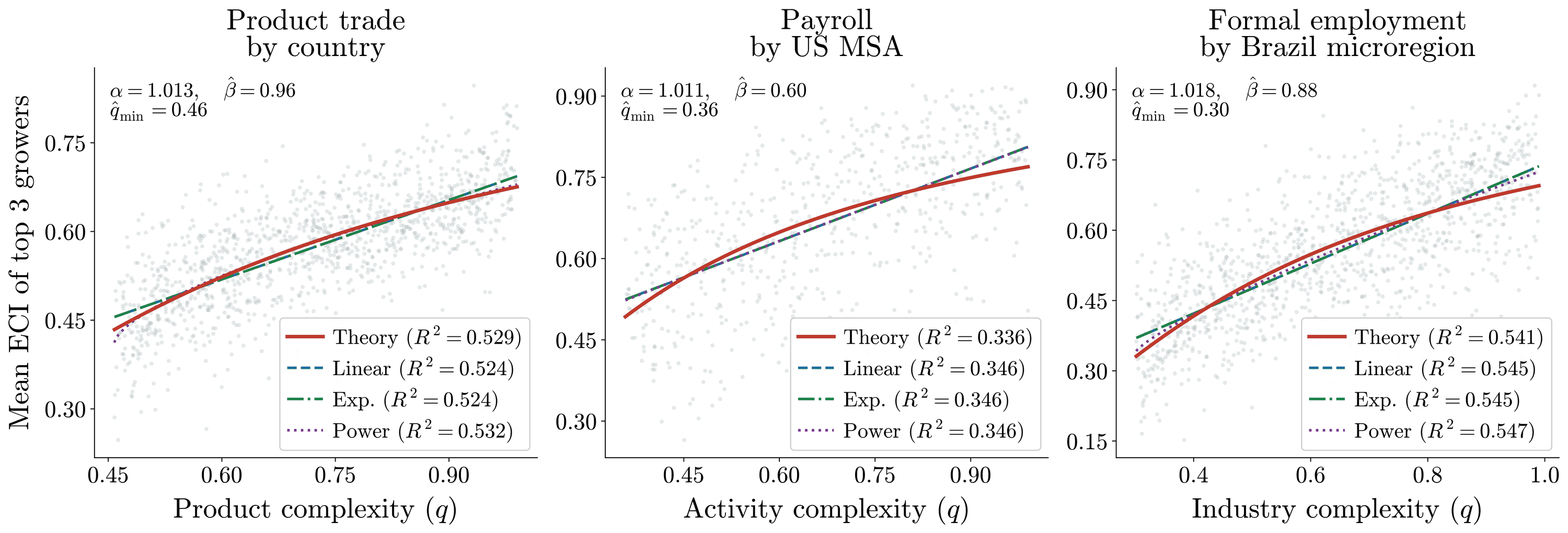}
    \caption{
        Sectoral growth with $Y_{cp}$ defined as activity output normalized by the economy's initial total output for international trade, US metropolitan areas, and Brazilian microregions. Each panel plots the mean initial ECI of the three economies experiencing the largest increases in $Y_{cp}$ against activity complexity. The data, estimation procedure, and curve specifications follow Figure~\ref{fig:oos_fastest_growers}, with $\beta$ and $q_{\min}$ re-estimated for this definition.
    }
    \label{fig:growth_initial_output}
\end{figure}

\begin{figure}[htbp]
    \centering
    \includegraphics[width=\linewidth]{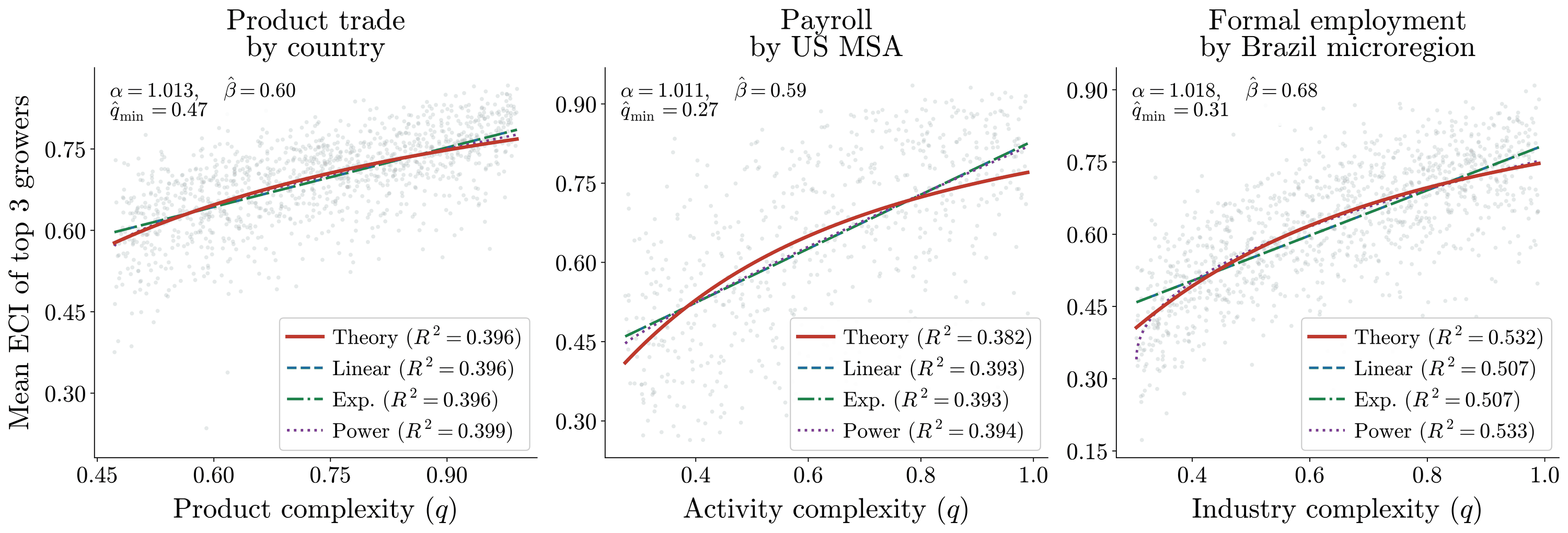}
    \caption{
     Sectoral growth with $Y_{cp}$ defined as activity output per capita for international trade, US metropolitan areas, and Brazilian microregions. Each panel plots the mean initial ECI of the three economies experiencing the largest increases in $Y_{cp}$ against activity complexity. The data, estimation procedure, and curve specifications follow Figure~\ref{fig:oos_fastest_growers}, with $\beta$ and $q_{\min}$ re-estimated for this definition.
    }
    \label{fig:growth_per_capita}
\end{figure}

Figure~\ref{fig:growth_initial_output} presents the results when activity output is normalized by initial total economy output, whereas Figure~\ref{fig:growth_per_capita} shows the results when activity output is expressed per capita. Across both specifications, we observe that for all three datasets the relationship is positive, as in the main analysis. The theory captures this relationship, and in each case its $R^2$ is close to the best benchmarks yield values of $0.532$, $0.346$, and $0.547$, respectively. The theory curve thus fits good even under alternate definitions of $Y_{cp}$ \\.

\end{document}